\documentclass[11pt]{article}

\usepackage[english]{babel}

\usepackage[letterpaper,top=2cm,bottom=2cm,left=3cm,right=3cm,marginparwidth=1.75cm]{geometry}
\usepackage[utf8]{inputenc}

\usepackage{fullpage}
\usepackage{hyperref,verbatim, enumerate, amsmath,amssymb,amsfonts,mathrsfs, amsthm, mathtools, bbm, algorithm, algpseudocode}
\usepackage[thmtools-compat]{keytheorems}
\usepackage[capitalize]{cleveref}
\usepackage{comment}
\usepackage{fullpage}
\usepackage{xcolor}
\usepackage{xurl}
\usepackage{lipsum}
\usepackage[font=itshape]{quoting}
\usepackage[bottom]{footmisc}
\usepackage{csquotes}
\usepackage{tikz}
\usepackage{array}
\usepackage{booktabs}
\usepackage{xltabular}
\usetikzlibrary{arrows.meta,positioning,calc}
\usepackage[normalem]{ulem}
\usepackage{paralist}
\usetikzlibrary{positioning,arrows.meta,matrix}
\usepackage{blindtext}
\usepackage{enumitem}
\usepackage{caption}
\usepackage{placeins}
\usepackage{graphicx}

\usepackage{fancybox}
\usepackage{hyperref}

\usepackage{amsmath,amsfonts,amsthm,amssymb,xcolor}
\usepackage{bm} %
\usepackage{bbm}

\usepackage{nicefrac}

\usepackage{float}

\newtheorem{theorem}{Theorem}[section]
\newtheorem{corollary}[theorem]{Corollary}
\newtheorem{lemma}[theorem]{Lemma}
\newtheorem{observation}[theorem]{Observation}

\newtheorem{claim}[theorem]{Claim}

\theoremstyle{definition}
\newtheorem{problem}[theorem]{Problem}
\theoremstyle{plain}

\newtheorem*{theorem*}{Theorem}
\newtheorem*{corollary*}{Corollary}
\newtheorem*{conjecture*}{Conjecture}
\newtheorem*{lemma*}{Lemma}
\newtheorem*{thm*}{Theorem}
\newtheorem*{prop*}{Proposition}
\newtheorem*{obs*}{Observation}

\newtheorem*{remark*}{Remark}
\newtheorem*{rec*}{Recommendation}

\theoremstyle{definition}
\newtheorem{definition}[theorem]{Definition}
\theoremstyle{plain}

\theoremstyle{definition}
\newtheorem*{definition*}{Definition}
\theoremstyle{plain}

\addtotheoremhook[definition]{postfoot}{\AddToHookNext{para/begin}{\OmitIndent}}
\addtotheoremhook[definition*]{postfoot}{\AddToHookNext{para/begin}{\OmitIndent}}

\newcommand{\newclass}[2]{\newcommand{#1}{{\text{\upshape\sffamily #2}}\xspace}}
\renewcommand{\P}{{\text{\upshape\sffamily P}}\xspace}
\newclass{\ZPP}{ZPP}
\newclass{\RL}{RL}
\newclass{\ZPTCo}{ZP-TC$^1$}
\newclass{\ipFL}{inplaceFL}
\newclass{\ipFCL}{inplaceFCL}
\newclass{\SCL}{SearchCL}
\newclass{\ipSCL}{inplaceSearchCL}
\newclass{\NP}{NP}
\newclass{\FNP}{FNP}
\newclass{\FZPP}{FZPP}
\newclass{\coNP}{coNP}
\newclass{\BPP}{BPP}
\newclass{\Logspace}{L}
\newclass{\NL}{NL}
\newclass{\coNL}{coNL}
\newclass{\UL}{UL}
\newclass{\Ppoly}{P/poly}
\newclass{\coUL}{coUL}
\newclass{\BPL}{BPL}
\newclass{\PL}{PL}
\newclass{\prBPL}{prBPL}
\newclass{\PSPACE}{PSPACE}
\newclass{\EXP}{EXP}
\newclass{\EXPSPACE}{EXPSPACE}
\newclass{\TIME}{TIME}
\newclass{\SPACE}{SPACE}
\newclass{\NSPACE}{NSPACE}
\newclass{\SC}{SC}
\newclass{\coNSPACE}{coNSPACE}
\newclass{\BPSPACE}{BPSPACE}
\newclass{\TFNP}{TFNP}
\newclass{\FP}{FP}
\newclass{\SCSPACE}{SearchCSPACE}
\newclass{\RSCSPACE}{RestorableSearchCSPACE}
\newclass{\RSCL}{RestorableSearchCL}
\newclass{\NC}{NC}
\newclass{\NCo}{NC$^1$}
\newclass{\ACz}{AC$^0$}
\newclass{\ACo}{AC$^1$}
\newclass{\SACo}{SAC$^1$}
\newclass{\TC}{TC}
\newclass{\TCz}{TC$^0$}
\newclass{\LUTCo}{L-uniform TC$^1$}
\newclass{\TCo}{TC$^1$}
\newclass{\NCt}{NC$^2$}
\newclass{\RNC}{RNC}
\newclass{\PSDNC}{pseudo-deterministic NC}
\newclass{\NUSPL}{non-uniform SPL}
\newclass{\RNCt}{RNC$^2$}
\newclass{\RNCtt}{RNC$^3$}
\newclass{\RNCo}{RNC$^1$}
\newclass{\QNC}{Quasi-NC}
\newclass{\VP}{VP}
\newclass{\CL}{CL}
\newclass{\CLP}{CLP}
\newclass{\CSPACE}{CSPACE}

\newclass{\GapL}{GapL}
\newclass{\SBPP}{SearchBPP}
\newclass{\psdBPP}{PseudodeterministicBPP}
\newclass{\StCL}{StableCL}
\newclass{\MATCH}{MATCH}
\newclass{\DET}{DET}
\newclass{\LOSSY}{LOSSY}
\newclass{\LOSSYNC}{LOSSY[NC]}
\newclass{\LOSSYC}{LOSSY[$\mathcal{C}$]}

\newclass{\ZPC}{ZP-$\mathcal{C}$}
\newclass{\ZPNC}{ZPNC}
\newclass{\FL}{FL}
\newclass{\SRSCL}{StableRestorableSearchCL}

\newcommand{\size}{\mathsf{size}}
\newcommand{\depth}{\mathsf{depth}}
\newcommand{\work}{\mathsf{work}}
\newcommand{\ttime}{\mathsf{time}}

\newcommand{\poly}{\mathrm{poly}}
\DeclareMathOperator{\polylog}{polylog}

\newcommand{\PO}{{\mathcal{O}_P}}
\newcommand{\upd}{\mathsf{UNPRED}}
\newcommand{\updc}[1]{\mathsf{UNPRED}^{#1}}

\newcommand{\updo}{\mathsf{UNPRED}^{\mathcal{O}_P}}

\newcommand{\nbupdc}[1]{\mathsf{NextBitUNPRED}^{#1}}

\newcommand{\lossyc}{\mathsf{LossyCoding}}
\newcommand{\weakex}{\mathsf{WeakExcess}}

\newcommand{\blossyc}{\mathsf{BlockLossyCoding}}
\newcommand{\bipmatch}{\mathsf{BipMATCH}}
\newcommand{\match}{\mathsf{MATCH}}
\newcommand{\lmi}{\mathsf{LinearMatroidIntersection}}
\newcommand{\graphspars}{\mathsf{GraphSparsification}}

\newcommand{\ind}{\mathsf{ind}}
\newcommand{\succs}{\text{ succeeds on }}
\newcommand{\adv}{\mathsf{adv}}
\newcommand{\succc}{\mathsf{succ}}

\newcommand{\nbupd}[3]{}
\newcommand{\str}{\mathsf{str}}

\newcommand{\lex}{\mathsf{lex}}
\newclass{\Comp}{Comp}
\newclass{\Decomp}{Decomp}

\newcommand{\iph}{\mathsf{input}\text{-}\mathsf{head}}

\newcommand{\lab}{\mathsf{lab}}
\newcommand{\fct}{\mathsf{FastCatTape}}
\newcommand{\ee}{\mathsf{EdgeExists}}

\newcommand{\start}{\mathsf{start}}

\newcommand{\nonsol}{\mathsf{NonSol}}

\newcommand{\ext}{\mathsf{ext}}
\newcommand{\advice}{\mathsf{advice}}

\newcommand{\len}{\mathsf{len}}

\newcommand{\load}{\mathsf{load}}
\newcommand{\compute}{\mathsf{Compute}}
\newcommand{\restore}{\mathsf{Restore}}

\newcommand{\eps}{\epsilon}

\newcommand\restr[2]{{%
  \left.\kern-\nulldelimiterspace %
  #1 %
  \vphantom{\big|} %
  \right|_{#2} %
  }}

\newcommand{\e}{\mathop{\mathbb{E}}}

\newcommand{\indeg}{\text{in-degree}}
\newcommand{\outdeg}{\text{out-degree}}

\newcommand{\emphdef}[1]{{\sf {#1}}}

\newenvironment{fminipage}%
  {\begin{Sbox}\begin{minipage}}%
  {\end{minipage}\end{Sbox}\fbox{\TheSbox}}

\newif\ifblind

\makeatletter
\newcommand{\formatcontents}{%
  \renewcommand{\@pnumwidth}{4em}%
  \renewcommand{\@tocrmarg}{4.9em}%
  \patchcmd{\l@section}{1.0em}{0.5em}{}{}%
}
\makeatother

\newcommand{\formatbibliography}{%
  \DeclareUrlCommand{\Doi}{\urlstyle{sf}}%
  \renewcommand{\path}[1]{\small\Doi{##1}}%
  \renewcommand{\url}[1]{\href{##1}{\small\Doi{##1}}}%
}

\allowdisplaybreaks[4]

\blindtrue
\title{The Complexity of Unpredictability}

\ifblind

\author{Aryan Agarwala\\Max-Planck-Institut f\"{u}r Informatik \\ \texttt{aryan@agarwalas.in} \and 
Dani Dorfman \\Max-Planck-Institut f\"{u}r Informatik \\ \texttt{ddorfman@mpi-inf.mpg.de}}
\fi

\date{}
\begin{document}

\setlength{\abovedisplayskip}{5pt}
\setlength{\belowdisplayskip}{5pt}

\pagenumbering{gobble}

\maketitle

\begin{abstract}
In this work, we initiate the study of middle-bit predictors: functions
that take as input a string with exactly one missing bit and attempt to
predict the value of that bit. These are closely related to next-bit
predictors, a classical object in pseudorandomness.

We study the deterministic complexity of the problem of computing an
unpredictable distribution ($\mathsf{UNPRED}$): given as input a list of
middle-bit predictors in circuit form and a parameter $\epsilon > 0$,
output a multiset of strings such that each predictor succeeds with
probability $< 1/2 + \epsilon$ on a uniformly sampled string from this
multiset. We denote the version in which the predictors are restricted
to be $\mathsf{TC}^1$ circuits by
$\mathsf{UNPRED}^{\mathsf{TC}^1}$.

We show that $\mathsf{UNPRED}$ offers a unified route towards resolving
several major open problems in \emph{catalytic computation}. The most
important of these is understanding the relationship between catalytic
logspace ($\mathsf{CL}$ and $\mathsf{SearchCL}$) and deterministic
polynomial time ($\mathsf{P}$ and $\mathsf{FP}$). It is widely believed
that $\mathsf{SearchCL} \subseteq \mathsf{FP}$. We prove the following
implications:
\[
\mathsf{UNPRED} \in \mathsf{FP}
\implies
\mathsf{SearchCL} \subseteq \mathsf{FP}
\implies
\mathsf{UNPRED}^{\mathsf{TC}^1} \in \mathsf{FP}.
\]
The questions of whether $\mathsf{UNPRED} \in \mathsf{FP}$ and whether
$\mathsf{SearchCL} \subseteq \mathsf{FP}$ are therefore almost
equivalent, up to the restriction to $\mathsf{TC}^1$ circuits.

Furthermore, we show that any deterministic polynomial-time algorithm
for $\mathsf{UNPRED}^{\mathsf{TC}^1}$ must exploit properties of Boolean
$\mathsf{TC}^1$ circuits, either in the algorithm itself or in its
analysis. We are not aware of any such algorithm for any natural
problem. Thus, a deterministic polynomial-time algorithm for
$\mathsf{UNPRED}^{\mathsf{TC}^1}$ would not only make progress in
catalytic computation but would also advance
our understanding of Boolean circuits.

Together, our results provide strong evidence that middle-bit
predictors could be the key to unlocking several major open problems
in catalytic computation, pseudorandomness, and Boolean circuit
complexity.
\end{abstract}
\pagebreak

\pagenumbering{roman}
\begingroup
\setcounter{tocdepth}{2}
\hypersetup{linktoc=all,hidelinks}
\formatcontents
\hyphenpenalty=10000
\tableofcontents
\endgroup
\clearpage

\pagenumbering{arabic}

\section{Introduction}

In this work, we initiate the study of middle-bit predictors. A middle-bit predictor over $n$-bit strings at an index $i \in [n]$ is a function $P:\{0, 1\}^{n-1} \rightarrow \{0, 1\}$. $P$ is said to \emph{succeed} on a string $s \in \{0, 1\}^n$ if, given as input the entirety of $s$ except its $i^{th}$ bit, $P$ guesses the $i^{th}$ bit of $s$ correctly. Middle-bit predictors are closely related to next-bit predictors, which are a classical object in pseudorandomness, going back to the work of Blum and Micali~\cite{BlumMicali84} and Yao~\cite{Yao82}. 

Understanding the strength of middle-bit predictors was explicitly posed as an open problem by Pyne and Tell in their recent survey on derandomizing low-space algorithms~\cite{PyneTell26}. The recent independent work of Doron, Moshkovitz, Oh, and Zuckerman~\cite{DoronMoshkovitzOhZuckerman26} also studies middle-bit predictors in a slightly different setting; we describe their work in more detail later.

In this work, we define the problem $\upd$: given as input a list of $m$ middle-bit predictors\footnote{Each predictor is given as a Boolean circuit, along with the index that it is predicting as an integer from $1$ to $n$.} over $\{0, 1\}^n$ and a parameter $0 < \epsilon < 1/2$, output a multiset $S \subseteq \{0, 1\}^n$ such that each of these $m$ predictors succeeds with probability less than $1/2 + \epsilon$ on a string $s$ sampled uniformly at random from $S$. Observe that a random multiset $S$ of size $\Theta(\log m/\epsilon^2)$ is a valid output with constant probability; thus, $\upd$ is easy with randomized algorithms, and in this work we seek deterministic algorithms.

The motivation for our work comes from two separate but deeply related fields: \emph{catalytic computation} and \emph{pseudorandomness.}

\subsection{Catalytic Computation}

The framework of catalytic computation was introduced by Buhrman, Cleve, Kouck\'{y}, Loff, and Speelman \cite{BuhrmanCleveKouckyLoffSpeelman14} in order to study the power of space that contains important but irrelevant information. In this model a space-bounded Turing machine is given access to a much longer read-write `catalytic' tape. The catalytic tape initially contains an arbitrary string chosen by an adversary, and upon termination it must contain \emph{exactly} the same string.

Catalytic computation has recently been at the center of several breakthroughs in space-bounded complexity. Examples include the Cook-Mertz algorithm for the \emph{tree evaluation problem}~\cite{CookMertz24}, Williams' simulation of time using low space \cite{Williams25}, and the OpenAI preprint claiming a proof that $\Logspace = \RL = \BPL$ \cite{OpenAI26b} \footnote{At this point it is unclear whether there is a human who understands OpenAI's proof of $\Logspace = \RL = \BPL$. However, catalytic computation seems to feature prominently in the preprint.}. Furthermore, techniques from catalytic computation have also recently been applied to \emph{hardness versus randomness} results in low-space derandomization~\cite{DoronPyneTell24, LiPyneTell24, DoronPyneTellWilliams25, PyneTell26a}. We refer interested readers to the survey of Pyne and Tell~\cite{PyneTell26}.  Thus, obtaining a complete understanding of catalytic computation could be crucial to future progress in space-bounded complexity theory.

\cite{BuhrmanCleveKouckyLoffSpeelman14} introduced the class \emph{catalytic logspace} ($\CL$), in which a logspace machine is augmented with a polynomial-length catalytic tape. Surprisingly, they gave strong evidence that $\CL$ is significantly more powerful than ordinary logspace ($\Logspace$), by showing that a $\CL$ machine can simulate log-depth threshold circuits ($\TCo$). $\TCo$ is believed to be stronger than $\Logspace$, and in particular contains nondeterministic logspace ($\NL$). On the other hand, while $\Logspace$ is contained in deterministic polynomial time ($\P$), \cite{BuhrmanCleveKouckyLoffSpeelman14} were unable to prove that $\CL \subseteq \P$. Instead, they showed that any catalytic logspace algorithm can be simulated by zero-error \emph{randomized} polynomial-time algorithms ($\ZPP$).

\begin{theorem}[\cite{BuhrmanCleveKouckyLoffSpeelman14}]
    \[\TCo \subseteq \CL \subseteq \ZPP.\]
\end{theorem}
The following problem naturally arises from their work.

\begin{problem}[\cite{Koucky16, Mertz23, CookLiMertzPyne25}]
\label{prob:CL-in-P}
$$\text{Is }\CL \subseteq \P?$$
\end{problem}

Standard derandomization assumptions~\cite{ImpagliazzoWigderson97} would imply that $\P = \ZPP$, and thus $\CL \subseteq \P$. Therefore, Problem~\ref{prob:CL-in-P} should be viewed as a derandomization problem. We show that the problem of designing a deterministic polynomial-time ($\FP$) algorithm for $\upd$ is almost equivalent to the search analogue of Problem~\ref{prob:CL-in-P} ($\SCL \subseteq \FP$).

\subsection{Our Results: $\upd$}
\subsubsection{$\SCL \subseteq \FP$ and $\upd \in \FP$ Are Almost Equivalent}
Our first result concerns an oracle version of $\upd$, which we call $\updo$. In this model, the predictors are not explicitly given to the algorithm as circuits, but can only be queried as black-box oracles (see Section~\ref{sec:oracle-model} for the catalytic oracle model). We show that there exists a catalytic logspace algorithm for $\updo$ in this oracle model.
\begin{theorem}[Theorem~\ref{thm:oracle-binunpred-in-searchcl}]
\label{intro-thm:updo-in-scl}
    \[\updo \in \SCL^{\PO}.\]
\end{theorem}

A corollary of this theorem is that the version of $\upd$ where the predictors are explicitly given as input in the form of $\TCo$ circuits can be solved in $\SCL$. We refer to this problem as $\upd^{\TCo}$.
\begin{corollary}[Corollary~\ref{cor:tco-unpred-in-searchcl}]
\label{intro-cor:upd-tco-in-scl}
    \[\updc{\TCo} \in \SCL.\]
\end{corollary}

This corollary follows by simply replacing each oracle query in the algorithm of Theorem~\ref{intro-thm:updo-in-scl} with an explicit evaluation of the $\TCo$ circuit of the predictor, using the $\CL$ algorithm of Buhrman, Cleve, Kouck\'{y}, Loff, and Speelman~\cite{BuhrmanCleveKouckyLoffSpeelman14} to do so.

Corollary~\ref{intro-cor:upd-tco-in-scl} shows that a deterministic polynomial-time algorithm for $\updc{\TCo}$ is necessary for establishing $\SCL \subseteq \FP$. The following theorem shows that a deterministic polynomial-time algorithm for $\upd$ is sufficient.

\begin{theorem}
\label{intro-thm:scl-reduces-to-upd}
    \[\SCL \leq_{\P} \upd.\]
\end{theorem}

Thus, Corollary~\ref{intro-cor:upd-tco-in-scl} and Theorem~\ref{intro-thm:scl-reduces-to-upd} together place the search version of the $\CL$ versus $\P$ problem between two versions of $\upd$:
$$\upd \in \FP \implies \SCL \subseteq \FP \implies \updc{\TCo} \in \FP.$$

It is very surprising to us that this weak equivalence is possible, since middle-bit predictors have no superficial connection to the catalytic condition of restoring space.

Of course, there is a gap between the two ends of this chain: one uses $\updc{\TCo}$, while the other uses $\upd$. This gap can be closed in two ways. First, one can define a uniform version of $\upd$ where the predictors are given as input in the form of polynomial-time Turing machines with certain complicated properties. One can then prove both Corollary~\ref{intro-cor:upd-tco-in-scl} and Theorem~\ref{intro-thm:scl-reduces-to-upd} using this complicated version of $\upd$, eliminating the gap entirely. An older draft of this paper indeed followed this approach. However, we concluded that this approach is overly complicated, and obfuscates the key technical arguments that make this reduction possible \footnote{If a reader is interested in this older draft, we would be happy to share it.}. In Section~\ref{sec:uniform-unpred}, we define this version and state the corresponding theorems without proof. Second, we can actually show that all problems \emph{currently} known to be in $\CL$ (or $\SCL$) reduce directly to $\updc{\TCo}$. We will do this in Sections~\ref{sec:nbp-and-ess-est} and~\ref{sec:block-lossy}.

Our takeaway from this set of results is straightforward: any future work on resolving the search analogue of Problem~\ref{prob:CL-in-P} can ignore the mysterious catalytic condition of restoring space, and simply try to design a deterministic polynomial-time algorithm for $\upd$. Such an algorithm would immediately imply $\SCL \subseteq \FP$ and thus $\CL \subseteq \P$. A natural starting point for this would be to design a deterministic polynomial-time algorithm for $\updc{\TCo}$, which is a necessary condition.

\subsubsection{A Deterministic Polynomial-Time Algorithm for $\upd$ Must Exploit Properties of Boolean Circuits}
By Theorem~\ref{intro-thm:scl-reduces-to-upd}, an $\FP$ algorithm for $\upd$ would suffice to prove that $\SCL \subseteq \FP$ and $\CL \subseteq \P$. The most direct approach towards this would be to design a deterministic polynomial-time oracle algorithm for $\updo$, similar to the catalytic logspace algorithm of Theorem~\ref{intro-thm:updo-in-scl}. Replacing each oracle query in such an algorithm with circuit evaluations would imply that $\upd \in \FP$ and thus $\CL \subseteq \P$. Unfortunately, the following theorem shows that this is unconditionally impossible. 
\begin{theorem}[Corollary~\ref{cor:updob-algo-exp-time}]
\label{intro-thm:updo-not-in-p}
Every deterministic oracle algorithm for $\updo$, on instances with $n$ predictors over $\{0, 1\}^n$ and $\epsilon = 1/10$, has worst-case running time $\exp(\Omega(\sqrt{n}))$. In particular,
    \[\updo \notin \P^{\PO}.\]
\end{theorem}
This shows that any deterministic algorithm for $\upd$ or $\updc{\TCo}$ which treats the input predictors purely as functions must take exponential time in the worst case. Therefore, any $\FP$ algorithm would have to exploit properties of Boolean circuits, either in the algorithm itself or in its analysis. We are not aware of any deterministic polynomial-time algorithm for any comparable problem which is able to exploit such properties of large circuit classes (as large as $\TCo$). Therefore, a deterministic polynomial-time algorithm for $\upd$ or $\updc{\TCo}$ would not only make substantial progress on (or completely resolve) Problem~\ref{prob:CL-in-P}, but would also significantly advance our understanding of Boolean circuits.

A more pessimistic interpretation of this result would be that this oracle barrier is too strong, and thus $\SCL \subseteq \FP$ is not solvable with current techniques. 

Both interpretations solidify $\updc{\TCo}$ as the eventual starting point towards attacking Problem~\ref{prob:CL-in-P}.

\subsection{Catalytic Computation: Comparison with Past Work and Connections to Other Problems}
\subsubsection{Comparison with Past Work}
There are two papers that we would like to compare our work with:
\begin{itemize}
    \item Cook, Ghentiyala, Mertz, Pyne, and Sheffield~\cite{CookGhentiyalaMertzPyneSheffield25} gave an oracle $O$ relative to which $\CL^O \nsubseteq \P^O$, ruling out relativizing proofs of $\CL \subseteq \P$. Theorems~\ref{intro-thm:updo-in-scl} and~\ref{intro-thm:updo-not-in-p} in our work can also be interpreted as a relativization barrier. However, we believe that ours is a more \emph{meaningful} barrier, since it identifies a precise problem, $\upd$, that requires substantially different techniques to be solved in $\CL$ and in $\P$. More importantly, our barrier gives a precise way forward: a polynomial-time algorithm for $\upd$ with explicitly given predictor circuits would imply $\CL \subseteq \P$.

    \item Cook, Li, Mertz, and Pyne~\cite{CookLiMertzPyne25} show that $\SCL$ reduces in polynomial time to $\lossyc$: the problem of finding a string that a compressor and decompressor circuit fail to accurately reconstruct (see Definition~\ref{def:lossy-coding}). Since $\upd \leq_{\P} \lossyc$ (by Lemma~\ref{lem:unpred-compression}), Theorem~\ref{intro-thm:scl-reduces-to-upd} is a strengthening of their result. To see why this strengthening matters, imagine a universe where $\CL$ can evaluate any circuit. That is, $\P \subseteq \CL$. Then Theorems~\ref{intro-thm:updo-in-scl} and~\ref{intro-thm:scl-reduces-to-upd} would imply that $\upd$ is $\SCL$-complete under polynomial-time reductions. In contrast, even in this extreme case, it is unclear whether $\lossyc \in \SCL$. More generally, for any circuit class $\mathcal{C}$ such that $\mathcal{C} \subseteq \CL$, Theorem~\ref{intro-thm:updo-in-scl} implies that $\updc{\mathcal{C}} \in \SCL$, whereas this is entirely unclear for $\lossyc$.  Thus, $\upd$ more accurately captures the strength of $\CL$.
\end{itemize}

\subsubsection{$\CL$ versus $\NC$}
As stated earlier, we view $\CL \subseteq \P$ (Problem~\ref{prob:CL-in-P}) as a derandomization problem. Such a viewpoint suggests that $\CL$ may itself be able to simulate certain randomized algorithms deterministically. Pyne~\cite{Pyne24} and Cook, Li, Mertz, and Pyne~\cite{CookLiMertzPyne25} demonstrated this derandomization ability of $\CL$ through a technique called \emph{compress-or-random}. At a high level, the idea is that a $\CL$ algorithm could use its catalytic tape as the random string for a zero-error randomized algorithm. If the catalytic tape is not a good random string for the randomized algorithm, then the $\CL$ algorithm could compress the string on the catalytic tape and create free space. This win-win argument was used to develop an understanding of the structure of catalytic computing~\cite{CookLiMertzPyne25, KouckyMertzPyneSami25, Pyne24}. However, these works did not place a `new' problem in $\CL$. That is, at this point, no problem was known to be in $\CL$ beyond the original $\TCo$ inclusion of \cite{BuhrmanCleveKouckyLoffSpeelman14}. $\TCo$ is contained in the class of problems solvable in parallel with polylogarithmic depth and polynomial work ($\NC$). This led to the following question.

\begin{problem}[\cite{Koucky16, Mertz23}]
\label{prob:CL-in-NC}
$$\text{Is }\CL \subseteq \NC?$$
\end{problem}

Unlike Problem~\ref{prob:CL-in-P}, there was no reason to believe that this is true, except for the fact that all problems known to be in $\CL$ were contained in $\NC$. Agarwala and Mertz~\cite{AgarwalaMertz25} broke this barrier by showing that bipartite maximum matching can be solved in $\CL$. Agarwala, Alekseev, and Vinciguerra~\cite{AgarwalaAlekseevVinciguerra26} then showed that linear matroid intersection can be solved in $\CL$. Arvind, Chakraborty, and Datta~\cite{ArvindChakrabortyDatta26} used the same techniques to show that the exact matching problem on planar graphs can be solved in $\CL$. Chakraborty, Datta, Kusre, Mukhopadhyay, and Sinhababu~\cite{ChakrabortyDattaKusreMukhopadhyaySinhababu26} generalized the bipartite matching algorithm of~\cite{AgarwalaMertz25} and showed that the maximum matching problem on general graphs can also be solved in $\CL$. Finally, Marcussen, Pyne, and Rubinfeld~\cite{MarcussenPyneRubinfeld26} recently showed that the graph spectral sparsification problem can be solved in $\CL$. 

At the time, none of these problems were known to be in $\NC$. Thus, these papers presented their algorithms as barriers towards proving that $\CL \subseteq \NC$ (Problem~\ref{prob:CL-in-NC}). Any proof of $\CL \subseteq \NC$ would immediately imply that all of these problems are contained in $\NC$. For many of these problems, this was a longstanding open question. In a recent breakthrough result, Chatterjee, Ghosh, Gurjar, Raj, and Thierauf~\cite{ChatterjeeGhoshGurjarRajThierauf26} showed that bipartite matching and linear matroid intersection can be solved in $\NC$, destroying the barriers of~\cite{AgarwalaMertz25, AgarwalaAlekseevVinciguerra26}. Unfortunately, their algorithm also did not give any insight about how to tackle Problem~\ref{prob:CL-in-NC}. This example shows that barriers to Problems~\ref{prob:CL-in-P} and~\ref{prob:CL-in-NC} based on individual graph problems can be fickle, and if solved do not necessarily provide any insight into the strength of $\CL$. Instead, we will give a barrier to Problem~\ref{prob:CL-in-NC} which we believe is both stronger and more meaningful than these graph problems. This barrier will be based on a subproblem of $\upd$. 

\subsubsection{Search Versus Decision for Catalytic Logspace}
\label{sec-intro:stablecl}
Many of our theorems concern the search analogues of Problem~\ref{prob:CL-in-P} and Problem~\ref{prob:CL-in-NC}. This naturally raises the following question: are the search and decision versions of catalytic logspace equivalent? For example, if one were to obtain the very strong result that $\Logspace = \CL$, would it then imply that $\FL = \SCL$? Formally,
\begin{problem}
\label{prob-intro:FLCLvsSCL}
    $$\text{Is }\FL^{\CL} = \SCL?$$
\end{problem}
Readers familiar with complexity theory may notice that this is similar to the question of whether $\SBPP$ equals $\FP^\BPP$. It is known that $\FP^{\BPP}$ is equal to the class of search problems solvable by polynomial time \emph{pseudodeterministic} algorithms~\cite{GatGoldwasser11}. These are randomized algorithms which, for any input, output the \emph{same} solution with high probability. These algorithms remove the additional instability in the output of search problems caused by randomness.

In our work, we define $\StCL$ to be the class of search problems solvable by catalytic logspace algorithms which, for any input, output the \emph{same} solution for all initial catalytic tape configurations. Thus, a stable algorithm removes the additional instability in the output of search problems caused by the catalytic tape. We prove that $\StCL$ has the same relationship with $\FL$ and $\CL$ that $\psdBPP$ has with $\FP$ and $\BPP$.

\begin{theorem}[Corollary~\ref{cor:flcl-equals-stcl}]
    \label{intro-thm:flcl-equals-stcl}
$$\FL^{\CL} = \StCL.$$
\end{theorem}

Therefore, Problem~\ref{prob-intro:FLCLvsSCL} is equivalent to the following.

\begin{problem}
\label{prob-intro:stablesearchcl}
    $$\text{Is }\StCL = \SCL?$$
\end{problem}

The $\SCL$ algorithms for bipartite matching~\cite{AgarwalaMertz25}, general matching~\cite{ChakrabortyDattaKusreMukhopadhyaySinhababu26}, linear matroid intersection~\cite{AgarwalaAlekseevVinciguerra26}, and graph sparsification~\cite{MarcussenPyneRubinfeld26} are \emph{not} stable, and it is unclear whether these algorithms can be made stable.

We prove in Sections~\ref{sec:nbp-and-ess-est} and~\ref{sec:block-lossy} that all four of these $\SCL$ algorithms can be interpreted as $\TCo$ reductions to $\upd^{\TCo}$. Therefore, the instability in these algorithms is due to the fact that our $\SCL$ algorithm for $\upd^{\TCo}$ (Corollary~\ref{intro-cor:upd-tco-in-scl}) is not stable.

It turns out that there is a good reason for this. Our $\SCL$ algorithm for $\upd^{\TCo}$ has a special property: there is a point in the algorithm where its entire output is explicitly written on the catalytic tape. We call such algorithms \emph{restorable}. In Corollary~\ref{cor:restorable-algos-cant-be-stable}, we show \emph{unconditionally} that $\updc{\TCo}$ does not have a restorable catalytic logspace algorithm which is additionally stable. Thus, a $\StCL$ algorithm for $\updc{\TCo}$ (which is a necessary condition for resolving Problem~\ref{prob-intro:stablesearchcl} positively) would require substantially new ideas: in particular, it must not be restorable. We emphasize that restorability is a natural by-product of the compress-or-random technique, and is not an artificial feature of our algorithm. We elaborate significantly in Section~\ref{sec:stablecl}.

Thus, $\upd$ provides a unified route towards resolving three major questions related to catalytic computation: $\P$ versus $\CL$ (Problem~\ref{prob:CL-in-P}), $\NC$ versus $\CL$ (Problem~\ref{prob:CL-in-NC}), and $\SCL$ versus $\StCL$ (Problem~\ref{prob-intro:stablesearchcl}). 

\subsection{Connection to Pseudorandomness}
\label{sec:intro-conn-to-psd}

A \emph{next-bit predictor} over $n$-bit strings at an index $i \in [n]$ is a function $P:\{0, 1\}^{i-1} \rightarrow \{0, 1\}$. $P$ is said to \emph{succeed} on a string $s \in \{0, 1\}^n$ if, given as input the length $i-1$ prefix of $s$, it produces as output the $i^{th}$ bit of $s$. Next-bit predictors are a classical object in pseudorandomness, going back to the work of Blum and Micali~\cite{BlumMicali84} and Yao~\cite{Yao82}. We refer to the version of $\upd$ where the input consists of next-bit predictors, instead of middle-bit predictors, as $\nbupdc{}$. Observe that every next-bit predictor is also a middle-bit predictor which simply ignores the suffix of its input after the index $i$. Thus, middle-bit predictors are a stronger object than next-bit predictors.

A central tool in the hardness-to-randomness paradigm~\cite{Yao82,BlumMicali84,NisanWigderson94,ImpagliazzoWigderson97}, which aims to show that $\BPP = \P$, is a \emph{distinguish-to-predict} (D2P) transformation. Yao's classical distinguish-to-predict lemma~\cite{Yao82} constructs, for any randomized polynomial-time algorithm $\mathcal{A}$ and input $x \in \{0, 1\}^n$, a list of next-bit predictors $\mathcal{P}$ and a parameter $\epsilon \geq 1/\poly(|x|)$ such that any polynomial-size solution to $\nbupdc{}$ on $(\mathcal{P},\epsilon)$ would suffice to derandomize $\mathcal{A}$ on input~$x$. This argument \emph{almost} reduces the problem of derandomizing $\mathcal{A}$ to $\nbupdc{}$, except for the fact that Yao's argument is \emph{probabilistic}, and a naive derandomization of his argument results in exponentially many next-bit predictors. Li, Pyne, and Tell~\cite{LiPyneTell24} provide an alternative argument that results in only polynomially many next-bit predictors; however, their construction of these next-bit predictors is non-uniform.

It makes intuitive sense that this reduction, which we call \emph{next-bit D2P}, is so difficult: Goldreich and Wigderson \cite{GoldreichWigderson00} and Li, Pyne, and Tell~\cite{LiPyneTell24} showed that $\nbupdc{}$ has a deterministic polynomial-time algorithm \footnote{In fact, they gave deterministic polynomial-time algorithms for the oracle version $\nbupdc{\PO}$.}. Thus, an efficient next-bit D2P algorithm would simply derandomize $\mathcal{A}$. Li, Pyne, and Tell~\cite{LiPyneTell24} formalize this framework of derandomization: for a fixed algorithm $\mathcal{A}$, given as input $x \in \{0, 1\}^*$, first solve next-bit D2P, and then solve $\nbupdc{}$. In this framework, however, the second step is easy: the deterministic polynomial-time algorithms for $\nbupdc{}$ in~\cite{GoldreichWigderson00, LiPyneTell24} are fairly straightforward and do not use any technical machinery. Thus, all of the hardness lies in the first step: next-bit D2P.

We believe that middle-bit predictors may offer a more promising route towards derandomization. Consider the same framework as above, but with middle-bit predictors instead of next-bit predictors. This would reduce derandomization to two problems: \emph{middle-bit D2P} and $\upd$. Since middle-bit predictors are more powerful than next-bit predictors, $\upd$ should be harder than $\nbupdc{}$, and thus middle-bit D2P should be easier than next-bit D2P. Intuitively, this framework with middle-bit predictors seems to be more balanced than the framework with next-bit predictors.

Our results support this conclusion: Theorem~\ref{intro-thm:updo-not-in-p} rules out deterministic polynomial-time oracle algorithms for $\updo$. This is in sharp contrast to $\nbupdc{\PO}$, which has simple deterministic polynomial-time oracle algorithms due to Goldreich and Wigderson~\cite{GoldreichWigderson00} and Li, Pyne, and Tell~\cite{LiPyneTell24}. Thus, middle-bit predictors are unconditionally stronger than next-bit predictors in the oracle versions of the problems. Similarly, Theorem~\ref{intro-thm:scl-reduces-to-upd} is precisely a middle-bit D2P for $\SCL$ algorithms, whereas a next-bit D2P for $\SCL$ is not known, and would imply that $\SCL \subseteq \FP$. Finally, we note that the smallest class of polynomial-time randomized algorithms that has been formally studied in the context of derandomization, as far as we are aware, is $\mathsf{FLOSSY}$~\cite{Korten22, CookLiMertzPyne25}. We do not define this class here, but we note that $\upd$ belongs to this class, since $\upd \leq_{\P} \lossyc$. Thus, while Theorems~\ref{intro-thm:updo-not-in-p} and~\ref{intro-thm:scl-reduces-to-upd} show that $\upd$ is harder than $\nbupdc{}$, $\upd$ lies in the smallest class of derandomization problems that has been studied, which shows that it is not unreasonably hard either.

The recent work of Doron, Moshkovitz, Oh, and Zuckerman~\cite{DoronMoshkovitzOhZuckerman26} also supports our conclusion. As far as we are aware, this is the only other work that studies middle-bit predictors. They give a middle-bit D2P for a class of randomized algorithms that they call `insensitive algorithms' with parameters that improve on the hybrid argument. A long line of work on `beating the hybrid argument' has aimed to achieve such an improvement for next-bit predictors~\cite{FeffermanShaltielUmansViola12, DoronMoshkovitzOhZuckerman22, ChenTell21b, ShaltielViola22}. Thus, their paper, along with ours, provides promising evidence that middle-bit predictors warrant further study in the context of pseudorandomness.

In this work, in addition to the new problem $\upd$, we also study the classical problem $\nbupdc{}$, and provide a complete characterization of it. 

\subsection{Our Results: $\nbupdc{}$}
\subsubsection{Sequential Algorithms Are Optimal for $\nbupdc{\PO}$}
The deterministic $\nbupdc{\PO}$ algorithms of~\cite{GoldreichWigderson00, LiPyneTell24} are inherently sequential: they construct the output strings one bit at a time, using $n$ rounds of queries. It remained open whether there are deterministic parallel algorithms for $\nbupdc{\PO}$. We answer this question negatively, showing that sequential algorithms are essentially optimal for $\nbupdc{\PO}$.

\begin{theorem}[Corollary~\ref{cor:nbupd-exp-lb}]
\label{intro-thm:nbupdo-not-in-nc}
Let $0 < \delta \leq 1$. Every deterministic parallel oracle algorithm for $\nbupdc{\PO}$ of depth $O(n^{1 - \delta})$, on instances with $n$ predictors over $\{0, 1\}^n$ with $\epsilon = 1/10$, has worst-case work $\exp(\Omega(n^{\delta}))$. In particular,
    \[\nbupdc{\PO} \notin \NC^{\PO}.\]
\end{theorem}

Now, while a $\SCL \subseteq \FP$ proof would have to show that $\updc{\TCo} \in \FP$, a $\SCL \subseteq \NC$ proof would have to additionally show that $\nbupdc{\TCo}\in \NC$. This gives a new barrier towards proving the search analogue of $\CL \subseteq \NC$ (Problem~\ref{prob:CL-in-NC}). Crucially, $\nbupdc{\TCo} \in \NC$ is incomparable to the problem of $\updc{\TCo} \in \FP$: $\upd$ is a harder problem than $\nbupdc{}$, but the parallel model is more restrictive than the sequential one. Therefore, future work on Problems~\ref{prob:CL-in-P} and~\ref{prob:CL-in-NC} could start with trying to design an $\NC$ algorithm for $\nbupdc{\TCo}$. Since next-bit predictors are classically studied, this problem may be more approachable than $\updc{\TCo} \in \FP$. Furthermore, a positive solution to this problem would give the first algorithm for $\nbupdc{}$ or $\upd$ which exploits properties of circuits. Thus, this would likely give valuable insight towards proving that $\updc{\TCo} \in \FP$.

\subsubsection{$\nbupdc{}$ Is Equivalent to Computing Approximate Solutions to Pessimistic Estimator Instances}
The $\SCL$ algorithms for bipartite matching~\cite{AgarwalaMertz25}, linear matroid intersection~\cite{AgarwalaAlekseevVinciguerra26}, and general matching~\cite{ChakrabortyDattaKusreMukhopadhyaySinhababu26} are very similar to one another. In particular, these three algorithms use the compress-or-random framework in a manner which allows the compression and decompression procedures to be implemented in $\NC$. This was specifically noted by Agarwala and Mertz~\cite{AgarwalaMertz25}.

However, the $\SCL$ algorithm for graph sparsification by Marcussen, Pyne, and Rubinfeld~\cite{MarcussenPyneRubinfeld26} is quite different from these. It is unclear whether the compression and decompression subroutines in~\cite{MarcussenPyneRubinfeld26} can be implemented in $\NC$. This is explicitly listed as an open problem in their work\footnote{They state this question as whether $\graphspars$ reduces to $\lossyc$ over $\NC$ circuits. We define and discuss the $\lossyc$ problem later in Definition~\ref{def:lossy-coding}.}.

Their algorithm works by defining a $\TCo$-computable \emph{pessimistic estimator} over strings of length $\poly(n)$, and then computing an approximate solution to this pessimistic estimator instance. We refer readers unfamiliar with pessimistic estimators to Section~\ref{sec:nbp-and-ess-est} for a definition. For now, it suffices to know that pessimistic estimators, introduced by Raghavan~\cite{Raghavan88}, are a generalization of the method of conditional expectation (also known as the method of conditional probabilities~\cite{ErdosSelfridge73,Spencer87}). It is a classical technique in derandomization and can be found in various textbooks, such as the ones by Alon and Spencer~\cite{AlonSpencer16} and by Motwani and Raghavan~\cite{MotwaniRaghavan95}.

We show that computing an approximate solution to a pessimistic estimator instance reduces to $\nbupdc{}$, and vice versa. Both reductions involve only minor changes in the parameters: the circuits grow polynomially in size and by a constant factor in depth. Thus, $\nbupdc{}$ is equivalent to the algorithmic task underlying this classical technique.
\begin{theorem}[Informal; Theorem~\ref{thm:pess-est-to-nbp}, Corollary~\ref{cor:nbp-reduces-to-pess-est}, and Observation~\ref{obs:nbp-to-pess-est-reduction-is-efficient}]
\label{intro-thm:nbupd-pess-est-equiv}
    $\nbupdc{}$ is equivalent to computing approximate solutions to pessimistic estimator instances.
\end{theorem}
Applying this theorem to the pessimistic estimator instance of Marcussen, Pyne, and Rubinfeld~\cite{MarcussenPyneRubinfeld26} immediately gives the following corollary.

\begin{corollary}[Theorem~\ref{thm:sparsification-reduces-to-nbp}]
\label{intro-cor:graph-spars-reduces-to-upd}
    \[\graphspars \leq_{\LUTCo} \nbupdc{\TCo}.\]
\end{corollary}

$\updc{\TCo}$ (and hence also $\nbupdc{\TCo}$) can be solved by a compress-or-random procedure in which both the compression and the decompression can be implemented in logspace-uniform $\TCo$ (and hence in $\NC$). We do not state this formally, but it follows from the observation that our compression and decompression procedures in Section~\ref{sec:upd-in-cl} (Lemma~\ref{lem:unpred-compression}) can be implemented in $\TCo$ when the predictors are $\TCo$ circuits. Together with Corollary~\ref{intro-cor:graph-spars-reduces-to-upd}, this answers the open question posed by Marcussen, Pyne, and Rubinfeld~\cite{MarcussenPyneRubinfeld26}. This is another example of how $\nbupdc{}$ and $\upd$ can shed light on catalytic computation.

We summarize our results in Tables~\ref{tab:characterization},~\ref{tab:reductions}, and~\ref{tab:oracle-hardness}. We omit the results related to the $\StCL$ versus $\SCL$ problem (Problem~\ref{prob-intro:stablesearchcl}), since they require additional definitions in order to be accurately stated. This will be done in Section~\ref{sec:stablecl}.

\begin{table}[!t]
\centering
\hypersetup{hidelinks}
\caption{Characterization of $\upd$ and $\nbupdc{}$.}
\label{tab:our-results}
\small
\renewcommand{\arraystretch}{1.2}
\renewcommand{\tabularxcolumn}[1]{m{#1}}
\setlength{\tabcolsep}{4pt}
\setlength{\fboxsep}{6pt}
\newlength{\unpredtablelabelwidth}
\settowidth{\unpredtablelabelwidth}{$\nbupdc{\PO}$}
\fbox{\begin{minipage}{\dimexpr\linewidth-2\fboxsep-2\fboxrule\relax}
\begin{tabularx}{\linewidth}{@{}>{\raggedright\arraybackslash}m{\unpredtablelabelwidth} >{\raggedright\arraybackslash}X@{}}
$\upd$
& Informal: $\upd$ is polynomial time equivalent to $\SCL$, up to circuit evaluation. \par\smallskip
{\centering
$\updo \in \SCL^{\PO} \quad \text{and} \quad \updc{\TCo} \in \SCL \leq_{\P} \upd$\par
} \\
\midrule
$\nbupdc{}$
& Informal: $\nbupdc{}$ is logspace equivalent to computing approximate solutions to pessimistic estimator instances. \\
\end{tabularx}
\end{minipage}}
\label{tab:characterization}
\end{table}

\begin{table}[!t]
\centering
\hypersetup{hidelinks}
\caption{All problems currently known to be in $\SCL$ reduce to $\updc{\TCo}$.}
\label{tab:reductions}
\small
\renewcommand{\arraystretch}{1.2}
\setlength{\tabcolsep}{8pt}
\setlength{\fboxsep}{6pt}
\fbox{\begin{minipage}{\dimexpr\linewidth-2\fboxsep-2\fboxrule\relax}
\begin{tabularx}{\linewidth}{@{}X@{}r@{\;}l@{}X@{}}
\multicolumn{4}{@{}c}{\textbf{Reduction}} \\
\midrule
& $\bipmatch$ & $ {}\leq_{\LUTCo} \updc{\TCo}$ & \\
& $\lmi$ & $ {}\leq_{\LUTCo} \updc{\TCo}$ & \\
& $\match$ & $ {}\leq_{\LUTCo} \updc{\TCo}$ & \\
\midrule
& $\graphspars$ & $ {}\leq_{\LUTCo} \nbupdc{\TCo}$ & \\
\end{tabularx}
\end{minipage}}
\end{table}

\begin{table}[!t]
\centering
\hypersetup{hidelinks}
\caption{Oracle hardness of $\upd$ and $\nbupdc{}$.}
\label{tab:oracle-hardness}
\small
\renewcommand{\arraystretch}{1.2}
\renewcommand{\tabularxcolumn}[1]{m{#1}}
\setlength{\tabcolsep}{4pt}
\setlength{\fboxsep}{6pt}
\fbox{\begin{minipage}{\dimexpr\linewidth-2\fboxsep-2\fboxrule\relax}
\begin{tabularx}{\linewidth}{@{}l >{\raggedright\arraybackslash}X c@{}}
 & \textbf{Lower bound} & \textbf{Separation} \\
\midrule
$\updo$
& Deterministic algorithm requires $\exp(\Omega(\sqrt{n}))$ time.
& $\updo \notin \P^{\PO}$ \\
\addlinespace[0.6em]
$\nbupdc{\PO}$
& Deterministic parallel algorithm with depth $O(n^{1-\delta})$ requires $\exp(\Omega(n^\delta))$ work.
& $\nbupdc{\PO} \notin \NC^{\PO}$ \\
\end{tabularx}
\par\smallskip
{\footnotesize Bounds are worst-case, for deterministic oracle algorithms with $n$ predictors over $\{0,1\}^n$, $\epsilon=1/10$, and fixed $0<\delta\leq1$.}
\end{minipage}}
\end{table}

\subsection{Future Directions and Open Problems}
We believe that our contribution in this work is providing strong evidence that $\upd$ (and middle-bit predictors generally) warrants further study, specifically because it may hold the key to solving several major open problems in catalytic computation and pseudorandomness. In this section we discuss directions that future work should explore. 

One ambitious goal to pursue is the following.
\begin{problem}
\label{prob:upd-in-fp}
    Is $\upd \in \FP$? 
\end{problem}
A positive solution to this problem would immediately imply $\CL \subseteq \P$ and $\SCL \subseteq \FP$ (Problem~\ref{prob:CL-in-P}). By Theorem~\ref{intro-thm:updo-not-in-p}, any deterministic polynomial-time $\upd$ algorithm would need to avoid the oracle barrier of $\updo \notin \P^{\PO}$. Thus, we begin by asking a simpler question.

\begin{problem}
\label{prob:upd-nco-in-fp}
Is $\updc{\NCo} \in \FP$?
\end{problem}

A positive solution to this weaker problem would provide the first $\upd$ algorithm that does not simply treat the predictors as black-box functions. In particular, it would need to use properties of Boolean circuits. It is likely that this would provide a substantial clue about how to solve Problem~\ref{prob:upd-in-fp}.

Similarly, future work could focus on studying parallel algorithms for $\nbupdc{}$. As we said earlier, these would be incomparable to sequential algorithms for $\upd$.

\begin{problem}
\label{prob:nbupd-in-nc}
Is $\nbupdc{\TCo} \in \NC$?
\end{problem}

A positive solution to this problem would immediately imply several $\NC$ inclusions which are not known, such as $\graphspars \in \NC$ (due to our reduction from Section~\ref{sec:graph-spars-to-pess-est}). 

Again, it would also be valuable to solve Problem~\ref{prob:nbupd-in-nc} for weak circuits. This would serve as a blueprint to overcome the oracle lower bound from Theorem~\ref{intro-thm:nbupdo-not-in-nc}.

\begin{problem}
\label{prob:nbupd-nco-in-fp}
Is $\nbupdc{\NCo} \in \NC$?
\end{problem}

Note that a positive solution to the search analogue of Problem~\ref{prob:CL-in-P} would require first solving Problem~\ref{prob:upd-nco-in-fp}. Similarly, a positive solution to the search analogue of Problem~\ref{prob:CL-in-NC} would require first solving Problem~\ref{prob:nbupd-nco-in-fp}.

Now, we state a problem related to the $\FL^{\CL}$ versus $\SCL$ problem (Problem~\ref{prob-intro:FLCLvsSCL}).

\begin{problem}
\label{prob:stableupd}
    Is $\updc{\TCo} \in \StCL$?
\end{problem}

A positive solution to this problem would immediately give $\StCL$ algorithms for \emph{every} problem currently known to be in $\SCL$. Corollary~\ref{cor:restorable-algos-cant-be-stable} shows that a $\StCL$ algorithm for $\updc{\TCo}$ cannot be restorable. Thus, we state the following subproblem.
\begin{problem}
    Construct a non-restorable $\SCL$ algorithm for $\updc{\TCo}$.
\end{problem}

Of course, any algorithm can be made intentionally non-restorable. The spirit of this problem is to find an algorithm that cannot be made restorable.

Finally, we look at future directions in pseudorandomness. In our work we show an efficient middle-bit D2P for $\SCL$ algorithms. Doron, Moshkovitz, Oh, and Zuckerman~\cite{DoronMoshkovitzOhZuckerman26} show a middle-bit D2P for insensitive algorithms that beats the hybrid argument. In both cases, a next-bit D2P seems hard. Future work should explore how far these new predictors can go.

\begin{problem}
    Construct efficient middle-bit D2Ps for classes of algorithms.
\end{problem}

Here, efficiency refers to both computational efficiency, as in our work, and parameter efficiency, as in~\cite{DoronMoshkovitzOhZuckerman26}. Since $\upd^{\PO} \in \SCL^{\PO}$, positive solutions to this problem may place new problems in $\SCL$. 
\subsection{Proof Overview}
We now give a brief overview of our main ideas. Each section additionally begins with a far more detailed discussion.

\usetikzlibrary{arrows.meta,decorations.pathreplacing}
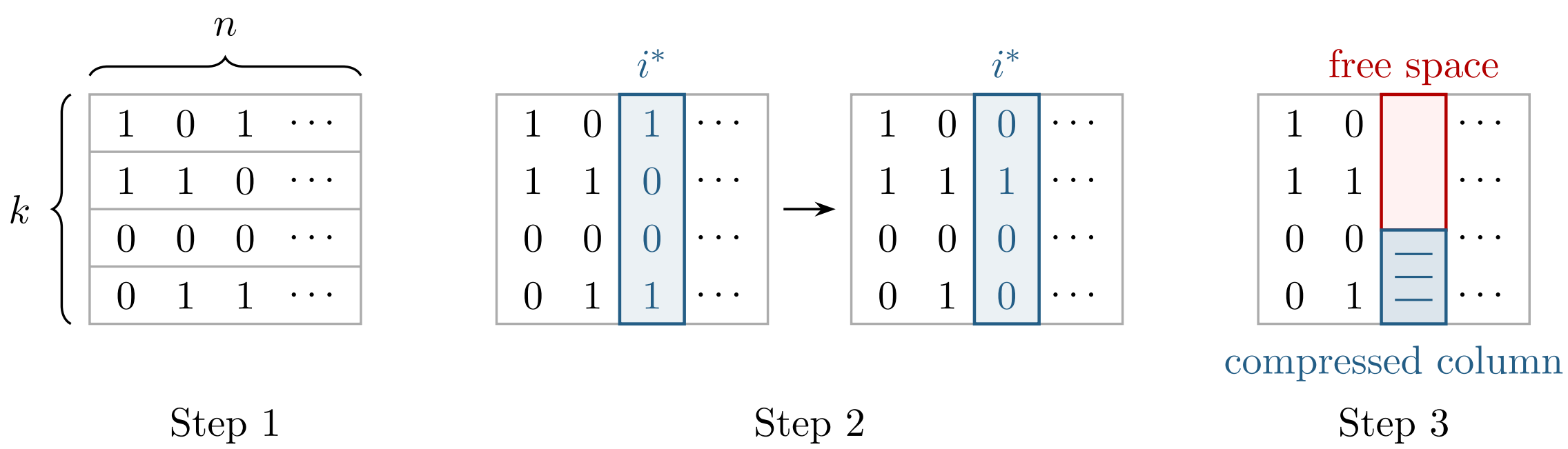
\begin{figure}[t]
    \centering
    \begin{tikzpicture}[
    x=1cm,y=1cm,
    every node/.style={font=\normalfont\normalsize,text=black},
    tape/.style={draw=gray!65,line width=.65pt}
]
  \definecolor{columnblue}{HTML}{245E86}

  \foreach \x/\stage in {0/0,3.9/1,7.3/2,11.2/3} {
    \begin{scope}[xshift=\x cm]
      \draw[tape] (0,0) rectangle (2.6,2.2);
      \ifnum\stage=0
        \foreach \y in {.55,1.1,1.65}
          \draw[tape] (0,\y) -- (2.6,\y);
      \fi
      \ifnum\stage>0
        \ifnum\stage<3
          \filldraw[fill=columnblue!9,draw=columnblue,line width=.85pt]
            (1.18,0) rectangle (1.8,2.2);
        \else
          \filldraw[fill=red!5,draw=red!70!black,line width=.85pt]
            (1.18,.9) rectangle (1.8,2.2);
          \filldraw[fill=columnblue!16,draw=columnblue,line width=.85pt]
            (1.18,0) rectangle (1.8,.9);
          \foreach \y in {.23,.45,.67}
            \draw[columnblue,line width=.6pt] (1.31,\y) -- (1.67,\y);
        \fi
      \fi

      \foreach \y/\a/\b/\c/\e in {
          1.92/1/0/1/0,1.37/1/1/0/1,.82/0/0/0/0,.27/0/1/1/0} {
        \node at (.35,\y) {$\a$};
        \node at (.92,\y) {$\b$};
        \node at (2.16,\y) {$\cdots$};
        \ifnum\stage=0
          \node at (1.49,\y) {$\c$};
        \fi
        \ifnum\stage=1
          \node[text=columnblue] at (1.49,\y) {$\c$};
        \fi
        \ifnum\stage=2
          \node[text=columnblue] at (1.49,\y) {$\e$};
        \fi
      }
    \end{scope}
  }

  \draw[decorate,decoration={brace,amplitude=5pt},line width=.65pt]
    (0,2.38) -- node[midway,above=7pt] {$n$} (2.6,2.38);
  \draw[decorate,decoration={brace,amplitude=5pt},line width=.65pt]
    (-.18,0) -- node[midway,left=7pt] {$k$} (-.18,2.2);

  \node[text=columnblue] (index-label) at (5.39,2.5) {$i^*$};
  \node[text=columnblue,anchor=base] at (8.79,0 |- index-label.base) {$i^*$};
  \node[text=red!70!black,anchor=base] at (12.69,0 |- index-label.base) {free space};
  \node[text=columnblue] at (12.5,-.38) {compressed column};

  \draw[-{Stealth[length=2mm,width=1.4mm]},line width=.7pt]
    (6.65,1.1) -- (7.15,1.1);
  \node at (1.3,-.98) {Step 1};
  \node at (6.9,-.98) {Step 2};
  \node at (12.5,-.98) {Step 3};
\end{tikzpicture}
\caption{A compress-or-random algorithm for $\updo$. Step 1: interpret the catalytic tape as a multiset $S$ of $k$ strings in $\{0, 1\}^n$. Step 2: If there is a predictor $P^*$ at index $i^*$ which succeeds on too many strings, replace the $(i^*)^{th}$ bit of each string $s$ with $0$ if $P^*$ succeeds on $s$, and with $1$ otherwise. This makes the $(i^*)^{th}$ column sparse. Step 3: compress the sparse string in the $(i^*)^{th}$ column.}
\label{fig:unpred-algo}
\end{figure}

\paragraph{Catalytic Logspace Algorithm for $\upd^{\PO}$.} Our catalytic logspace algorithm for $\upd$ in the oracle model, i.e., $\updo$, follows the \emph{compress-or-random} framework briefly described in the introduction. The idea is as follows: the algorithm interprets the catalytic tape itself as a multiset of $k = \Theta(\log(2m)/\epsilon^2)$ $n$-bit strings. First, the algorithm checks whether this multiset is a valid output, by counting, for each predictor, the number of strings on which it succeeds. If every count is below $(1/2 + \epsilon)k$, the algorithm outputs the multiset and halts. Otherwise, some predictor $P^*$, at an index $i^* \in [n]$, succeeds on too many strings. The algorithm replaces the $(i^*)^{th}$ bit of each string with $0$ if $P^*$ succeeds on the string, and $1$ otherwise. The amount of information on the catalytic tape is exactly the same as before, since the $(i^*)^{th}$ bit of each string can be recovered by evaluating $P^*$ on the string and flipping its output if the bit stores $1$. Since $P^*$ succeeds on too many strings, the $(i^*)^{th}$ bit of most strings is set to $0$. This sparsity allows us to compress the catalytic tape. Thus, we simply iterate this process until either the catalytic tape contains a good output for the problem or until this repeated compression frees up enough space for the algorithm to perform a brute-force search over all multisets of size $k$. Finally, the algorithm decompresses, restoring the catalytic tape. This is our entire algorithm. Replacing our oracle queries with circuit evaluations directly gives our $\upd^{\TCo}$ algorithm. See Figure~\ref{fig:unpred-algo} for a visual representation of the compression step.

\begin{figure}
    \centering
\usetikzlibrary{arrows.meta,shapes.geometric}
\begin{tikzpicture}[
    x=1cm,y=1cm,
    every node/.style={font=\normalfont\normalsize},
    edge/.style={-{Stealth[length=2.2mm]},line width=.85pt,
                 shorten <=2pt,shorten >=2pt},
    vertex/.style={circle,draw=black,line width=.7pt,fill=white,
                   minimum size=4mm,inner sep=0pt},
    endpoint/.style={ellipse,draw=black,line width=.75pt,
                     minimum width=2cm,minimum height=9mm,inner sep=4pt}
]
  \definecolor{walkstart}{HTML}{087F64}
  \definecolor{walkend}{HTML}{BC2727}

  \node[endpoint] (c0) at (0,1.8) {$C(y)\circ 0$};
  \node[endpoint] (c1) at (0,-1.8) {$C(y)\circ 1$};
  \node[circle,draw=black,line width=.8pt,fill=white,
        minimum size=10mm,inner sep=0pt] (y) at (8,0) {$y$};

  \node[vertex] (a1) at (2.4,2.4) {};
  \node[vertex] (a2) at (4.1,1.7) {};
  \node[vertex] (a3) at (5.8,2.1) {};
  \draw[edge,walkstart] (c0.east) -- node[above left] {$l_0$} (a1);
  \draw[edge] (a1) -- (a2);
  \draw[edge] (a2) -- (a3);
  \draw[edge,walkend] (a3) -- (y.north west);

  \node[vertex] (b1) at (2.4,-2.4) {};
  \node[vertex] (b2) at (4.1,-1.7) {};
  \node[vertex] (b3) at (5.8,-2.1) {};
  \draw[edge,walkstart] (c1.east) -- node[below left] {$l_1$} (b1);
  \draw[edge] (b1) -- (b2);
  \draw[edge] (b2) -- (b3);
  \draw[edge,walkend] (b3) -- (y.south west);

  \draw[edge] (6.1,0) -- (y.west);
  \draw[edge] (y.north east) -- (9.05,1.45);
  \node at (5.75,0) {$\cdots$};
  \node at (9.25,1.75) {$\cdots$};

  \draw[edge,walkstart] (y.south east) -- node[right] {$l_y$} (9.05,-1.45);
  \node at (9.25,-1.75) {$\cdots$};
\end{tikzpicture}

    \caption{A non-solution $y \in \{0, 1\}^n$ in the graph $G$. Vertex labels correspond to the string on the input tape of $M$ and edge labels correspond to the rest of its Turing machine configuration. The first edge of each walk is coloured green, the last edge is coloured red, and intermediate edges are coloured black. Each vertex has at most one outgoing green edge. Non-solutions, such as $y$, have at least two incoming red edges. Each vertex has an equal number of incoming and outgoing black edges.}
    \label{fig:lossy-to-weakex}
\end{figure}
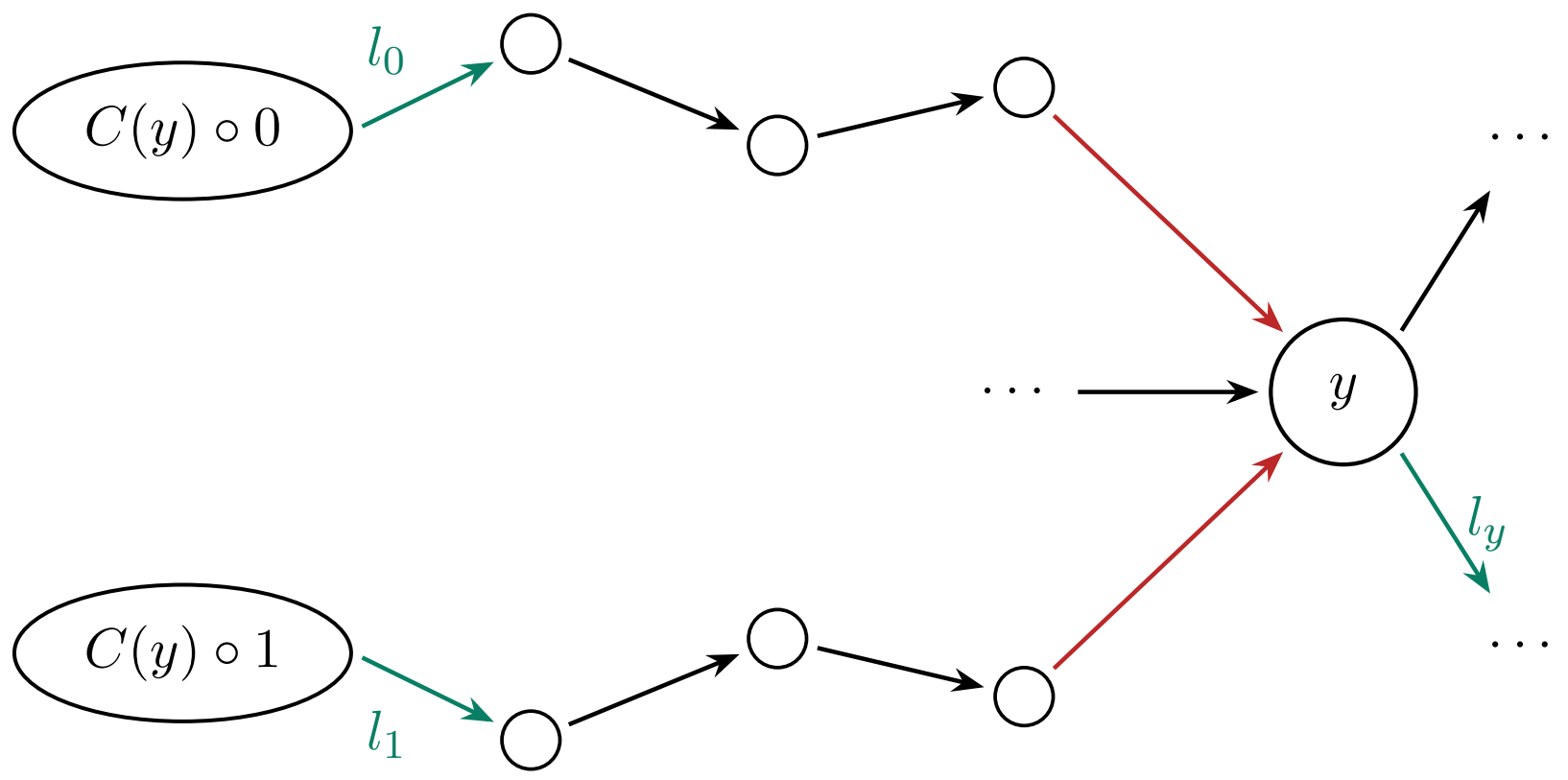

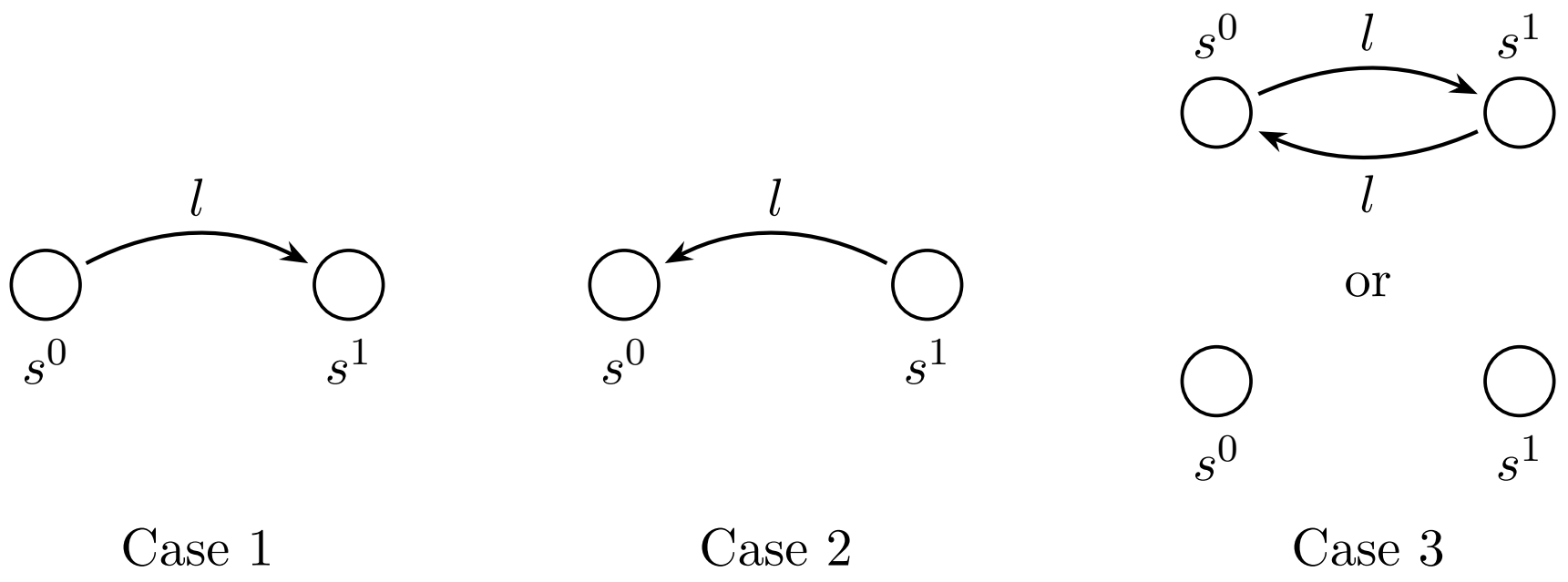
\begin{figure}
    \centering
\usetikzlibrary{arrows.meta}
\begin{tikzpicture}[
    x=1cm,y=1cm,
    every node/.style={font=\normalfont\normalsize},
    vertex/.style={circle,draw=black,line width=.7pt,fill=white,
                   minimum size=5mm,inner sep=0pt},
    edge/.style={-{Stealth[length=2mm]},line width=.8pt,
                 shorten <=2pt,shorten >=2pt}
]
  \begin{scope}[yshift=-9.82225pt]
  \node[vertex,label=below:{$s^0$}] (a0) at (0,.65) {};
  \node[vertex,label=below:{$s^1$}] (a1) at (2.2,.65) {};
  \draw[edge] (a0) to[bend left=28] node[above] {$l$} (a1);

  \node[vertex,label=below:{$s^0$}] (b0) at (4.2,.65) {};
  \node[vertex,label=below:{$s^1$}] (b1) at (6.4,.65) {};
  \draw[edge] (b1) to[bend right=28] node[above] {$l$} (b0);

  \node[vertex,label=above:{$s^0$}] (c0) at (8.5,1.9) {};
  \node[vertex,label=above:{$s^1$}] (c1) at (10.7,1.9) {};
  \node at (9.6,.65) {or};
  \node[vertex,label=below:{$s^0$}] (d0) at (8.5,-.05) {};
  \node[vertex,label=below:{$s^1$}] (d1) at (10.7,-.05) {};
  \draw[edge] (c0) to[bend left=24] node[above] {$l$} (c1);
  \draw[edge] (c1) to[bend left=24] node[below] {$l$} (c0);
  \end{scope}

  \node at (1.1,-1.6) {Case 1};
  \node at (5.3,-1.6) {Case 2};
  \node at (9.6,-1.6) {Case 3};
\end{tikzpicture}

    \caption{Reduction from $\weakex$ to $\upd$. Case 1: the predictor outputs $1$. Case 2: the predictor outputs $0$. Case 3: the predictor outputs $0$ or $1$ with equal probability.}
    \label{fig:weakex-to-unpred}
\end{figure}

\paragraph{Reduction from $\SCL$ to $\upd$.} Cook, Li, Mertz, and Pyne~\cite{CookLiMertzPyne25} show that $\SCL$ reduces to $\lossyc$: given a compressor $C:\{0,1\}^n \rightarrow \{0,1\}^{n-1}$ and a decompressor $D:\{0, 1\}^{n-1} \rightarrow \{0, 1\}^n$, find $x$ such that $D(C(x)) \neq x$. In their instances, $C$ is computable in polynomial time and $D$ is computable in polynomial time by an $\ipFL$ machine~\cite{CookGhentiyalaMertzPyneSheffield25} $M$. Such a machine has a read-write input tape and a read-write work tape of logarithmic size, and must halt with its output written on the read-write input tape. We exploit this in-place structure. To ensure that the input and output of $M$ have the same size, the input of $M$ is padded with an extra bit which is simply ignored by the computation. Consider the multigraph $G$ on $\{0,1\}^n$ which contains, for every non-solution $y$ (i.e., $y$ such that $D(C(y)) = y$) and every bit $b$, the walk traced by the input tape of the in-place machine when it is run on $C(y) \circ b$: every step of the computation adds an edge, which is labelled by the $O(\log n)$-bit internal state of the machine (and $b$). This walk starts at the vertex $C(y) \circ b$ and ends at the vertex $D(C(y)) = y$. Since the machine is deterministic, a non-solution $y$ is the start vertex of at most one walk but the end vertex of two: the walks starting at $C(y) \circ 0$ and $C(y) \circ 1$. Hence, $y$ has strictly larger in-degree than out-degree. Thus, every vertex of $G$ with out-degree $\geq$ in-degree is a solution; we call the problem of finding such a vertex $\weakex$. See Figure~\ref{fig:lossy-to-weakex} for a visual representation of this reduction.

We now briefly describe our reduction from $\weakex$ to $\upd$. For every edge $e$ of the graph $G$, the two endpoints of the edge differ in at most one bit. Moreover, the position of this bit is determined by the label $l$ of $e$. For a label $l$, we refer to this position as $\ind(l) \in [n]$. For each label $l$, we define a middle-bit predictor $P_{l}$ over $\{0, 1\}^n$ at index $\ind(l)$ as follows. Given as input a string $s \in \{0, 1\}^{n-1}$, $P_l$ must output either $0$ or $1$. Let $s^0 \in \{0, 1\}^n$ be the string obtained from $s$ by inserting a $0$ in the $\ind(l)^{th}$ position, and $s^1$ be the string obtained by inserting a $1$. If $P_l$ outputs $0$ then it would succeed on $s^0$, and if it outputs $1$ then it would succeed on $s^1$. In order to decide between the two cases, the predictor $P_l$ looks at the edges between $s^0$ and $s^1$ with label~$l$ in the graph $G$. If there are more such edges directed from $s^0$ to $s^1$ than from $s^1$ to $s^0$, then $P_l$ outputs $1$. Else, if there are more such edges directed from $s^1$ to $s^0$, then $P_l$ outputs $0$. Else, if the numbers of edges in both directions are equal, $P_l$ outputs $0$ or $1$, each with probability half \footnote{We require predictors to be deterministic. Thus, we simulate this random guessing deterministically by constructing two predictors for each label, one which always outputs $1$ in this equality case, and the other which always outputs $0$.}. See Figure~\ref{fig:weakex-to-unpred} for a visual representation of these predictors. Since the labels $l$ have $O(\log n)$ many bits, there are a total of $\poly(n)$ many predictors. A solution to $\upd$ on these middle-bit predictors (with a small enough parameter $\epsilon$) is a set of vertices such that each middle-bit predictor succeeds on very few vertices from this set. An averaging argument shows that this set must contain a single vertex on which very few of these middle-bit predictors succeed. Intuitively, these predictors are defined so that they are more likely to succeed on the head of an edge than its tail. Thus, if a vertex has higher in-degree than out-degree, many of these predictors will succeed on that vertex. The contrapositive of this statement is that a vertex on which very few of these predictors succeed should have out-degree $\geq$ in-degree. Thus, it would be a solution to $\weakex$. This is an informal description. It remains to analyze the computational complexity of these predictors, to give exact thresholds for the terms `very few' and `many' in our argument, and to show that a solution to $\weakex$ can actually be efficiently extracted from a solution to $\upd$. We address these details in Section~\ref{sec:weakex-to-upd}.

\begin{figure}[t]
    \centering
    \begin{tikzpicture}[
    x=1cm,y=1cm,
    every node/.style={font=\normalfont\normalsize},
    cube edge/.style={draw=black,line width=.6pt},
    vertex/.style={circle,draw=black,fill=white,
                   line width=.6pt,minimum size=4mm,inner sep=0pt},
    diagram label/.style={inner sep=0pt,outer sep=0pt},
    directed edge/.style={-{Stealth[length=2.2mm]},draw=edgeorange,
                          line width=1.2pt,shorten <=2pt,shorten >=2pt}
]
  \definecolor{edgeorange}{HTML}{F27200}
  \definecolor{successyellow}{HTML}{FFDA31}
  \definecolor{failureblue}{HTML}{004D80}
  \definecolor{vertexgreen}{HTML}{2E8B57}

  \foreach \panel/\offset in {0/0,1/5.5} {
    \begin{scope}[xshift=\offset cm]
      \ifnum\panel=0
        \node[vertex,fill=successyellow] (A) at (0,0) {};
        \node[vertex,fill=failureblue] (B) at (2.6,0) {};
        \node[vertex] (G) at (3.7,2.95) {};
      \else
        \node[vertex] (A) at (0,0) {};
        \node[vertex] (B) at (2.6,0) {};
        \node[vertex,fill=vertexgreen] (G) at (3.7,2.95) {};
      \fi
      \node[vertex] (C) at (2.6,2.1) {};
      \node[vertex] (D) at (0,2.1) {};
      \node[vertex] (E) at (1.1,.85) {};
      \node[vertex] (F) at (3.7,.85) {};
      \node[vertex] (H) at (1.1,2.95) {};

      \foreach \u/\v in {A/D,D/C,H/D,H/E,E/A,E/F,B/C,B/F}
        \draw[cube edge] (\u) -- (\v);

      \ifnum\panel=0
        \foreach \u/\v in {C/G,G/H,F/G}
          \draw[cube edge] (\u) -- (\v);
        \draw[directed edge] (A) --
          node[diagram label,above=3.6pt] {$s$} (B);
        \node[diagram label,anchor=north,yshift=-3pt] at (A.south) {$s^0$};
        \node[diagram label,anchor=north,yshift=-3pt] at (B.south) {$s^1$};
      \else
        \draw[cube edge] (A) -- (B);
        \draw[directed edge] (C) -- (G);
        \draw[directed edge] (H) -- (G);
        \draw[directed edge] (G) -- (F);
        \node[diagram label,anchor=south,yshift=3pt] at (G.north) {$v$};
      \fi
    \end{scope}
  }
\end{tikzpicture}
    \caption{Hypercube interpretation of $\updo$. The diagram on the left is a visual representation of the case when the predictor outputs $0$ on the query string $s$. The edge $s$ is oriented from $s^0$ to $s^1$, indicating that the predictor fails on the string $s^1$. The diagram on the right is a visual representation of the vertex $v$ that the $\updo$ algorithm wants to find: a vertex with high in-degree.}
    \label{fig:updo-lower-bound}
\end{figure}
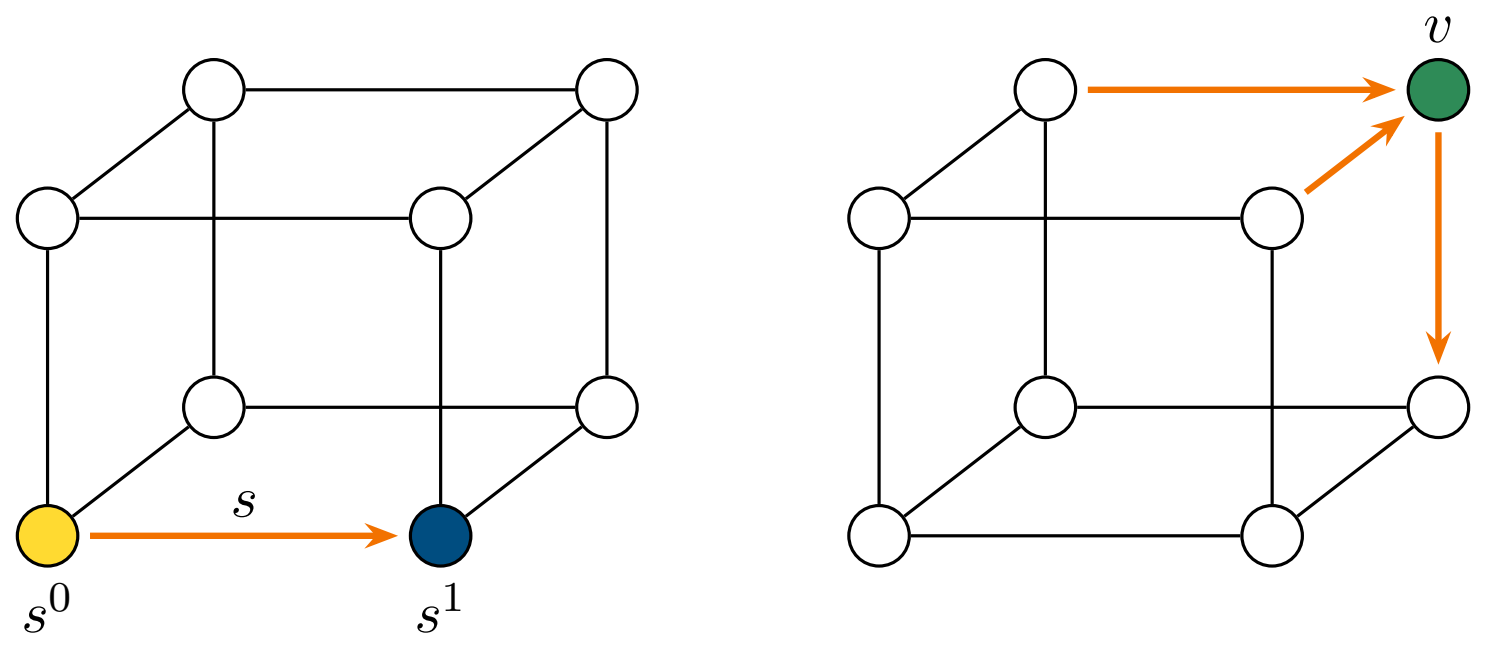

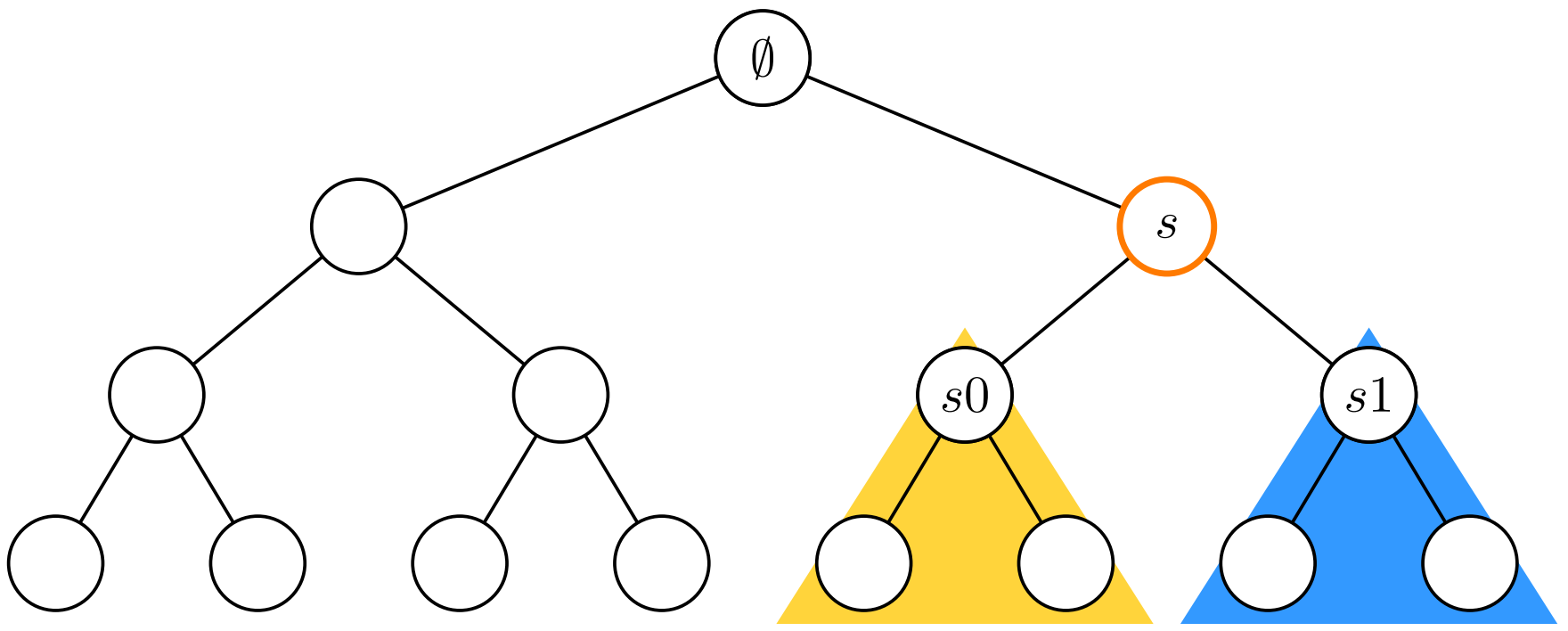
\begin{figure}
    \centering
   \begin{tikzpicture}[
    x=1cm,y=1cm,
    every node/.style={font=\normalfont\normalsize},
    vertex/.style={circle,draw=black,fill=white,line width=.7pt,
                   minimum size=7mm,inner sep=0pt},
    branch/.style={draw=black,line width=.7pt}
]
  \definecolor{queryorange}{HTML}{FF7A00}
  \definecolor{leftyellow}{HTML}{FFD43B}
  \definecolor{rightblue}{HTML}{3399FF}

  \fill[leftyellow] (7.5,-2) -- (6.1,-4.2) -- (8.9,-4.2) -- cycle;
  \fill[rightblue] (10.5,-2) -- (9.1,-4.2) -- (11.9,-4.2) -- cycle;

  \foreach \depth in {0,...,3} {
    \pgfmathtruncatemacro{\last}{2^\depth-1}
    \foreach \index in {0,...,\last} {
      \coordinate (v-\depth-\index)
        at ({12*(\index+.5)/2^\depth},{-1.25*\depth});
    }
  }

  \foreach \depth in {1,...,3} {
    \pgfmathtruncatemacro{\previous}{\depth-1}
    \pgfmathtruncatemacro{\last}{2^\depth-1}
    \foreach \index in {0,...,\last} {
      \pgfmathtruncatemacro{\parent}{floor(\index/2)}
      \draw[branch] (v-\previous-\parent) -- (v-\depth-\index);
    }
  }

  \foreach \depth in {0,...,3} {
    \pgfmathtruncatemacro{\last}{2^\depth-1}
    \foreach \index in {0,...,\last}
      \node[vertex] at (v-\depth-\index) {};
  }
  \node[vertex] at (v-0-0) {$\emptyset$};
  \node[vertex,draw=queryorange,line width=1.3pt] at (v-1-1) {$s$};

  \node[vertex] at (v-2-2) {$s0$};
  \node[vertex] at (v-2-3) {$s1$};
\end{tikzpicture}
    \caption{Binary tree interpretation of $\nbupdc{\PO}$. A query $s$ to a predictor corresponds to an internal vertex of this tree. Depending on the answer of the predictor, it either fails on the entire subtree of $s0$ or the entire subtree of $s1$.}
    \label{fig:nbupdo-lower-bound}
\end{figure}

\paragraph{Lower Bounds for $\updo$ and $\nbupdc{\PO}$.} We obtain both lower bounds by constructing predictor oracles as adversaries. Consider an $\upd$ instance over $\{0, 1\}^n$ which has exactly one predictor $P_i$ at each index $i \in [n]$. Let $s \in \{0, 1\}^{n-1}$ be a query to the predictor $P_i$. Let $s^0 \in \{0, 1\}^n$ be the string obtained from $s$ by inserting a $0$ in its $i^{th}$ position, and similarly $s^1$ be the string obtained by inserting a $1$. If $P_i$ outputs $0$ on the query $s$, then it succeeds on $s^0$ and fails on $s^1$. On the other hand, if $P_i$ outputs $1$, then it fails on $s^0$ and succeeds on $s^1$. This can be thought of as the predictor $P_i$ assigning an orientation to the edge $(s^0, s^1)$ in the $n$-dimensional hypercube. If it fails on $s^0$ then the edge is oriented from $s^1$ to $s^0$, and if it fails on $s^1$ then the edge is oriented from $s^0$ to $s^1$. The goal of an $\updo$ algorithm is to compute a set of strings such that each predictor succeeds on very few strings from this set. A subproblem of this is to identify a single string such that very few predictors succeed on that string. In our hypercube interpretation, this is analogous to computing a single vertex with large in-degree. Thus, the adversary wants to answer queries in a manner which maximizes the number of queries that the algorithm needs to make in order to reveal a vertex with large in-degree. We formalize the problem that the adversary needs to solve as an \emph{online} problem: given as input a sequence of hypercube edges, assign an orientation to each edge before seeing the next one, such that the orientation minimizes the maximum in-degree of any vertex in the subgraph formed by queried edges. We give a simple greedy algorithm for this online problem which, when implemented by the adversary, results in the algorithm needing to make $\exp(\Omega(\sqrt{n}))$ queries in total to identify a single vertex with large in-degree. Our greedy algorithm is very simple: it can answer each query in $\poly(n)$ time (note that the query itself has $n-1$ bits). This concludes our lower bound on $\updo$. This idea is formalized in Section~\ref{sec:lower-bound-for-updo}. See Figure~\ref{fig:updo-lower-bound} for a visual representation of this hypercube argument.

We use a similar approach for our lower bound on $\nbupdc{\PO}$. As $\updo$ naturally corresponds to the $n$-dimensional hypercube, $\nbupdc{\PO}$ naturally corresponds to the complete binary tree of depth $n$. We label each vertex at the $i^{th}$ level of this tree with a string $s \in \{0, 1\}^i$, and the two children of this vertex with the strings $s0$ and $s1$ (the root corresponds to the empty string $\emptyset$). A query $s \in \{0, 1\}^{i-1}$ to the $i^{th}$ next-bit predictor $P_i$ naturally corresponds to the vertex in the $(i-1)^{th}$ level of this tree with label $s$. If $P_i$ outputs $1$ on input $s$, then it succeeds on every string in the subtree of $s1$ and fails on every string in the subtree of $s0$. Else, if $P_i$ outputs $0$, then it fails on every string in the subtree of $s1$ and succeeds on every string in the subtree of $s0$. This can be viewed as $P_i$ placing a \emph{token} on the child of vertex $s$ that it fails on. The number of predictors that fail on a string $s \in \{0, 1\}^n$ is equal to the number of tokens on the path from the root to the leaf $s$ in the binary tree. Thus, an adversary wants to answer queries in a manner which minimizes the maximum number of tokens on any root-to-leaf path. We once again formulate the problem that the adversary wants to solve as an online problem, and then give a simple greedy algorithm for this online problem which results in the exponential lower bound that we want. This is done in Section~\ref{sec:lower-bound-for-nbupdc}. See Figure~\ref{fig:nbupdo-lower-bound} for a visual representation of this binary tree argument.

\paragraph{$\StCL$ versus $\SCL$.} Our main contribution here is to show that $\updc{\TCo}$ unconditionally does not have a stable restorable catalytic logspace algorithm. Our proof is via a simple counting argument. Assume for the sake of contradiction that $\updc{\TCo}$ is computable by a stable restorable catalytic logspace Turing machine $M$. Fix an input $x$ to $M$ describing an $\updc{\TCo}$ instance over $\{0, 1\}^n$, where $|x| \leq \poly(n)$. For all initial configurations of the catalytic tape, $M$ reaches a state where its output is explicitly written on the catalytic tape. This follows from the fact that $M$ is restorable. Due to the fact that $M$ is stable, this output must be the \emph{same} for all initial catalytic tape configurations. The output consists of at least one string, and thus consumes $\geq n$ bits on the catalytic tape. Let the catalytic tape have length $c$. At the point when the output is explicitly written on the catalytic tape, the machine $M$ could have at most $2^{c-n} \cdot 2^{O(\log n)}$ different configurations. This is due to the fact that $n$ bits of the catalytic tape are fixed, so the catalytic tape could have at most $2^{c-n}$ different configurations, and the work tape has length $O(\log |x|) = O(\log n)$. The remaining state of $M$, i.e., tape head pointers, etc., can also be described using $O(\log n)$ many bits (assuming $c \leq \poly(n)$, which is true because a catalytic logspace machine has a polynomial-length catalytic tape). Since $M$ is a catalytic machine, it must eventually revert the catalytic tape back to its original configuration. There are $2^{c}$ different initial configurations of the catalytic tape. Thus, by injectivity, we must have $2^{c-n + O(\log n)} \geq 2^c$. For large enough $n$, this is a contradiction. Thus, $\updc{\TCo}$ does not have a stable restorable catalytic logspace algorithm. This argument is formalized in Section~\ref{sec:stablecl}.

\paragraph{Next-Bit Predictors and Pessimistic Estimators.} In Section~\ref{sec:nbp-and-ess-est} we show that the problem of computing an approximate solution to a pessimistic estimator instance is equivalent to $\nbupdc{}$. The reduction from $\nbupdc{}$ to the problem of computing an approximate solution to a pessimistic estimator instance is very technical, and closely resembles an argument of Wigderson and Xiao~\cite{WigdersonXiao08}. We defer an explanation of this reduction to Section~\ref{sec:nbp-to-pess-est}. The intuition for the other direction is as follows. A pessimistic estimator $F$ over $\{0, 1\}^n$ is a function which takes as input a string $s$ of length at most $n$, and produces a positive real number. The function has the property that for any string $s$ of length less than $n$, $(F(s0) + F(s1))/2 \leq F(s)$. Therefore, appending a uniformly random bit to $s$ \emph{does not increase} the value of $F$ in expectation. The goal of this problem is to compute a string $s \in \{0, 1\}^n$ such that $F(s) \leq \alpha F(\emptyset)$, where $\alpha \geq 1 + 1/\poly(n)$ is an approximation parameter.  Given a pessimistic estimator $F$, we construct a $\nbupdc{}$ instance over $\{0, 1\}^n$ such that any solution contains an approximate solution to the pessimistic estimator instance. This instance consists of one next-bit predictor~$P_i$ at each index $i \in [n]$. Given as input a string $s \in \{0, 1\}^{i-1}$, $P_i$ outputs $1$ with probability $F(s1)/(F(s0) + F(s1))$, and outputs $0$ with probability $F(s0)/(F(s0) + F(s1))$ \footnote{We require predictors to be deterministic. We approximate this random guessing by constructing $T$ deterministic copies of each predictor, where $T$ is a positive integer. Copy $j \in [T]$ guesses $1$ exactly when $F(s1)/(F(s0) + F(s1)) \geq (2j-1)/(2T)$.}. Thus, for a string $s \in \{0, 1\}^n$, if $F(s_{\leq i}) > F(s_{< i})$, the predictor $P_i$ is likely to succeed on $s$. Any string $s$ such that $F(s) > \alpha F(\emptyset)$ must have many such indices, or have a few indices on which the inequality is very strong, i.e., on which $F(s_{\leq i})$ is much larger than $F(s_{< i})$. Thus, on any string that is not an $\alpha$-approximate solution to the pessimistic estimator instance, these predictors are likely to succeed often. As a contrapositive, any string $s \in \{0, 1\}^n$ on which these predictors do not succeed often is an approximate solution to the pessimistic estimator instance. This is exactly what we wanted from our reduction. This argument is formalized in Section~\ref{sec:pess-est-to-nbp}. As a corollary of this reduction, along with the algorithm by Marcussen, Pyne, and Rubinfeld~\cite{MarcussenPyneRubinfeld26}, we obtain a reduction from graph sparsification to $\nbupdc{\TCo}$.

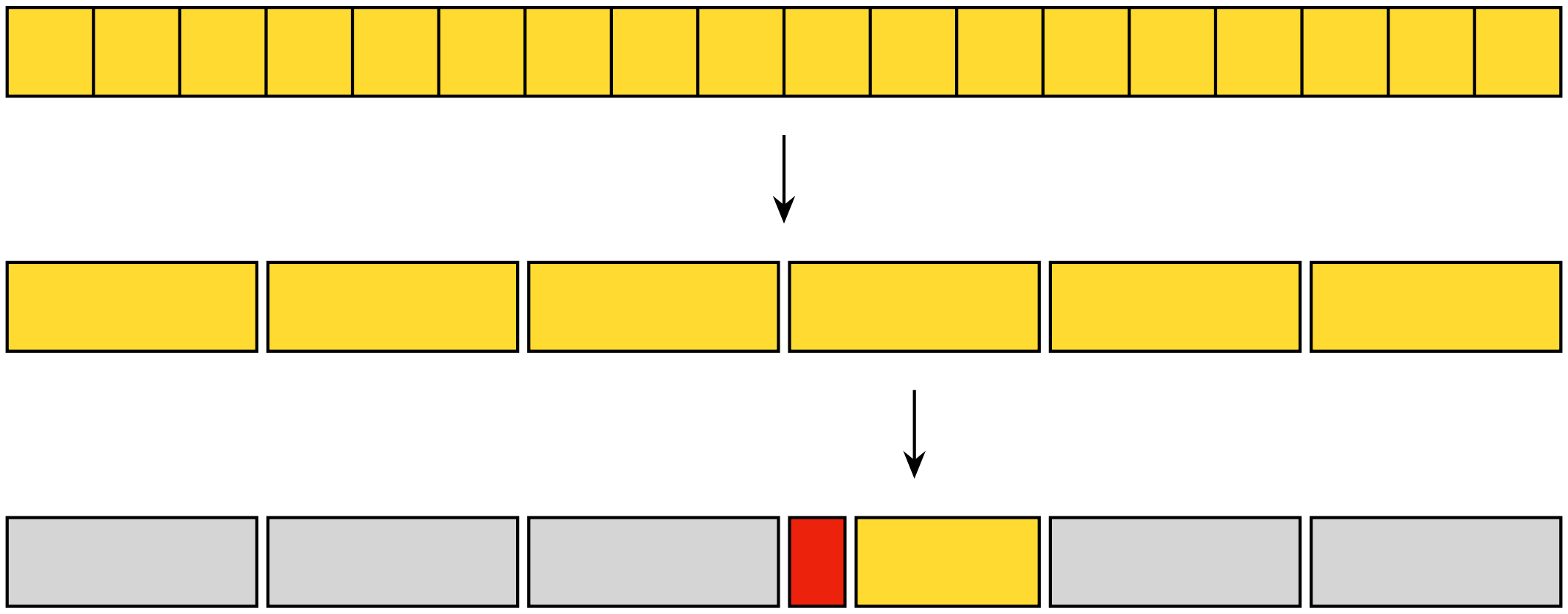
\begin{figure}
    \centering
    \begin{tikzpicture}[
    x=1cm,
    y=1cm,
    line width=.8pt,
    line join=miter,
    block/.style={draw=black},
    arrow/.style={
        draw=black,
        -{Stealth[length=2.5mm,width=2mm]}
    }
]
    \definecolor{blockyellow}{HTML}{FFDA31}
    \definecolor{blockred}{HTML}{ED220D}
    \definecolor{blockgray}{HTML}{D5D5D5}

    \draw[block,fill=blockyellow]
        (0,4.6) rectangle (14,5.4);
    \foreach \i in {1,...,17}
        \draw ({14*\i/18},4.6) -- ({14*\i/18},5.4);

    \foreach \i in {0,...,5}
        \draw[block,fill=blockyellow]
            ({2.35*\i},2.3)
            rectangle ({2.35*\i+2.25},3.1);

    \foreach \i in {0,1,2,4,5}
        \draw[block,fill=blockgray]
            ({2.35*\i},0)
            rectangle ({2.35*\i+2.25},.8);

    \draw[block,fill=blockred]
        (7.05,0) rectangle (7.55,.8);
    \draw[block,fill=blockyellow]
        (7.65,0) rectangle (9.30,.8);

    \draw[arrow] (7,4.25) -- (7,3.45);
    \draw[arrow] (8.175,1.95) -- (8.175,1.15);
\end{tikzpicture}
    \caption{$\blossyc$. The input is split into blocks. A single block is chosen and compressed. The red region represents free space.}
    \label{fig:blocklossyc}
\end{figure}

\paragraph{Reductions from Bipartite Matching, Linear Matroid Intersection, and General Matching to $\updc{\TCo}$.} The compress-or-random algorithms of \cite{AgarwalaMertz25,AgarwalaAlekseevVinciguerra26,ChakrabortyDattaKusreMukhopadhyaySinhababu26} share a common structure: the catalytic tape is split into blocks of $\ell = O(\log n)$ bits each, and the compressor selects one block, depending on the tape, and compresses only that block ($\blossyc$, Definition~\ref{def:block-lossy}, Figure~\ref{fig:blocklossyc}). We give a simple reduction from $\blossyc$ to $\upd$. This reduction first reduces $\blossyc$ to a version of $\upd$ over alphabets of size $\poly(n)$. That is, instead of binary strings we have strings over an alphabet set $A$, and the predictors try to guess a character of this string with probability $\geq 1/|A| + \epsilon$. We show that this large alphabet version of $\upd$ reduces to the binary version. This is done in Section~\ref{sec:block-lossy}.

\subsection{Organization of the Paper}
In Section~\ref{sec:prelims} we give preliminaries. In Section~\ref{sec:upd-in-cl} we give catalytic logspace algorithms for $\updo$ and $\updc{\TCo}$ (Theorem~\ref{intro-thm:updo-in-scl} and Corollary~\ref{intro-cor:upd-tco-in-scl}). In Section~\ref{sec:CL-equiv} we reduce $\SCL$ to $\upd$ (Theorem~\ref{intro-thm:scl-reduces-to-upd}). In Section~\ref{sec:unpred-hardness} we prove oracle lower bounds for $\updo$ and $\nbupdc{\PO}$. In Section~\ref{sec:stablecl} we discuss $\StCL$ versus $\SCL$. In Section~\ref{sec:nbp-and-ess-est} we prove the equivalence between $\nbupdc{}$ and pessimistic estimators, and its consequence for graph sparsification. In Section~\ref{sec:block-lossy} we reduce bipartite matching, linear matroid intersection, and general matching to $\updc{\TCo}$. Section~\ref{sec:ai-discussion} contains our AI disclosure.

\section{Preliminaries}
\label{sec:prelims}

\subsection{Notation}
\paragraph{Numbers and Logic.}$\mathbb{N}$ refers to the natural numbers $\{1, \dots\}$, $\mathbb{Z}$ refers to the integers, and $\mathbb{R}$ refers to the real numbers. We use notation such as $\mathbb{Z}_{\geq0}$ to refer to the set of nonnegative integers. For $n \in \mathbb{Z}_{\geq 0}$, we define $[n] \coloneqq \{1, \dots, n\}$. We use $\log$ to refer to logarithms in base $2$, and $\ln$ for logarithms in base $e$. $\exp(x)$ refers to the function $e^x$. For a Boolean proposition $p$, we use $\mathbbm{1}[p]$ as an indicator function, which is $1$ if $p$ is true and $0$ otherwise.

\paragraph{Strings, Tuples, and Functions.}We use $\emptyset$ to refer to the empty string. Let $s$ be a string of length $n$ and $i \in [n]$ be an index. We use $s_i$ to refer to the $i^{th}$ character of $s$. We use $s_{-i} \coloneqq s_1 \dots s_{i-1}s_{i+1}\dots s_n$ to refer to the string $s$ without its $i^{th}$ character. We use $s_{\leq i} \coloneqq s_1\dots s_i$ to refer to the length $i$ prefix of $s$, and $s_{< i} \coloneqq s_1\dots s_{i-1}$ to refer to the length $i-1$ prefix of $s$. We set $s_{\leq0}\coloneqq\emptyset$ for every string $s$. For two strings $s$ and $t$, $s \circ t$ is the concatenation of $s$ and $t$. We identify tuples of strings with their concatenation. Similarly, for a tuple $s$ of $n$ elements, $s_i$ refers to the $i^{th}$ element in the tuple. For a function $f:X \rightarrow X$ and $i \in \mathbb{Z}_{\geq 0}$, we use $f^{(i)}$ to refer to the function $f$ composed with itself $i$ times. That is, on an input $x$, $f^{(i)}(x) = f(f(\dots f(x)))$, where $f$ is applied $i$ times.

\paragraph{Graphs.}All graphs in this paper are finite. For a graph $G$, we use $V(G)$ to refer to its vertex set, and $E(G)$ for its edge set. A simple undirected graph has edges that are unordered pairs of distinct vertices, with no repeated pairs. A simple directed graph has edges that are ordered pairs of distinct vertices, with no repeated pairs. A directed multigraph may have multiple edges with the same ordered endpoints, including self-loops. In a multigraph, in-degrees and out-degrees count edge copies, and each self-loop contributes one to both.

\paragraph{Multisets.} All multisets in this paper are finite. For a multiset $S$, $|S|$ counts the number of elements in $S$ \emph{with} multiplicity, and $s\in S$ means that $s$ occurs in $S$ with non-zero multiplicity. For a set $X$, we write $S\subseteq X$ if every element of $S$ belongs to $X$. For a non-empty multiset $S$, $U(S)$ is the uniform distribution over the elements of $S$, where an element of $S$ is sampled with probability proportional to its multiplicity. We also write $s\in S$ in probability subscripts as shorthand for $s\sim U(S)$.

\subsection{Standard Complexity Classes}

\paragraph{Languages and Search Problems.} A language $L$ is a subset of $\{0, 1\}^*$. A search problem $P$ is defined to be a subset of $\{0, 1\}^* \times \{0, 1\}^*$. An algorithm solves a search problem $P$ if, on every input $x$, it computes $y$ such that $(x, y) \in P$.

\paragraph{Machine Classes.} $\P$ is the class of languages decidable by a Turing machine in deterministic polynomial time. $\Logspace$ is the class of languages decidable by a Turing machine with logarithmic work space. $\FP$ and $\FL$ are their search problem counterparts.

\paragraph{Randomized Classes.}$\BPP$ and $\BPL$ are the classes of languages decidable by randomized Turing machines with error at most $1/3$ on every input, with $\BPP$ machines restricted to run in polynomial time, and $\BPL$ machines restricted to use at most logarithmic work space. $\ZPP$ is the class of languages decidable by randomized Turing machines with zero error in expected polynomial time. $\SBPP$ consists of search problems solvable by randomized polynomial-time algorithms that output a valid solution with probability at least $2/3$ on every input.

\paragraph{Circuits.}Unless stated otherwise, circuits use $\mathsf{AND}$, $\mathsf{OR}$, $\mathsf{NOT}$, and $\mathsf{THRESHOLD}$ gates. $\mathsf{AND}$, $\mathsf{OR}$, and $\mathsf{THRESHOLD}$ gates have unbounded fan-in, and $\mathsf{NOT}$ gates have fan-in $1$. Circuits may have multiple output bits. For a circuit $C$, $\size(C)$ is the number of gates of $C$, including input gates. $\depth(C)$ is the maximum number of gates on any directed path of $C$. $\TCo$ is the class of problems solvable by logspace-uniform families of polynomial-size, $O(\log n)$-depth circuits. We write $\LUTCo$ when making this uniformity explicit. $\NC$ is the class of problems solvable by logspace-uniform families of polynomial-size, $\polylog(n)$-depth circuits.

\paragraph{Measures of Computation.} For a sequential algorithm $\mathcal{A}$, $\ttime(\mathcal{A})$ is the maximum number of steps that $\mathcal{A}$ takes on any given input, as a function of its input size. For a fixed input $x \in \{0, 1\}^*$, $\ttime(\mathcal{A}(x))$ is the number of steps that $\mathcal{A}$ takes on input $x$. For a parallel algorithm $\mathcal{A}$, $\depth(\mathcal{A})$ denotes the maximum number of sequential steps of $\mathcal{A}$ on any given processor, and $\work(\mathcal{A})$ denotes the total number of operations of $\mathcal{A}$ across all processors.

\paragraph{Search Reductions.}Let $\mathcal{C}$ be a computational class. For search problems $A$ and $B$, $A\leq_{\mathcal{C}}B$ means that there is (1) an instance map $f$ which maps instances of $A$ to instances of $B$; and (2) a solution map $g$ which takes as input an instance of $A$ and a solution of $B$, and outputs a solution of $A$, such that (1) both $f$ and $g$ are computable in $\mathcal{C}$; and (2) for every $x \in \{0, 1\}^*$, if $(f(x), y) \in B$; that is, $y$ is a solution to the instance $f(x)$ for the search problem $B$, then $(x, g(x, y)) \in A$; that is, $g(x, y)$ is a solution to the instance $x$ for the search problem $A$. This definition traces a reduction as follows: in order to solve the problem $A$, a machine can compute $f(x)$, solve $B$ on $f(x)$ to obtain a solution $y$, and then compute $g(x, y)$ to obtain a solution for $x$. In particular, $\mathcal{C}=\P$ means that both maps are computable in deterministic polynomial time, and $\mathcal{C}=\LUTCo$ means that both maps are computable by logspace-uniform $\TCo$ circuit families.

\subsection{\texorpdfstring{Predictors and $\upd$}{Predictors and UNPRED}}

\begin{definition}[$\text{Next-Bit Predictor}$]
Let $n \in \mathbb{N}$. A next-bit predictor over $\{0, 1\}^n$ at index $\ind \in [n]$ is a function $P:\{0, 1\}^{\ind-1} \rightarrow \{0, 1\}$. $P$ succeeds on a string $s \in \{0, 1\}^n$ if
\[P(s_{<\ind}) = s_\ind.\]
That is, given a length $\ind-1$ prefix of the string, $P$ predicts its next bit. 
\end{definition}

\begin{definition}[$\text{Middle-Bit Predictor}$]
Let $n \in \mathbb{N}$. A middle-bit predictor over $\{0, 1\}^n$ at index $\ind \in [n]$ is a function $P:\{0, 1\}^{n-1} \rightarrow \{0, 1\}$. $P$ succeeds on a string $s \in \{0, 1\}^n$ if
\[P(s_{-\ind}) = s_{\ind}.\]
That is, given the entire string except its $\ind^{th}$ bit, $P$ predicts the remaining bit. 
\end{definition}

\begin{definition}[Predictor Success and Advantage]
Let $n \in \mathbb{N}$, let $P$ be a (next or middle)-bit predictor over $\{0, 1\}^n$, and let $\mathbf{D}$ be a probability distribution over $\{0, 1\}^n$. The \emph{success} of $P$ on $\mathbf{D}$ is
\[\succc_\mathbf{D}(P) \coloneqq \Pr_{s \sim \mathbf{D}}[P \succs s].\]
The \emph{advantage} of $P$ on $\mathbf{D}$ is
\[\adv_\mathbf{D}(P) \coloneqq \max \{0, \succc_{\mathbf{D}}(P) - \frac{1}{2}\}.\]
\end{definition}

Let $\mathcal{P}$ be a non-empty family of predictors. We overload the notation above by defining
\[\succc_{\mathbf{D}}(\mathcal{P}) \coloneqq \max_{P \in \mathcal{P}} \succc_{\mathbf{D}}(P).\]
\[\adv_{\mathbf{D}}(\mathcal{P}) \coloneqq \max_{P \in \mathcal{P}} \adv_{\mathbf{D}}(P).\]

\begin{definition}[$\upd$]
\label{def:unpred}
$\upd$ is the search problem defined as follows:
\begin{enumerate}
    \item \textbf{Input:} $n, m \in \mathbb{N}$, $0 < \epsilon \leq 1$, Boolean circuits  $P_1, \dots, P_m$, and $\ind_1, \dots, \ind_m \in [n]$.
    
    \noindent
    For all $i \in [m]$, $P_i:\{0, 1\}^{n-1} \rightarrow \{0, 1\}$ is a middle-bit predictor over $\{0, 1\}^n$ at index $\ind_i \in [n]$.
    \item \textbf{Output:} A non-empty multiset $S \subseteq \{0, 1\}^n$ such that
    \[\adv_{U(S)}(\{P_1, \dots, P_m\}) < \epsilon.\]
\end{enumerate}
\end{definition}

\noindent
To avoid precision issues with $\epsilon$, we assume that we are given an integer $r$ \textbf{in unary} and $\epsilon \coloneqq 1/r$.\footnote{In all reductions to $\upd$ we implicitly convert $\epsilon$ to $1/\lceil 1 / \epsilon \rceil$.} We assume that $n$ is also given in unary.

\paragraph{Variants of $\upd$.}Let $\mathcal{C}$ be a circuit class, such as $\TCo$. $\updc{\mathcal{C}}$ refers to a version of $\upd$ where the predictor circuits belong to the class $\mathcal{C}$. $\updo$ refers to the version where the predictors can only be accessed through queries to a predictor oracle $\PO$; we define this oracle model formally in Section~\ref{sec:oracle-model}. We also consider the subproblem of $\upd$ where each predictor circuit, instead of being a middle-bit predictor, is a next-bit predictor. We refer to the unrestricted version of this as $\nbupdc{}$, the version restricted to circuit class $\mathcal{C}$ as $\nbupdc{\mathcal{C}}$, and the version with oracles as $\nbupdc{\PO}$.

\paragraph{Output Size.}Definition~\ref{def:unpred} does not bound the output size of $\upd$. As is the case with hitting sets, there is typically no need to bound it beyond the resource bounds of the algorithm: for a time-bounded algorithm, the size of the output is naturally bounded by the runtime. Catalytic logspace algorithms, however, are not bounded by polynomial runtime in the worst case. Thus, we will be explicit about output size bounds in our algorithms in this case. In fact, our catalytic algorithms write their entire output on the catalytic tape itself, whose size is polynomially bounded.

\paragraph{Parameters.}If $\epsilon > 1/2$, then the problem is trivial, since $\adv_{\mathbf{D}}(P) \leq 1/2$ for every distribution $\mathbf{D}$ and predictor $P$; thus, we will assume that $\epsilon \leq 1/2$ when convenient. We require $\epsilon > 0$ because we view $\upd$ as a total derandomization problem: if $0 < \delta < 1$, then the Chernoff bound guarantees that a random multiset of size $\geq \lceil\ln(m/\delta)/(2\epsilon^2)\rceil$ (where elements of $\{0, 1\}^n$ are sampled independently with replacement) constitutes a valid output with probability at least $1 - \delta$. At $\epsilon = 0$, no valid output exists\footnote{Even the natural alternative requirement $\succc_{U(S)}(\mathcal{P}) < 1/2$ cannot be met if $\mathcal{P}$ contains both constant predictors at some index.}. We assume that $\epsilon$ is encoded in unary so that the output size of this randomized algorithm, $O(\ln(m/\delta)/\epsilon^2)$, is polynomial in the input size. We assume that $n$ is encoded in unary since, in the oracle version, the input size could otherwise be much smaller than $n$, and a logspace algorithm could not even index over the length of the strings that it needs to output.

\paragraph{Averaging Argument.} The following is a simple fact that is used many times in this paper. It states that for any family of predictors $\mathcal{P}$ and any multiset $S$ such that $\adv_{U(S)}(\mathcal{P})$ is small, there exists a string $s \in S$ such that not too many predictors in $\mathcal{P}$ succeed on $s$.
\begin{lemma}
\label{lem:yao-for-unpred}
Let $n, m \in \mathbb{N}$ and $\epsilon > 0$. Let $\mathcal{P} \coloneqq \{P_1, \dots, P_m\}$ be a family of (middle or next)-bit predictors over $\{0, 1\}^n$. Let $S \subseteq \{0, 1\}^n$ be a non-empty multiset such that
\[\adv_{U(S)}(\mathcal{P}) < \epsilon.\]
Then there exists $s \in S$ such that $\Pr_{P \in \mathcal{P}}[P \succs s] < 1/2 + \epsilon$.
\end{lemma}
\begin{proof}
    Assume to the contrary that for all $s \in S$, $\Pr_{P \in \mathcal{P}}[P \succs s] \geq 1/2 + \epsilon$. Then 
    \begin{align*}
        \Pr_{s \in S, P \in \mathcal{P}}[P \succs s] &\geq 1/2 + \epsilon \\
        \implies \exists P \in \mathcal{P} \text{ such that } \Pr_{s \in S}[{P \succs s}] &\geq 1/2 + \epsilon\\
        \implies \adv_{U(S)}(\mathcal{P}) &\geq \epsilon.
    \end{align*}
    This completes the proof.
\end{proof}

\subsection{Catalytic and In-Place Computation}

Our main computational model in this paper is the catalytic space model:
\begin{definition}[Catalytic Machines]
    Let $s \coloneqq s(n)$ and $c \coloneqq c(n)$. A \emphdef{catalytic Turing machine} with
    space $s$ and \emphdef{catalytic space} $c$ is a Turing machine $M$ with a
    read-only input tape of length $n$, a write-only output tape, a read-write work tape of length $s$,
    and a second read-write work tape of length $c$ called the \emphdef{catalytic tape},
    which will be initialized to an adversarial string $\tau$.
\end{definition}
\begin{definition}[Catalytic Computation of a Language]
\label{def:cat-decision}
    A catalytic Turing machine $M$ decides a language $L \subseteq \{0, 1\}^*$ if for every $x \in \{0, 1\}^n$ and $\tau \in \{0, 1\}^c$, the result of executing $M$ on input $x$ with initial catalytic tape $\tau$ (i.e., $M(x, \tau)$) fulfils two properties: 1) $M$ terminates with $1$ written on the output tape if $x \in L$, and with $0$ otherwise; and 2) $M$ halts with $\tau$ written on the catalytic tape.
\end{definition}

\begin{definition}[Catalytic Decision Classes]
    We define $\CSPACE[s,c]$ to be the family of languages decidable by
    catalytic Turing machines with space $s$ and catalytic space $c$.
    We also define \emphdef{catalytic logspace} as
    $$\CL \coloneqq \bigcup_{d \in \mathbb{N}} \CSPACE[d \log n, d n^d].$$
\end{definition}

Past work defines the search class $\SCL$ with the following semantics:

\begin{definition}[Catalytic Computation of a Search Problem]
\label{def:compute-search-problem}
    A catalytic Turing machine $M$ computes a search problem $P \subseteq \{0, 1\}^* \times \{0, 1\}^*$ if for every $x \in \{0, 1\}^n$ and $\tau \in \{0, 1\}^c$, the result of executing $M$ on input $x$ with initial catalytic tape $\tau$ (i.e., $M(x, \tau)$) fulfils two properties: 1) $M$ terminates with $y$ such that $(x, y) \in P$ written on the output tape; and 2) $M$ halts with the catalytic tape in state $\tau$.
\end{definition}

\begin{definition}[Catalytic Search Classes]
    We define $\SCSPACE[s,c]$ to be the family of search problems computable by
    catalytic Turing machines with space $s$ and catalytic space $c$.
    We also define $\SCL$ as
    $$\SCL \coloneqq \bigcup_{d \in \mathbb{N}} \SCSPACE[d \log n, d n^d].$$
\end{definition}

In this work, we introduce a stronger definition. To see why we need it, suppose that a catalytic logspace machine $M_1$ decides a language $L$ on input $x$ whenever the string on its catalytic tape belongs to a set $A(x)$, and that a catalytic logspace machine $M_2$ computes the search problem $\{(x, y) \mid y \in A(x)\}$. To decide $L$, a catalytic machine $M$ could try to use $M_2$ to compute some $\tau \in A(x)$, write $\tau$ on its catalytic tape, and then run $M_1$. However, Definition~\ref{def:compute-search-problem} does not guarantee that $M$ can afterwards restore its catalytic tape to its original state. We therefore introduce `restorable computation', which requires the catalytic computation to explicitly write the output to the search problem on its catalytic tape, and then restore it back to its original configuration.

\begin{definition}[Restorable Catalytic Computation of a Search Problem]
\label{def:restorable-comp}
A restorable catalytic Turing machine $M$ consists of two procedures: $\compute_M$ and $\restore_M$, which share a read-only input tape, a read-write work tape, and a read-write catalytic tape. $M$ computes a search problem $P \subseteq \{0, 1\}^* \times \{0, 1\}^*$ if, for every input $x$ and every initial catalytic tape $\tau$, it satisfies the following two properties: 1) $\compute_M(x, \tau)$ halts with $y \in \{0, 1\}^*$ such that $(x, y) \in P$ written on the first $|y|$ bits of its catalytic tape; and 2) $\restore_M$ starts from the work tape and catalytic tape configuration that $\compute_M(x, \tau)$ halted with, and halts with $\tau$ written on the catalytic tape.
We refer to the string $y$ as $M(x, \tau)$.
\end{definition}
Similar to the classes $\SCSPACE$ and $\SCL$, we define the classes $\RSCSPACE$ and $\RSCL$. 
\begin{definition}
    We define $\RSCSPACE[s,c]$ to be the family of search problems restorably computable by
    catalytic Turing machines with space $s$ and catalytic space $c$.
    We also define $\RSCL$ as
    $$\RSCL \coloneqq \bigcup_{d \in \mathbb{N}} \RSCSPACE[d \log n, d n^d].$$
\end{definition}

Continuing the example above Definition~\ref{def:restorable-comp}, if the machine $M_2$ \textit{restorably} computed $\tau$, then indeed $M$ could first use $\compute_{M_2}$ to obtain $\tau \in A(x)$ on the catalytic tape, then run the machine $M_1$ to decide $L$, and finally use the $\restore_{M_2}$ procedure to obtain the original catalytic tape again. Previous work on catalytic computation~\cite{CookLiMertzPyne25, AgarwalaMertz25, AgarwalaAlekseevVinciguerra26, ChakrabortyDattaKusreMukhopadhyaySinhababu26} used pairs of `catalytic subroutines' which compress and decompress the catalytic tape, which can be seen as restorable catalytic computation of compression functions. However, this previous work required all the information needed for restoration to be written on the catalytic tape. Our definition allows this information to be stored on the work tape, since $\restore_M$ has access to both the work tape and the catalytic tape of $\compute_M$, not just the catalytic tape.

Definition~\ref{def:restorable-comp} does not specify how the machine knows \emph{which} prefix of the catalytic tape contains $y$. In a catalytic logspace machine $|y|$ would be polynomially bounded, so this could be done in many ways, for example by writing $|y|$ on the work tape. This detail does not affect any of our results.

\paragraph{In-Place Computation.}
\label{sec:in-place-comp}
Recently, Cook, Ghentiyala, Mertz, Pyne, and Sheffield~\cite{CookGhentiyalaMertzPyneSheffield25} introduced in-place computation, in which the input tape is read-write, instead of being read-only. This model captures transformations that a catalytic Turing machine might make on the catalytic tape. For example, suppose the catalytic tape has a matrix $M$ written on it, and the algorithm needs to replace it with $M^{-1}$. Even though each entry of $M^{-1}$ can be computed in catalytic logspace, the algorithm cannot simply overwrite the entries of $M$ one at a time, since each overwritten entry may be needed to compute the remaining entries of $M^{-1}$, and $M$ is much larger than the work tape. In order to study this situation, Cook et al. introduced the class $\ipFL$.

\begin{definition}[~\cite{CookGhentiyalaMertzPyneSheffield25}; $\ipFL$]
An in-place Turing machine $M$ has a read-write input tape and a read-write work tape. $M$ is defined to compute a search problem $P \subseteq \{0, 1\}^* \times \{0, 1\}^*$ if, for every $x \in \{0, 1\}^*$, the Turing machine $M$ when executed on input $x$ terminates with $y$ such that $(x, y)\in P$ written on the first $|y|$ bits of the input tape.

Moreover, we say that $P \in \ipFL$ if there exist an in-place Turing machine $M$ computing $P$ and $d \in \mathbb{N}$ such that, on any input $x$, $M$ uses at most $d \log |x|$ work space.
\end{definition}

We emphasize that, in the above definition, the input tape has size exactly $|x|$, and thus the output, $y$, must have size at most $|x|$ as well.

\subsection{Oracle Model}
\label{sec:oracle-model}
We will describe both algorithms and lower bounds in a query access model, where the predictors are not available to the algorithm in the form of their circuit representations. In this case, an algorithm is explicitly given as input only $n, m, \epsilon$, and $\ind_1, \dots, \ind_m$. For any $i \in [m]$, the algorithm may access the predictor $P_i$ at a specific input $s \in \{0, 1\}^{n-1}$ by producing the tuple $(i, s)$. In response, the algorithm receives the bit $P_i(s)$ on a length $1$ read-only tape. Every time there is a query, this tape is overwritten with the result. This tape only provides $1$ bit of additional space to the algorithm, which is insignificant for the purpose of this work. We refer to this version of the problem simply as $\updo$, and define $\nbupdc{\PO}$ analogously. $\PO$ here refers to the predictor oracle, i.e., the function which takes as input the tuple $(i, s)$ and produces as output $P_i(s)$.

It remains to specify where the tuple $(i, s)$ is written. For time-bounded computation, we have an infinite-length work tape. Thus, the oracle query can simply be written on this work tape. For circuit computation, we define an extra `oracle gate', where input wires to this gate carry the input to the oracle, and the wires leaving this gate carry the output of the oracle. In the case of space-bounded computation, the tuple can be encoded using $\lceil\log(m+1)\rceil + n-1$ bits, which will typically be too large to explicitly write on the work tape. In the catalytic computation model, \textbf{we require any oracle query to be explicitly written on read-write tape(s)}.

Thus, we do not allow an imaginary oracle query tape in our catalytic models. We are being deliberately vague about what it means for the query to be written down. Formally, there should be a logspace-computable injective function which takes as input an index $i$, and produces as output the location of the $i^{th}$ bit of the oracle query. However, for the sake of readability, we will not consider this extra detail. In all of our oracle algorithms, it will be clear where the query is written. Moreover, none of our oracle lower bounds are in space-bounded models, and thus this detail is never exploited.
\section{\texorpdfstring{Catalytic Logspace Algorithms for $\upd$}{Catalytic Logspace Algorithms for UNPRED}}
\label{sec:upd-in-cl}
This section contains our catalytic logspace algorithms for $\updo$ and $\updc{\TCo}$.

We will first give a $\RSCL^{\PO}$ algorithm for $\updo$ in Theorem~\ref{thm:oracle-binunpred-in-searchcl}. This is the main algorithm for the section and contains our key ideas. Then, by replacing the oracle $\PO$ with explicit evaluation of $\TCo$ circuits, we obtain a $\RSCL$ algorithm for $\updc{\TCo}$ in Corollary~\ref{cor:tco-unpred-in-searchcl}.

Now, let us describe our $\RSCL^{\PO}$ algorithm for $\upd^{\PO}$. Our algorithm uses the compress-or-random framework in a way similar to Pyne's algorithm for simulating $\BPL$ in $\CL$~\cite{Pyne24}.

In the $\upd^{\PO}$ problem, we receive as input an integer $n$, indicating that the predictors are over $\{0, 1\}^n$, an integer $m$, indicating that there are $m$ predictors, $m$ integers $\ind_1, \dots, \ind_m \in [n]$ where $\ind_i$ is the index that the $i^{th}$ predictor is trying to predict, and finally the accuracy parameter $\epsilon$. We have query access to the predictors $\mathcal{P} = \{P_1, \dots, P_m\}$.
 
We know that a uniformly random multiset of $ k \coloneqq \Theta(\ln(2m)/\epsilon^2)$ many strings in $\{0, 1\}^n$ constitutes a valid output for $\upd^{\PO}$ with constant probability. The compress-or-random framework suggests that we should treat the catalytic tape as a random string. Thus, we first check whether the bits on the catalytic tape constitute a valid output for the problem. That is, we interpret the first $kn$ bits on the tape as a multiset $S \subseteq \{0, 1\}^n$, compute $\adv_{U(S)}(\mathcal{P})$, and thus check whether $\adv_{U(S)}(\mathcal{P}) < \epsilon$. If this is true, then we already have a valid output on the catalytic tape, so we do not need to do anything more.

What if it turns out that $S$ is \emph{not} a valid output? The compress-or-random framework dictates that in such a case one should try to compress the representation of $S$. The fact that $S$ is not a valid output for this instance means that we have a predictor $P_i$ such that $\adv_{U(S)}(P_i) \geq \epsilon$. In other words, the predictor $P_i$ performs \textit{too} well on $S$. Our approach towards compression is as follows: let $S = \{s^1, \dots, s^k\}$ where each $s^j$ is a string in $\{0, 1\}^n$. For every $j \in [k]$, we replace the $(\ind_i)^{th}$ bit of $s^j$, i.e., $s^j_{\ind_i}$, with a bit $s'_j$ which we define to be $0$ if and only if $P_i$ succeeds on the string $s^j$. Observe that the original bit $s^{j}_{\ind_i}$ can be recovered by first computing the result of $P_i$ on the remaining bits of $s^j$, and then taking the $\mathsf{XOR}$ of the result with $s'_j$.

Clearly, what we just did is not compression. For each $j \in [k]$, we replaced a single bit $s^j_{\ind_i}$ with another bit $s'_j$. Moreover, in order to recover the original bit $s^j_{\ind_i}$, we now need to remember $i$ as well. Thus, not only did we not save space, but we now need extra space to store $i$!

The key observation here is as follows: consider the length $k$ string $s' \coloneqq s'_1 \dots s'_k$. The fact that the predictor $P_i$ had $\adv_{U(S)}(P_i) \geq \epsilon$ implies that $P_i$ failed on at most $(1/2 - \epsilon)k$ many strings from $S$. Thus, the string $s'$ has at most $(1/2 - \epsilon)k$ many $1$s in it. This shows that $s'$ has very low Hamming weight. Thus, although the string $s'$ is not shorter than the string it replaced, it has significantly more structure.

We prove in Lemma~\ref{lem:low-hamming-weight-compression} that such a string $s'$ with low Hamming weight can be efficiently compressed. Moreover, it can be compressed enough to also store the index $i$ and still be shorter than $k$ bits. This makes intuitive sense: Hoeffding's inequality shows that $\textit{most}$ strings of length $k$ have Hamming weight approximately $k/2$. Thus, if a string has Hamming weight far lower than $k/2$, one should be able to represent the string using significantly fewer than $k$ many bits.

Therefore, whenever the catalytic tape contains a multiset $S \subseteq \{0, 1\}^n$ which is \textit{not} a solution to the $\updo$ instance, we can compress the catalytic tape. We do this repeatedly until either the catalytic tape contains a string which \textit{is} a solution, or until we have freed up $kn$ bits on the catalytic tape. In the former case, we have a solution! In the latter case, we now have enough free space on the catalytic tape to run an algorithm which brute forces over all multisets of size $k$ until it finds a solution. In either case, we have the solution to the input instance \textit{explicitly written on the catalytic tape} at some point. We can then decompress everything and thus restore the catalytic tape back to its initial state. Thus, we obtain a $\RSCL^{\PO}$ algorithm for $\updo$.

This concludes our overview. We now prove our lemma which shows that strings with unexpectedly low Hamming weight can be compressed.

\begin{lemma}[Low Hamming Weight Strings Can Be Compressed]
\label{lem:low-hamming-weight-compression}
There exists a pair of $\ipFL$ algorithms $\mathsf{HamComp}$ and $\mathsf{HamDecomp}$ as follows. Let $n \in \mathbb{N}$.
\begin{enumerate}
    \item $\mathsf{HamComp}$ takes as input $s \in \{0, 1\}^n$ with $\sum_{i=1}^n s_i < n/2$. Let $\delta \coloneqq 1 - 2/n \sum_{i=1}^n s_i$. $\mathsf{HamComp}$ replaces $s$ with a string $(0^y, t)$ where $y \coloneqq \lfloor \delta^2n\log_2(e)/4\rfloor$ and $t \in \{0, 1\}^{n-y}$.
    \item For any string $s \in \{0, 1\}^n$ with $\sum_{i=1}^n s_i < n/2$, \[\mathsf{HamDecomp}(\mathsf{HamComp}(s)) = s.\]
    That is, $\mathsf{HamDecomp}$ takes as input the output of $\mathsf{HamComp}(s)$, $(0^y, t)$ above, and replaces it with the original string $s$.
\end{enumerate}
\end{lemma}
\begin{proof}
The proof relies on a re-indexing of length $n$ binary strings which orders strings by their Hamming weight (in increasing order), and breaks ties lexicographically. Let $\lex_n(s)$ denote the lexicographic rank of $s$ amongst all length $n$ binary strings which have the same Hamming weight as $s$. Let $s \coloneqq b \circ s'$ be a length $n$ binary string with Hamming weight $k$ where $s' \in \{0, 1\}^{n-1}$ and $b \in \{0, 1\}$. Note that
\begin{equation}
\label{eq:lex-recursion}
    \lex_n(s) = \begin{cases}
        \lex_{n-1}(s'),& \text{if }b = 0,\\
        \lex_{n-1}(s') + \binom{n-1}{k},& \text{otherwise,}
    \end{cases}
\end{equation}
and $\lex_0(\emptyset) \coloneqq 0$. We define $\mathsf{reind}_n(s) \coloneqq \lex_n(s) + \sum_{i=0}^{k-1} \binom{n}{i}$.

The algorithm for $\mathsf{HamComp}$ is as follows:
\begin{enumerate}
    \item Initialize $k_0 = 0$ on the work tape.
    \item Iterate $i$ from $1$ to $n$:
    \begin{enumerate}
        \item Let $s^i$ be the length $i$ suffix of $s$, and let $k_i$ be the Hamming weight of $s^i$.
        \item Replace $k_{i-1}$ on the work tape with $k_i$. To do this, we only need to add $1$ to $k_{i-1}$ if $s_{n-i+1} = 1$. 
        \item Replace $s^i$ with $\lex_i(s^i)$ on the input tape\footnote{Note that this is possible due to the fact that $\lex_i(s^i) < 2^i$, and thus fits inside $i$ bits of space occupied by $s^i$.}. $\lex_{i-1}(s^{i-1})$ is already written on the input tape. Using Equation \ref{eq:lex-recursion}, computing $\lex_i(s^i)$ only involves adding $\binom{i-1}{k_i}$ to $\lex_{i-1}(s^{i-1})$,~if $s_{n-i+1} = 1$. Computing the binomial coefficient can be done in logspace \cite{HesseAllenderMixBarrington02, BuhrmanCleveKouckyLoffSpeelman14}, and the addition is easy to do in place. For this step, $s_{n-i+1}$ can first be temporarily written on the work tape and replaced with $0$ on the input tape.
    \end{enumerate}
    \item Now, $\lex_n(s)$ is written on the input tape and $k \coloneqq k_n$ is written on the work tape. Replacing $\lex_n(s)$ with $\mathsf{reind}_n(s)$ just involves adding $\sum_{i=0}^{k-1}\binom{n}{i}$ to $\lex_n(s)$. Since each binomial coefficient can be computed in logspace and each addition can be done in place, this can be done in $\ipFL$.
\end{enumerate}
By definition, $\mathsf{reind}_n(s) < \sum_{i=0}^k \binom{n}{i}$. For $k \leq (1-\delta)n/2$, by the Chernoff bound, this is $\leq e^{-\delta^2n/4} \cdot 2^n$. This implies that 
\[\mathsf{reind}_n(s) < 2^n \cdot 2^{-\delta^2n\log_2(e)/4} \leq 2^{n-y}.\]
Thus, the binary encoding of $\mathsf{reind}_n(s)$ has a prefix of at least $y$ many $0$s.

We now proceed to the decompression algorithm. We define the following function:
\begin{equation}
\label{eq:lex-inv-recursion}
    \mathsf{lex}_{n, k}^{-1}(r) \coloneqq 
    \begin{cases}
        0 \circ \lex_{n-1, k}^{-1}(r), &\text{if } r < \binom{n-1}{k},\\
        1 \circ \lex^{-1}_{n-1, k-1}(r - \binom{n-1}{k}), &\text{otherwise,}
    \end{cases}
\end{equation}
and $\lex^{-1}_{0, 0}(0) \coloneqq \emptyset.$
Simply observe that for a string $s \in \{0, 1\}^n$ of Hamming weight $k$, we have $\lex_{n, k}^{-1}(\lex_n(s)) = s$.

The algorithm for $\mathsf{HamDecomp}$ is as follows:
\begin{enumerate}
    \item First, from $r = \mathsf{reind}_n(s)$ on the input tape, write down the Hamming weight $k$ of $s$ on the work tape. This is easy to compute: it is simply the minimum integer $k$ such that $\sum_{i=0}^k \binom{n}{i} > r$. The sum of binomial coefficients can be computed in logspace, as can the comparison of $n$-bit integers.
    \item Once we have $k$, subtract $\sum_{i=0}^{k-1} \binom{n}{i}$ from $r$ on the input tape. Now, we simply want to compute $\lex_{n, k}^{-1}(r)$, where $r$ is the string on the input tape, and $k$ is written on the work tape.
    \item Use Equation~\ref{eq:lex-inv-recursion} to replace $r$ with $\lex_{n, k}^{-1}(r)$ by iteratively replacing the prefix in rounds, similar to the algorithm for $\mathsf{HamComp}$, where we replaced the suffix.
\end{enumerate}
This completes the proof.
\end{proof}

The following lemma uses the compression of low Hamming weight strings from Lemma~\ref{lem:low-hamming-weight-compression} in order to compress a multiset $S \subseteq \{0, 1\}^n$ which does not constitute a valid output to a $\updo$ instance.

\begin{lemma}[Non-Solutions of $\upd$ Can Be Compressed]
\label{lem:unpred-compression}
Let ($n$, $m$, $\epsilon$, $\mathcal{P} = \{P_1, \dots, P_m\}$, $\ind_1, \dots, \ind_m$) be a $\updo$ instance, and let $k \in \mathbb{N}$ such that $\lfloor \epsilon^2k\log_2(e)\rfloor \geq \lceil\log_2 m\rceil + 1$. Let $\tau \in \{0, 1\}^*$ be a string of length $\geq kn$ written on the catalytic tape, where the first $kn$ bits of $\tau$ are interpreted as $k$ binary strings of length $n$. Let $S$ be the multiset formed by these $k$ strings. There exists a pair of $\CL^{{\mathcal{O}}_P}$ subroutines $\mathsf{Comp}$ and $\mathsf{Decomp}$ such that, if $\adv_{U(S)}(\mathcal{P}) \geq \epsilon$, then:
\begin{enumerate}
    \item $\mathsf{Comp}$ replaces $S$ on the catalytic tape with $(0, t)$, where $t \in \{0, 1\}^{nk - 1}$.
    \item $\mathsf{Decomp}$ replaces $(0, t)$ on the catalytic tape with $S$.
\end{enumerate}
\end{lemma}
\begin{proof}
Interpret the first $kn$ bits of $\tau$ as $s^1_1s^2_1 \dots s^k_1 \dots s^1_n\dots s^k_n$. Let $S \coloneqq \{s^1, s^2, \dots, s^k\}$. That is, we assume that the strings are written in column-major order: the first $k$ bits of $\tau$ are the first bit of each of the $k$ strings, and so on. Of course, we can easily reorder strings between column-major and row-major order. We choose column-major ordering because it is convenient for us.

We will refer to the positions of the catalytic tape containing $s^1_i \dots s^k_i$ as the $i^{th}$ \emph{block} of the catalytic tape.
The algorithm for $\mathsf{Comp}$ is as follows:
    \begin{enumerate}
        \item\label{step:comp-find-predictor} If $\adv_{U(S)}(\mathcal{P}) \geq \epsilon$, then there exists $i^* \in [m]$ such that $\adv_{U(S)}(P_{i^*}) \geq \epsilon$. Such an $i^*$ can be computed in catalytic logspace by iterating over all $i \in [m]$ and explicitly computing $\adv_{U(S)}(P_{i})$ for all $i$.
        \item\label{step:comp-record-failures} Let $j \coloneqq \ind_{i^*}$. Store $i^*$ and $j$ on the work tape. For all $i \in [k]$, replace $s^i_j$ on the catalytic tape with $s'_i$ defined as follows:
        \[s'_i \coloneqq \begin{cases}
            0, &\text{if }P_{i^*} \succs s^i,\\
            1, &\text{otherwise.}
        \end{cases}\]Let $s' \coloneqq s'_1 \dots s'_k \in \{0, 1\}^k$.
        \item Due to the fact that $\adv_{U(S)}(P_{i^*}) \geq \epsilon$, the string $s'$ has Hamming weight $\leq (1 - 2\epsilon) \cdot k/2$.
        \item Apply $\mathsf{HamComp}$ from Lemma~\ref{lem:low-hamming-weight-compression} on input $s'$ to replace $s'$ on the catalytic tape with $(0^{\lceil \log m\rceil + 1}, \bar{s})$, where $\bar{s} \in \{0, 1\}^{k - \lceil \log m\rceil - 1}$.
        \item Replace $(0^{\lceil \log m\rceil + 1}, \bar{s})$ on the catalytic tape with $(0, i^* - 1, \bar{s})$.
        \item\label{step:comp-swap-blocks} Swap the $j^{th}$ block of the catalytic tape, now containing $(0, i^* - 1, \bar{s})$, with the $1^{st}$ block.
    \end{enumerate}
The algorithm for $\mathsf{Decomp}$ is as follows:
\begin{enumerate}
    \item We have $(0, i^* - 1, \bar{s})$ written in the $1^{st}$ block of the catalytic tape, and want to replace it with $s^1_{j} \dots s^k_{j}$.
    \item Let $j \coloneqq \ind_{i^*}$. Write down $j$ and $i^*$ on the work tape.
    \item In order to undo Step~\ref{step:comp-swap-blocks} of $\mathsf{Comp}$, swap the $j^{th}$ block on the catalytic tape with the $1^{st}$ block, which contains $(0, i^* - 1, \bar{s})$.
    \item Replace $(0, i^* - 1, \bar{s})$ in the $j^{th}$ block of the catalytic tape with $(0^{\lceil \log m\rceil + 1}, \bar{s})$. That is, just replace $i^* - 1$ with $0$s.
    \item Now, use $\mathsf{HamDecomp}$ from Lemma~\ref{lem:low-hamming-weight-compression} to replace $(0^{\lceil \log m\rceil + 1}, \bar{s})$ with $s'$. 
    \item\label{step:decomp-restore-bits} Note that $s'_i = 0$ if and only if $P_{i^*}$ succeeds on $s^i$. Moreover, for each $i \in [k]$, we have $s^i_{-j}$ written on the catalytic tape. Thus, for each $i \in [k]$, we can re-obtain $s^i_j \coloneqq s'_i \oplus P_{i^*}(s^i_{-j})$.
\end{enumerate}
Note that the algorithms $\mathsf{Comp}$ and $\mathsf{Decomp}$ use the predictor oracles only in Steps~\ref{step:comp-find-predictor} and~\ref{step:comp-record-failures} of $\mathsf{Comp}$, and in Step~\ref{step:decomp-restore-bits} of $\mathsf{Decomp}$. In all three cases, the input to the predictor is explicitly written down on the catalytic tape. Thus, these algorithms fit within our catalytic oracle model. This completes the proof.
\end{proof}

Finally, we can combine our ideas to obtain a $\RSCL^{\PO}$ algorithm for $\updo$. The idea is straightforward: at each stage either the string on the catalytic tape constitutes a valid output, or, by Lemma~\ref{lem:unpred-compression}, it can be further compressed. In the former case, we are done. In the latter case, we can go ahead and compress. Eventually, either the catalytic tape will contain a valid output, or we will obtain a sufficient amount of free space on the catalytic tape to perform a brute-force search.

\begin{theorem}
\label{thm:oracle-binunpred-in-searchcl}
\[\updo \in \RSCL^{\PO} \subseteq \SCL^{\PO}.\]
Moreover, for an instance with parameters $(n, m, \epsilon)$, the algorithm outputs a multiset of size 
\[\lceil \frac{\log (8m)}{\epsilon^2} \rceil = O(\log (2m)/\epsilon^2).\]
\end{theorem}
\begin{proof}
We will describe the $\RSCL^{\mathcal{O}_P}$ algorithm that solves a $\updo$ instance. Let the input instance be $(n, m, \epsilon, \mathcal{P} = \{P_1, \dots, P_m\}, \ind_1, \dots, \ind_m)$. Let \[k \coloneqq \lceil \frac{\log (8m)}{\epsilon^2} \rceil.\]
The multiset that we output will have size $k$. It's easy to see that $\lfloor \epsilon^2k\log (e)\rfloor \geq \lceil\log_2 m\rceil + 1$, and that $k = O(\log (2m)/\epsilon^2)$.
We divide the catalytic tape $\tau$ into two parts:
\begin{enumerate}
    \item The first part has length $kn$. We will call this the `output section'.
    \item The second part has length $kn$. We will call this the `reserve section'.
\end{enumerate}
The algorithm works as follows:
\begin{enumerate}
    \item Interpret the $kn$ bits of the output section of $\tau$ as $S = s^1_1s^2_1 \dots s^k_1 \dots s^1_n\dots s^k_n$. Each $s^i$ is a string in $\{0, 1\}^n$, and $S$ is thus a size $k$ multiset of strings in $\{0, 1\}^n$.
    \item Maintain a counter $x$ on the work tape, which is initialized with value $1$.
    \item\label{step:unpred-check-solution} Check whether $S$ is a valid output for the input $\updo$ instance. To do this, for each predictor $P_i$ and each $s \in S$, check whether $P_i \succs s$. Count the number of strings on which $P_i$ succeeds, and simply check whether this number is at least $k(1/2 + \epsilon)$. All of this can be done in logspace, treating the catalytic tape as read-only. 
    \item If $S$ constitutes a valid output for this instance, simply halt the compute procedure and skip to Step~\ref{step:unpred-restore} for restoration. 
    \item Else, run $\mathsf{Comp}$ from Lemma~\ref{lem:unpred-compression} to compress $S$ down to a $0$ followed by $kn - 1$ bits. Swap this $0$ with the $x^{th}$ bit in the reserve section. Increment $x$.
    \begin{enumerate}[ref=\theenumi(\alph*)]
        \item\label{step:unpred-bruteforce} If $x > kn$, then the reserve section is now a blank string of $kn$ $0$s. Use this reserve section to brute force over all possible size $k$ multisets of length $n$ binary strings, and check whether each of them is a valid solution for this instance (using the algorithm in step~\ref{step:unpred-check-solution}). We know that at least one of them will be, so as soon as we find one, swap the reserve section with the output section and halt the compute procedure. This swap is to ensure that the output, which is in the reserve section, occupies the first $kn$ bits of the catalytic tape. As soon as the restoration procedure begins, swap the reserve section and the output section again, revert the reserve section to all $0$s, and move to Step~\ref{step:unpred-restore}.
        \item Else, return to Step~\ref{step:unpred-check-solution}.
    \end{enumerate}
    \item\label{step:unpred-restore} While $x > 1$, swap the $(x-1)^{th}$ bit in the reserve section with the first bit of the output section, run $\mathsf{Decomp}$ from Lemma~\ref{lem:unpred-compression} on the output section, and decrement $x$.
\end{enumerate}
By the compression and decompression properties of Lemma~\ref{lem:unpred-compression}, we are guaranteed to have the original catalytic tape upon termination of the restoration procedure. This algorithm uses the predictor oracle in Steps~\ref{step:unpred-check-solution} and~\ref{step:unpred-bruteforce}, and in the Lemma~\ref{lem:unpred-compression} subroutines. In each of these uses, the input to the predictor oracle is explicitly written on the catalytic tape. Thus, this algorithm fits within our catalytic oracle model, as defined in Section~\ref{sec:oracle-model}. This completes the proof.
\end{proof}

Replacing oracle queries in the algorithm of Theorem~\ref{thm:oracle-binunpred-in-searchcl} with explicit circuit evaluations implies the following corollary.

\begin{corollary}
\label{cor:tco-unpred-in-searchcl}
    \[\updc{\TCo} \in \RSCL \subseteq \SCL.\]
Moreover, for an instance with parameters $(n, m, \epsilon)$, the algorithm outputs a multiset of size 
\[\lceil \frac{\log (8m)}{\epsilon^2} \rceil = O(\log (2m)/\epsilon^2).\]
\end{corollary}
\begin{proof}
Buhrman, Cleve, Kouck\'{y}, Loff, and Speelman~\cite{BuhrmanCleveKouckyLoffSpeelman14} showed that\[\TCo \subseteq \CL.\]
Thus, if the predictors are given as input in the form of $\TCo$ circuits, then one can directly use the oracle algorithm in Theorem~\ref{thm:oracle-binunpred-in-searchcl}, and substitute every oracle call to the predictors with an explicit circuit evaluation using the algorithm from~\cite{BuhrmanCleveKouckyLoffSpeelman14}. This completes the proof. 
\end{proof} 

\section{\texorpdfstring{Catalytic Logspace Reduces to $\upd$}{Catalytic Logspace Reduces to UNPRED}}
\label{sec:CL-equiv}
In this section we show that $\upd \in \FP$ would imply $\SCL \subseteq \FP$, which would of course imply that $\CL \subseteq \P$.

It was shown by Buhrman, Cleve, Kouck\'{y}, Loff, and Speelman~\cite{BuhrmanCleveKouckyLoffSpeelman14} that $\CL \subseteq \ZPP$. Their proof follows from the fact that for any catalytic logspace machine $M$, there exists a logspace-computable polynomial $p_M:\mathbb{N} \rightarrow \mathbb{N}$ such that for any input $x \in \{0, 1\}^*$, 
\[\mathbb{E}_{\tau}[\ttime(M(x, \tau))] \leq p_M(|x|).\]
Thus, a randomized algorithm could sample a uniformly random catalytic tape $\tau$ and run $M(x, \tau)$. If $M$ terminates within $2p_M(|x|)$ steps, then this is a polynomial-time simulation. Else, after $2p_M(|x|)$ many steps it could sample a fresh catalytic tape and start again. By Markov's inequality, the probability over a uniformly random catalytic tape $\tau$ that $M(x, \tau)$ takes at most $2p_M(|x|)$ time is $\geq 1/2$. Therefore, excluding the time required to sample $\tau$, the expected runtime of this algorithm is $O(p_M(|x|))$, which implies that it is a $\ZPP$ algorithm. 

A natural approach towards derandomizing this algorithm is to solve the following problem deterministically in polynomial time:
\begin{definition}[Fast Catalytic Tape Construction]
Let $M$ be a catalytic logspace machine which, on an input of length $n$, uses at most $q(n)$ catalytic space. Let $p:\mathbb{N} \rightarrow \mathbb{N}$ be a function such that $p(n) = \poly(n)$. $\fct(M, p)$ is a search problem defined as follows:
\begin{enumerate}
    \item \textbf{Input:} $x \in \{0, 1\}^*$.
    \item \textbf{Output:} $\tau \in \{0, 1\}^{q(|x|)}$ such that $\ttime(M(x, \tau)) \leq p(|x|)$.
\end{enumerate}
\end{definition}

We will show in this section (Theorem~\ref{thm:fct-to-upd}) that for every catalytic logspace machine $M$ there exists a logspace-computable polynomially bounded function $p_M$ such that $\fct(M, p_M) \leq_{\P} \upd$. Thus, $\upd \in \FP$ would imply $\fct(M, p_M) \in \FP$, which would imply that~$M$ can be simulated by a polynomial-time machine.

Our reduction from $\fct(M, p_M)$ to $\upd$ goes through reductions to two problems that we have not formally introduced yet. The first is the $\lossyc$ problem (Definition~\ref{def:lossy-coding}). The second is the $\weakex$ problem (Definition~\ref{def:weakex}). These are both total derandomization problems.

In the $\lossyc$ problem you are given a pair of functions: one which compresses an $n$-bit string to $n-1$ bits, and one which decompresses an $n-1$-bit string to $n$ bits. You are asked to produce as output an $n$-bit string which is \emph{not} compressed and decompressed correctly by this pair of functions. Cook, Li, Mertz, and Pyne (Theorem~\ref{thm:clmp}) proved that $\fct(M, p_M)$ reduces to an instance of $\lossyc$ where both functions are computable in polynomial time. Moreover, both functions are actually computable in-place using polynomial time and logarithmic work space.

In Section~\ref{sec:lossy-to-weakex}, we reduce such instances of $\lossyc$ to the $\weakex$ problem. In this problem you are given as input a highly structured yet exponentially large directed graph in the form of its compressed circuit representation. You are required to produce as output a vertex of this graph whose out-degree is at least as large as its in-degree. We defer further intuition about this problem until after we have formally defined the class of graphs and the problem, in Section~\ref{sec:weakex-to-upd}. 

Thus, we walk backwards through our reductions. First, we reduce $\weakex$ to $\upd$ in Section~\ref{sec:weakex-to-upd}. Then, we reduce certain families of $\lossyc$ to $\weakex$ in Section~\ref{sec:lossy-to-weakex}. Finally, we complete our reduction from $\fct$ to $\upd$ in Section~\ref{sec:cl-to-upd-conc}.

\subsection{\texorpdfstring{$\weakex$ Reduces to $\upd$}{WeakExcess Reduces to UNPRED}}
\label{sec:weakex-to-upd}

Before we can define the $\weakex$ problem, we need to define a well-structured family of graphs that we work with. We call these \emph{deterministic hypercube digraphs}. These are directed multigraphs over the vertex set $\{0, 1\}^n$. For every edge $(u, v)$ in such a graph, $u$ and $v$ differ in at most one bit. Finally, each edge of the graph has a \emph{label} in $\{0, 1\}^\ell$. For each vertex $v$ and label $l \in \{0, 1\}^\ell$, there must be at most one edge leaving $v$ which has the label $l$.

Although our statements and proofs are general, we focus on the parameter regime~$\ell = O(\log n)$.

\begin{definition}[Deterministic Hypercube Digraph]
\label{def:det-hypercube-digraph}
Let $n, \ell \in \mathbb{N}$. A deterministic hypercube digraph of dimension $n$ with $\ell$-bit labels is defined to be a directed multigraph \[G = (V, E, \lab:E \rightarrow \{0, 1\}^{\ell}, \ind:\{0, 1\}^{\ell} \rightarrow [n]),\] where $V$ is the vertex set, $E$ a multiset of edges, $\lab$ is an edge-labelling function on $E$ with labels in the set $\{0, 1\}^{\ell}$, and $\ind$ is a function which maps labels in $\{0, 1\}^{\ell}$ to indices in $[n]$. The constraints of this definition are as follows:
\begin{enumerate}
    \item $V \coloneqq \{0, 1\}^n$.
    \item For each vertex $v \in V$ and label $l \in \{0, 1\}^{\ell}$, there is at most one edge $e$ leaving $v$ with $\lab(e) = l$.
    \item For all edges $e = (v, v') \in E$ we have $|\{i \mid v_i \neq v'_i\}| \leq 1$.
    
     Moreover, $\{i \mid v_i \neq v'_i\} \subseteq \{\ind(\lab(e))\}$. That is, for each edge label $l \in \{0, 1\}^{\ell}$, the value $\ind(l)$ tells you which index of the vertex string could change when you traverse an edge labelled with~$l$.
\end{enumerate}
\end{definition}

Now we define the $\weakex$ problem. In this problem you are given a deterministic hypercube digraph $G$ of dimension $n$ with $\ell$-bit labels and you need to compute a vertex $v \in \{0, 1\}^n$ such that $\outdeg_G(v) \geq \indeg_G(v)$. The catch of course is that the graph has $\exp(n)$ size. Thus, you are given this graph with its representation compressed into a Boolean circuit. You may send the circuit queries of the form $(v, l)$, where $v \in \{0, 1\}^n$ and $l \in \{0, 1\}^{\ell}$. In response, the circuit accepts if and only if there exists a non-self-loop edge in $G$ leaving $v$ with label $l$. We refer to this as an $\ee(v, l)$ query.

\begin{definition}[$\weakex$]
\label{def:weakex}
$\weakex$ is a total search problem defined as follows.
\begin{enumerate}
    \item \textbf{Input:} The input consists of integers $n \in \mathbb{N}$ and $\ell \in \mathbb{N}$, given in unary. Then, a circuit $C:\{0, 1\}^n \times \{0, 1\}^{\ell} \rightarrow \{0, 1\}$. Finally, $\ind:\{0, 1\}^{\ell} \rightarrow [n]$, explicitly given as a table. 
    
    There is a hidden deterministic hypercube digraph of dimension $n$ with $\ell$-bit labels. Let this be $G = (V, E, \lab, \ind)$. $G$ can be accessed via the circuit $C$. $C(v, l) = 1$ if and only if there exists a non-self-loop edge $e \in E$ leaving $v$ with label $l$. Else, $C(v, l) = 0$. Let this value be denoted by $\ee(v, l)$.
    \item \textbf{Output:} The output consists of $v \in \{0, 1\}^n$ such that
    \[\outdeg_G(v) \geq \indeg_G(v).\]
\end{enumerate}
\end{definition}

$\weakex$ resembles the problem $\mathsf{Excess}$, introduced by Beame, Cook, Edmonds, Impagliazzo, and Pitassi~\cite{BeameCookEdmondsImpagliazzoPitassi98}, in which one is given access to a graph $G$, along with a vertex $v^*$ where $\outdeg_G(v^*) > \indeg_G(v^*)$, and is asked to compute a vertex $v$ such that $\indeg_G(v) > \outdeg_G(v)$. It is unclear to us whether there is any connection between these problems beyond the superficial.

$\weakex$ is an auxiliary problem that we use for the purpose of simplifying our exposition. We do not claim that it is worth studying in its own right, though it may well be. Nevertheless, in order to provide intuition for this problem we first show that $\weakex$ is a total derandomization problem. That is, a randomly sampled multiset $S \subseteq V$ of size $\Theta(2^\ell)$ contains a solution to $\weakex$ with constant probability. 

\begin{observation}
Let $G = (V, E, \lab, \ind)$ be a deterministic hypercube digraph of dimension $n$ with $\ell$-bit labels, let $k \in \mathbb{N}$ be a positive integer, and let $S \subseteq \{0, 1\}^n$ be a random multiset of size $k$, where each string in $S$ is drawn independently and uniformly at random from $\{0, 1\}^n$ with replacement. Then
\[\Pr[S \text{ contains a solution to }\weakex] \geq 1 - (1 - \frac{1}{2^{\ell}+1})^k \geq 1 - \frac{1}{e^{k/(2^{\ell}+1)}}.\]
Thus, by setting $k \geq 2(2^{\ell}+1)$, $\weakex$ becomes a derandomization problem.
\end{observation}
\begin{proof}
This is due to the fact that for all $v \in V$, we have $\outdeg_G(v) \leq 2^{\ell}$. Thus, for every $2^{\ell}$ vertices that have $\indeg_G(v) > \outdeg_G(v)$, there must exist at least one vertex which has $\outdeg_G(v) \geq \indeg_G(v)$. Therefore, 
\begin{align*}
    \Pr_{v \sim U(\{0, 1\}^n)}[\outdeg_G(v) \geq \indeg_G(v)] &\geq \frac{1}{2^\ell + 1}\\
    \implies \Pr_{S \subseteq \{0, 1\}^n}[\nexists v \in S \text{ with }\outdeg_G(v) \geq \indeg_G(v)] &\leq (1 - \frac{1}{2^\ell + 1})^k\\
    \implies \Pr_{S \subseteq \{0, 1\}^n}[\exists v \in S \text{ with }\outdeg_G(v) \geq \indeg_G(v)] &\geq 1 - (1 - \frac{1}{2^\ell + 1})^k\\
    &\geq 1 - \frac{1}{e^{k/(2^{\ell}+1)}}.
\end{align*}
This completes the proof.
\end{proof}
We will now start proving our reduction from $\weakex$ to $\upd$. 

First, we introduce a bit of notation that will make the rest of this section far easier to read. For a string $s \in \{0, 1\}^{n-1}$, an index $i \in [n]$, and a bit $b \in \{0, 1\}$, we define $s^{i, b}$ to be the string in $\{0, 1\}^n$ obtained by adding the bit $b$ in the $i^{th}$ position of $s$. That is, $s^{i, b} \coloneqq s_1s_2 \dots s_{i-1} \circ b \circ s_{i} \dots s_{n-1}$.

In the following definition we will formally define the predictors that will be the result of our reduction. Before we do this, we briefly discuss the intuition underlying this definition. 

Imagine the following family of middle-bit predictors: for each label $l \in \{0, 1\}^{\ell}$, there is a predictor $P_{l}$ over $\{0, 1\}^n$ at index $\ind(l)$. Given a string $s \in \{0, 1\}^{n-1}$, $P_l$ has two options: if it outputs $0$, then it is guessing that the completion of $s$ is the string $s^{\ind(l), 0}$, and otherwise it is guessing that the completion is $s^{\ind(l), 1}$. Thus, $P_l$ needs to decide between both of these strings. In order to do this, $P_l$ looks at the set of edges between the vertices $s^{\ind(l), 0}$ and $s^{\ind(l), 1}$ which have the label $l$.

If there are strictly more edges going from $s^{\ind(l), 0}$ to $s^{\ind(l), 1}$, then $P_l$ outputs $1$. Else, if there are strictly more edges from $s^{\ind(l), 1}$ to $s^{\ind(l), 0}$, then $P_l$ outputs $0$. In the remaining case $P_l$ simply guesses an output uniformly at random. 

In this construction, intuitively, predictors tend to guess the heads of edges more often than they do the tails. Thus, for a vertex $v \in \{0, 1\}^n$, if $v$ is the head of an edge far more often than it is the tail (i.e., if $v$ has $\outdeg_G(v) < \indeg_G(v)$), then these predictors are likely to perform well on it. As a corollary of this, if the predictors do not perform well on a vertex $v$, it implies that $v$ must be a valid output for $\weakex$. This intuition is formalized in Lemma~\ref{lem:pred-from-weakex}.

Now we formally define these predictors. In order to make the predictors deterministic, we have to simulate random guessing by constructing \textit{two} predictors for each label $l$.

\begin{definition}[Predictors Corresponding to a $\weakex$ Instance]
\label{def:pred-from-weakex}
    Let $G = (V, E, \lab, \ind)$ be a deterministic hypercube digraph of dimension $n$ with $\ell$-bit labels.
    
    We define a family of middle-bit predictors $\mathcal{P}(G) \coloneqq (P_{lb}(G))_{l \in \{0, 1\}^{\ell}, b \in \{0, 1\}}$. For $l \in \{0, 1\}^{\ell}$ and $b \in \{0, 1\}$, $P_{lb}(G)$ is a middle-bit predictor over $\{0, 1\}^n$ at index $\ind(l)$. For $s \in \{0, 1\}^{n-1}$, we define
    \begin{equation*}
    P_{lb}(G)(s) \coloneqq \begin{cases}
        1, &\text{ if } \ee(s^{\ind(l), 0}, l) \land \lnot \ee(s^{\ind(l), 1}, l),\\
        0, &\text{ if } \ee(s^{\ind(l), 1}, l) \land \lnot \ee(s^{\ind(l), 0}, l),\\
        b, &\text{otherwise.}
    \end{cases}
    \end{equation*}
\end{definition}

\begin{lemma}
\label{lem:pred-from-weakex}
Let $G = (V, E, \lab, \ind)$ be a deterministic hypercube digraph of dimension $n$ with $\ell$-bit labels.
    For any string $s \in \{0, 1\}^n$,
    \begin{equation}
        \label{eq:pred-from-weakex}
        \Pr_{l \in \{0, 1\}^{\ell}, j \in \{0, 1\}}[P_{lj}(G) \succs s] = \frac{1}{2} +  \frac{\indeg_G(s) - \outdeg_G(s)}{2^{\ell + 1}}.
    \end{equation}
\end{lemma}
\begin{proof}
This follows directly from the following facts:
\begin{enumerate}
    \item For every label $l \in \{0, 1\}^{\ell}$ such that there exists a non-self-loop edge $e$ with $\lab(e) = l$ with vertex $s$ as the \textbf{head}, but not a non-self-loop edge $e'$ with $\lab(e') = l$ with vertex $s$ as the \textbf{tail}, both $P_{l0}$ and $P_{l1}$ succeed on $s$. 
    \item For every label $l \in \{0, 1\}^{\ell}$ such that there exists a non-self-loop edge $e$ with $\lab(e) = l$ with vertex $s$ as the \textbf{tail}, but not a non-self-loop edge $e'$ with $\lab(e') = l$ with vertex $s$ as the \textbf{head}, neither $P_{l0}$ nor $P_{l1}$ succeeds on $s$.
    \item For every other label $l \in \{0, 1\}^{\ell}$, only $P_{l s_{\ind(l)}}$ succeeds on $s$.
\end{enumerate}

Let's say we have $a$ labels of the first type, $b$ of the second, and $c$ of the third. Of course, $a+b+c = 2^{\ell}$. Each label of type $a$ contributes $1$ to $\indeg_G(s) - \outdeg_G(s)$, each label of type $b$ contributes $-1$, and each label of type $c$ contributes $0$. Therefore, the right-hand side of Equation~\ref{eq:pred-from-weakex} can be written as $1/2 + (a-b)/2^{\ell + 1}$. On the other hand, each label of type $a$ contributes $1/2^{\ell}$ to the left-hand side of Equation~\ref{eq:pred-from-weakex}, each label of type $b$ contributes $0$, and each label of type $c$ contributes $1/2^{\ell + 1}$. Therefore, the left-hand side can be written as $a/2^{\ell} + c/2^{\ell + 1}$. By substituting $c$ with $2^{\ell} - (a+b)$, the left-hand side becomes $a/2^\ell + 1/2 - a/2^{\ell + 1} - b/2^{\ell + 1} = 1/2 + (a-b)/2^{\ell + 1}$, which is exactly the right-hand side. This completes the proof.
\end{proof}
\begin{corollary}
\label{cor:pred-from-weakex}
Let $G = (V, E, \lab, \ind)$ be a deterministic hypercube digraph of dimension $n$ with $\ell$-bit labels, and let $S \subseteq \{0, 1\}^n$ be a non-empty multiset such that
\[\adv_{U(S)}(\mathcal{P}(G)) < 1/2^{\ell + 1}.\]
$S$ must contain a solution to $\weakex$ on $G$.
\end{corollary}
\begin{proof}
By Lemma~\ref{lem:yao-for-unpred}, there exists a string $s \in S$ such that $\Pr_{l \in \{0, 1\}^\ell, j \in \{0, 1\}}[P_{lj}(G) \succs s] < 1/2 + 1/2^{\ell + 1}.$ Combining this with Lemma~\ref{lem:pred-from-weakex} shows that $\indeg_G(s) - \outdeg_G(s) < 1 \implies \indeg_G(s) \leq \outdeg_G(s)$, since both the in-degree and out-degree are integers. This shows that $S$ contains a solution to $\weakex$.
This completes the proof.
\end{proof}

\begin{theorem}
\label{thm:weakex-to-upd}
    \[\weakex \leq_{\P} \upd.\]
\end{theorem}
\begin{proof}
    Let $(n, \ell, C, \ind)$ be a $\weakex$ instance. Let $G$ be the deterministic hypercube digraph corresponding to this instance. Consider the $\upd$ instance described by the predictors $\mathcal{P}(G)$ and $\epsilon \coloneqq 1/2^{\ell + 1}$. By Corollary~\ref{cor:pred-from-weakex}, any solution $S$ to this $\upd$ instance contains a solution to the $\weakex$ instance. 

    It remains to verify the efficiency of this reduction. First, $1/\epsilon = 2^{\ell + 1}$ is polynomial in the size of the $\weakex$ instance, and so it can be output in unary format in polynomial time. Then, the $\upd$ instance is over $\{0, 1\}^n$ and contains $2^{\ell + 1}$ predictors. The circuit $C$ has $n+\ell$ input gates, which implies that $|C| \geq n$. Therefore, $n$ can be produced as unary output, as can $2^{\ell + 1}$. Next, we need to verify that the circuit computing each predictor can be computed in polynomial time. The predictors are indexed by $l \in \{0, 1\}^{\ell}$ and $b \in \{0, 1\}$. On an input $s$, a predictor makes two simple $\ee$ queries and then computes a function over the resulting two outputs of $C$, and $b$. Thus, each predictor is essentially $2$ copies of $C$ with the input bits hardwired to $l$, along with the $\ind(l)^{th}$ input bit in the two copies hardwired to $0$ and $1$ respectively. Clearly, these predictors can be computed in polynomial time. Finally, the index of a predictor corresponding to $l \in \{0, 1\}^{\ell}$ is just $\ind(l)$, which is explicitly given as input and thus can be copied directly. 

    Now, let $S$ be a polynomial-size solution to this $\upd$ instance. For every $v \in S$, one can check whether $v$ is a solution to the $\weakex$ instance using the following equation:
    \[\indeg_G(v) -\outdeg_G(v) = \sum_{l \in \{0, 1\}^{\ell}} (\ee(v^{\oplus \ind(l)}, l) - \ee(v, l)),\]
    where $v^{\oplus \ind(l)}$ refers to the string $v$ with the $\ind(l)^{th}$ bit flipped. Clearly, this can be computed in polynomial time for each $v$, since it only requires $2^{\ell + 1}$ many queries to $C$. By computing this value for each $v \in S$, we can simply output the first $v$ for which this value is non-positive. Such a $v$ must exist, and is a solution to the $\weakex$ instance.
    This completes the proof.
\end{proof}

This completes the reduction from $\weakex$ to $\upd$. We will now describe the reduction from $\lossyc$ to $\weakex$.

\subsection{\texorpdfstring{$\lossyc$ Reduces to $\weakex$}{LossyCoding Reduces to WeakExcess}}
\label{sec:lossy-to-weakex}

The $\lossyc$ problem is a total derandomization problem introduced by Korten~\cite{Korten22}.

\begin{definition}[Lossy Coding]
\label{def:lossy-coding}
Let $C:\{0, 1\}^{n} \rightarrow \{0, 1\}^{n-1}$ and $D:\{0, 1\}^{n-1} \rightarrow \{0, 1\}^n$. A string $x \in \{0, 1\}^n$ is a solution to $\lossyc$ on $(C, D)$ if $D(C(x)) \neq x$.
\end{definition}

Similar to $\upd$ and $\weakex$, it is easy to see that $\lossyc$ is a total derandomization problem.

\begin{observation}[\cite{Korten22}]
For any $C:\{0, 1\}^{n} \rightarrow \{0, 1\}^{n-1}$ and $D:\{0, 1\}^{n-1} \rightarrow \{0, 1\}^n$,
    \[\Pr_{x \in \{0, 1\}^n}[x \text{ is a solution to }\lossyc \text{ on }(C, D)] \geq 1/2.\]
Therefore, $\lossyc$ is a derandomization problem.
\end{observation}

$\lossyc$ has been studied in many regimes with various restrictions on the functions $C$ and $D$. We will describe our own restrictions on $C$ and $D$ in a moment. First, for a $\lossyc$ instance $(C, D)$, we define a function $D^{\ext}$ which is just $D$ with an extra input bit that is completely ignored by the function. This only serves the purpose of turning $D$ into a length-preserving function.

\begin{definition}[Extension of $D$]
\label{def:d-ext}
Let $(C, D)$ be a $\lossyc$ instance with $C:\{0, 1\}^n \rightarrow \{0, 1\}^{n-1}$ and $D:\{0, 1\}^{n-1} \rightarrow \{0, 1\}^n$. We define the function $D^{\ext}:\{0, 1\}^n \rightarrow \{0, 1\}^n$ as follows. For $x \in \{0, 1\}^n$,
\begin{equation*}
    D^{\ext}(x) \coloneqq D(x_{<n}).
\end{equation*}
\end{definition}

\begin{definition}[$\lossyc$ Family]
    Let $(C_x, D_x)_{x \in \{0, 1\}^*}$ be a family of $\lossyc$ instances. We refer to $\lossyc$ on $(C_x, D_x)_{x \in \{0, 1\}^*}$ as the search problem where the input consists of $x \in \{0, 1\}^*$ and the output must consist of a solution to $\lossyc$ on $(C_x, D_x)$.
\end{definition}

Let $(C_x, D_x)_{x \in \{0, 1\}^*}$ be a family of $\lossyc$ instances. Our reduction from $\lossyc$ on this family to $\weakex$ is efficient when two conditions hold: (1) $C_x$ and $D_x$ can be computed in polynomial time, and (2) $D^{\ext}_x$ can be computed by a polynomial-time $\ipFL$ machine with read-only access to advice $x$. Note that the measures of efficiency, i.e., `polynomial' time and `logarithmic' space, are in terms of the length of $x$. Thus, an implicit requirement is also that the instances $(C_x, D_x)$ are over strings of polynomial size, and this size is computable in polynomial time.

For an in-place Turing machine $M$ which has read-only access to advice $x$ and read-write input $y$, we will refer to the machine's computation on $(x, y)$ as $M_x(y)$ instead of $M(x, y)$. This is to avoid any confusion between the roles of the two strings.

Now, for a $\lossyc$ instance $(C, D)$, we use $\nonsol(C, D)$ to refer to the set of strings which are $\textit{not}$ a solution to the instance.

\begin{definition}
Let $(C:\{0, 1\}^n \rightarrow \{0, 1\}^{n-1}, D:\{0, 1\}^{n-1} \rightarrow \{0, 1\}^n)$ be a $\lossyc$ instance. We define the set $\nonsol(C, D) \subseteq \{0, 1\}^n$ as
\[\nonsol(C, D) \coloneqq \{x \in \{0, 1\}^n \mid D(C(x)) = x\}.\]
\end{definition}

We now define some additional notation for the remainder of this section. For an in-place Turing machine $M$, we denote the configuration of $M$ by a tuple $(y, \sigma)$, where $y$ is the string on the read-write input tape, and $\sigma$ is all remaining information about the configuration of $M$: the work tape string, the tape-head locations, and the state of the finite-control automaton. We refer to the second element of the tuple, $\sigma$, as the \emph{state} of $M$. In cases where $M$ has access to read-only advice, we include the tape head location in the state, but not the advice string itself. Thus, observe that the length of $\sigma$ is $O(\log (\text{input length}) + \log (\text{advice length})  + \text{work space})$. We will refer to this as the \emph{state size} of $M$. 

For a configuration $(y, \sigma)$, we use $M(y, \sigma)$ to denote the configuration reached by running $M$ for $1$ step starting at $(y, \sigma)$. Following our function composition notation, $M^{(t)}(y, \sigma)$ denotes the configuration reached after running $M$ for $t$ steps. We denote by $\iph(\sigma)$ the current location of the input tape head. We use $\sigma_{\start}$ to denote the initial state of $M$. Following standard notation, $\ttime(M(y))$ denotes the number of steps $M$ runs on input $y$ before halting.

\usetikzlibrary{positioning,arrows.meta,calc}

\begin{figure}[t]
    \centering
    \begin{tikzpicture}[
        >=Stealth,
        basebox/.style={
            draw,
            minimum height=9mm,
            outer sep=0pt,
            inner ysep=0pt,
            font=\small,
            text height=1.8ex,
            text depth=.3ex
        },
        topchunk/.style={
            basebox,
            draw=teal!65!black,
            fill=teal!9,
            minimum width=6cm
        },
        bit/.style={
            basebox,
            draw=orange!75!black,
            fill=orange!20,
            minimum width=1cm
        },
        vertex/.style={
            basebox,
            draw=teal!65!black,
            fill=teal!9,
            minimum width=7cm
        },
        finalvertex/.style={
            basebox,
            draw=green!60!black,
            fill=green!12,
            minimum width=7cm
        },
        edgelabel/.style={
            right=5mm,
            font=\scriptsize,
            inner sep=1pt
        }
    ]

        \node[topchunk, anchor=west] (v0left)
            at (-3.5cm,0) {$C(y)$};

        \node[bit, anchor=west] (v0right)
            at (v0left.east) {$b$};

        \coordinate (v0mid) at
            ($(v0left.south west)!0.5!(v0right.south east)$);

        \node[vertex] (v1)
            at (0,-2.3)
            {$M^{(1)}(C(y)\circ b,\sigma_{\start})_1$};

        \node[vertex] (v2) at (0,-4.6)
            {$M^{(2)}(C(y)\circ b,\sigma_{\start})_1$};

        \node (dots) at (0,-6.2) {$\dots$};

        \node[finalvertex] (vt) at (0,-7.8)
            {$M^{(t)}(C(y)\circ b,\sigma_{\start})_1
              = D(C(y)) = y$};

        \draw[->, shorten <=2pt, shorten >=2pt] (v0mid) --
            node[edgelabel] {
                \colorbox{violet!13}{$\sigma_{\start}$}
                $\circ$
                \colorbox{orange!20}{$b$}
            } (v1.north);

        \draw[->, shorten <=2pt, shorten >=2pt] (v1.south) --
            node[edgelabel] {
                \colorbox{violet!13}{
                    $M^{(1)}(C(y)\circ b,\sigma_{\start})_2$
                }
                $\circ$
                \colorbox{orange!20}{$b$}
            } (v2.north);

        \draw[->, shorten <=2pt, shorten >=2pt] (v2.south) --
            node[edgelabel] {
                \colorbox{violet!13}{
                    $M^{(2)}(C(y)\circ b,\sigma_{\start})_2$
                }
                $\circ$
                \colorbox{orange!20}{$b$}
            } (dots.north);

        \draw[->, shorten <=2pt, shorten >=2pt] (dots.south) --
            node[edgelabel] {
                \colorbox{violet!13}{
                    $M^{(t-1)}(C(y)\circ b,\sigma_{\start})_2$
                }
                $\circ$
                \colorbox{orange!20}{$b$}
            } (vt.north);

    \end{tikzpicture}

    \caption{
        A computation path of $M$ starting with input
        $C(y)\circ b$, represented in the graph $G$.
        The boxes correspond to vertices, the arrows to edges, and the text on the right side of each arrow is the label of the corresponding edge. Here $t \coloneqq \ttime(M(C(y)\circ b))$. The label of each vertex corresponds to the string on the read-write input tape of $M$. The label of each edge corresponds to its state and the bit $b$.
    }
    \label{fig:graphconstruction}
\end{figure}
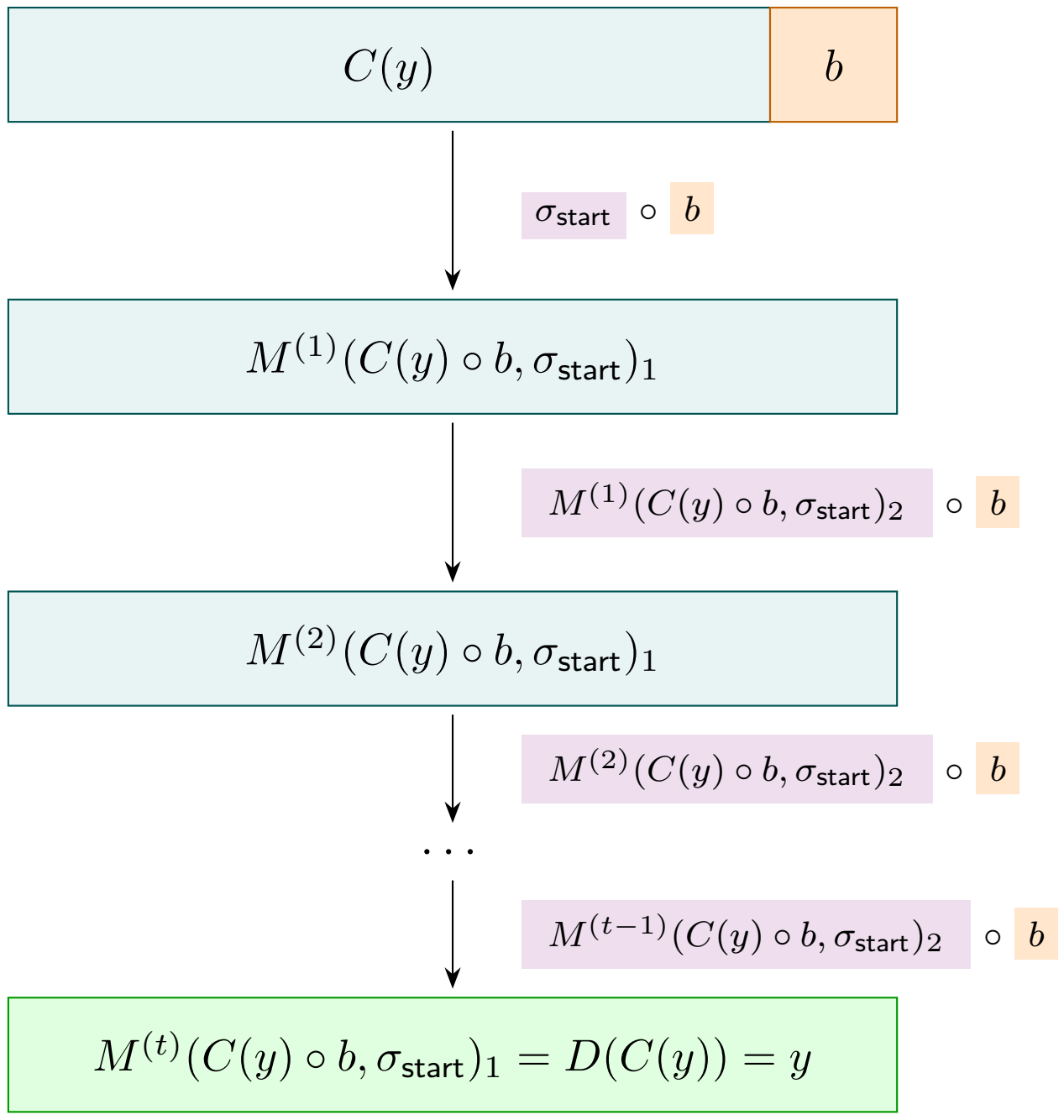

Now, we informally describe our reduction to $\weakex$ and the intuition behind it. We fix a lossy coding instance $(C, D)$ and an in-place Turing machine $M$ which computes $D^{\ext}$. We construct a deterministic hypercube digraph $G \coloneqq G(C, D, M)$ as follows. The vertices of $G$ correspond to strings on the input tape of $M$. The edges of $G$ will carry labels which correspond to the remaining configuration of $M$, along with an additional bit which only serves to make the graph deterministic. Thus, for any edge of $G$, the tail of the edge along with its label defines a total configuration of~$M$. We will define $G$ to be the union of all directed computation paths of $M$ on inputs of the form $C(y) \circ b$ where $y \in \nonsol(C, D)$ and $b \in \{0, 1\}$. Refer to Figure~\ref{fig:graphconstruction} for a visual representation of such a path in $G$. We note that a computation path of $M$, when projected to $G$, forms a directed walk. 

The intuition behind this construction is as follows: a solution to $\lossyc$ on $(C, D)$ is any string $y \notin \nonsol(C, D)$. For all strings $y \in \nonsol(C, D)$, since $M$ is deterministic, $y$ will be the source of at most one computation path in $G$: the one whose sink is $D^{\ext}(y)$. On the other hand, $y$ will be the sink of at least two computation paths: the ones whose sources are $C(y) \circ 0$ and $C(y) \circ 1$ respectively. Every other occurrence of $y$ in a computation path of $G$ will be as an intermediate vertex; such occurrences do not affect $\outdeg_G(y) - \indeg_G(y)$. Thus, for any $y \in \nonsol(C, D)$, we will have $\indeg_G(y) > \outdeg_G(y)$. The contrapositive of this statement is that for any vertex $y$ of $G$ such that $\outdeg_G(y) \geq \indeg_G(y)$, it must be the case that $y \notin \nonsol(C, D)$. Thus, a solution for $\weakex$ on $G$ is a solution for $\lossyc$ on $(C, D)$.

\begin{definition}
\label{def:hypercube-from-lossy}
Let $(C, D)$ where $C:\{0, 1\}^n \rightarrow \{0, 1\}^{n-1}$ and $D:\{0, 1\}^{n-1} \rightarrow \{0, 1\}^n$ be a $\lossyc$ instance, and $M$ be an in-place Turing machine which computes $D^{\ext}$. Let $\ell$ be the state size of $M$.
We define a deterministic hypercube digraph of dimension $n$ with $\ell + 1$-bit labels, $G(C, D, M) \coloneqq (V, E, \lab, \ind)$, as follows:
\begin{enumerate}
    \item $V \coloneqq \{0, 1\}^n$.
    \item $E \coloneqq \{e_{x, b, t} \mid x \in \nonsol(C, D), b \in \{0, 1\}, 0 \leq t < \ttime(M(C(x) \circ b))\}$,
    where
    \[e_{x, b, t} \coloneqq (M^{(t)}(C(x) \circ b, \sigma_{\start})_1, M^{(t+1)}(C(x) \circ b, \sigma_{\start})_1).\]
    Note that this is a multigraph, and thus we may have $(x, b, t)$ and $(x', b', t')$ with $e_{x, b, t} = e_{x', b', t'}$. We retain both distinct copies in the graph.
    \item For every $x \in \nonsol(C, D)$, $b \in \{0, 1\}$, and $0 \leq t < \ttime(M(C(x) \circ b))$, we define
    \[\lab(e_{x,b,t}) \coloneqq M^{(t)}(C(x) \circ b, \sigma_{\start})_2 \circ b.\]
    \item For any $l \in \{0, 1\}^{\ell + 1}$ we define 
    \[\ind(l) \coloneqq \iph(l_{\le \ell}).\]
\end{enumerate}
\end{definition}
Since $G(C, D, M)$ has $\ell + 1$-bit labels, the input size of the corresponding $\weakex$ instance is at least $2^{\ell+1}$ bits. In order for this to be at most polynomial, we need $\ell$ to be at most logarithmic. $\ell$, being the state size of $M$, is at least linear in the size of its work tape. This is precisely why we need $M$ to be an $\ipFL$ machine. 
\begin{observation}
\label{obs:reductionlossyweakex}
Let $C,D,M,n,\ell$, and $G(C,D,M)$ be as in Definition~\ref{def:hypercube-from-lossy}. $G(C, D, M) = (V, E, \lab, \ind)$ is indeed a deterministic hypercube digraph of dimension $n$ with $\ell + 1$-bit labels. 
\end{observation}
\begin{proof}
First, for any vertex $v$ of $G$ and label $l \in \{0, 1\}^{\ell + 1}$, there is at most one edge leaving $v$ with label $l$. This follows from the fact that the function $D$ is injective when restricted to inputs in the image of $C$ over $\nonsol(C, D)$. That is, for all $y \neq y' \in \nonsol(C, D)$, $D(C(y)) = y \neq y'= D(C(y'))$. If there were two edges $e_{x, b, t}$ and $e_{x', b', t'}$ leaving the same vertex and carrying the same label, their corresponding computation paths must end in the same output, $y$. This is only acceptable in two cases: (1) When these edges belong to different computation paths starting at $C(y) \circ 0$ and $C(y) \circ 1$ respectively; however, this would imply that $b \neq b'$, and thus the labels of these edges would differ; or (2) When these edges belong to the same computation path. However, this would mean that the computation of $M$ is cyclic, which violates the assumption that $M$ computes $D^{\ext}$. Thus, the graph $G$ is indeed `deterministic' (every vertex has at most one edge leaving it with any given label).

Second, the index of the bit of a vertex $v$ that can change when you traverse an edge leaving~$v$ with label $l$ can be determined only from the label. This is easy to see: at any point in the computation of $M$, the only input tape bit that can change is the one at the location of the input tape head. This location is exactly $\ind(l) = \iph(l_{\leq \ell})$.

This completes the proof.
\end{proof}

\begin{lemma}
\label{lem:lossy-to-weakex}
Let $(C, D)$ where $C:\{0, 1\}^n \rightarrow \{0, 1\}^{n-1}$ and $D:\{0, 1\}^{n-1} \rightarrow \{0, 1\}^n$ be a $\lossyc$ instance, and $M$ be an in-place Turing machine which computes $D^{\ext}$. Let $\ell$ be the state size of $M$. Let $G(C, D, M) = (V, E, \lab, \ind)$. Then any vertex $v \in \{0, 1\}^n$ such that
\[\outdeg_G(v) \geq \indeg_G(v)\]
satisfies $D(C(v)) \neq v$. That is, any solution to $\weakex$ on $G$ is a solution to $\lossyc$ on~$(C, D)$.
\end{lemma}
\begin{proof}
    Observe that for any $y \in \nonsol(C, D)$ and $b \in \{0, 1\}$, the family of edges
    \[(e_{y, b, t})_{t=0}^{\ttime(M(C(y) \circ b)) - 1}\]
    forms a directed walk (possibly with self-loops) from the vertex $C(y) \circ b$ to the vertex $y$. Thus, for all intermediate vertices $v'$ on this walk, the contribution of this walk to
    \[\outdeg_G(v') - \indeg_G(v')\]
    is $0$. The first vertex of the walk, $C(y) \circ b$, receives a contribution of $+1$, and the last vertex of the walk, $y$, receives a contribution of $-1$.

    Therefore, for any vertex $v$ which satisfies $\outdeg_G(v) \geq \indeg_G(v)$, we have
    \[|\{(y \in \nonsol(C, D), b \in \{0, 1\}) \mid C(y) \circ b = v\}| \geq |\{(y \in \nonsol(C, D), b \in \{0, 1\}) \mid y = v\}|.\]
    Note that the left-hand side of this inequality is at most $1$, since any $y \in \nonsol(C, D)$ and $b \in \{0, 1\}$ with $C(y) \circ b = v$ would imply that $y = D(v_{<n})$ and $b = v_n$. The right-hand side of this inequality is either $0$ or $2$. The inequality thus implies that the right-hand side of the inequality is $0$, which implies that $v \notin \nonsol(C, D)$. This implies that $D(C(v)) \neq v$.

    This completes the proof.
\end{proof}

\begin{observation}
\label{obs:ee-from-ipt}
Let $(C, D)$ where $C:\{0, 1\}^n \rightarrow \{0, 1\}^{n-1}$ and $D:\{0, 1\}^{n-1} \rightarrow \{0, 1\}^n$ be a $\lossyc$ instance, and $M$ be an in-place Turing machine which computes $D^{\ext}$. Let $\ell$ be the state size of $M$. Let $v \in \{0, 1\}^n$ be a vertex of $G(C, D, M)$ and $l \in \{0, 1\}^{\ell + 1}$ be a label. The answer to an $\ee(v, l)$ query on $G(C, D, M)$ is $\mathsf{true}$ if and only if:
\begin{enumerate}
    \item There exists $y \in \nonsol(C, D)$ and $0 \leq t < \ttime(M(C(y) \circ l_{\ell + 1}))$ such that
    \[M^{(t)}(C(y) \circ l_{\ell + 1}, \sigma_{\start}) = (v, l_{\le \ell}),\] and
    \item $M(v, l_{\leq \ell})_1 \neq v$. This is to ensure that the edge is not a self-loop.
\end{enumerate}
\end{observation}

This follows directly from definitions. Now, the following theorem observes that when $C$ is polynomial-time computable, and $M$ is a polynomial-time $\ipFL$ machine computing $D^{\ext}$, these $\ee$ queries can be answered by a polynomial-size circuit.

\begin{theorem}
\label{thm:lossy-to-weakex}
    Let $n:\{0, 1\}^* \rightarrow \mathbb{N}$ be a function and $(C_x, D_x)_{x \in \{0, 1\}^*}$ be a family of $\lossyc$ instances where $C_x:\{0, 1\}^{n(x)} \rightarrow \{0, 1\}^{n(x) - 1}$ and $D_x:\{0, 1\}^{n(x) - 1} \rightarrow \{0, 1\}^{n(x)}$ such that
    \begin{enumerate}
        \item The function $n$ is computable in polynomial time and is polynomially bounded. That is, $n(x) \leq \poly(|x|)$.
        \item There exists a polynomial-time Turing machine $M$ which, given as input $x \in \{0, 1\}^*$ and $y \in \{0, 1\}^{n(x)}$, computes $C_x(y)$.
        \item There exists a polynomial-time $\ipFL$ Turing machine $M'$ which, given read-only access to advice $x$ and read-write input $y \in \{0, 1\}^{n(x)}$, computes $D^{\ext}_x(y)$.
    \end{enumerate}
Then
\[\lossyc \text{ on }(C_x, D_x)_{x} \leq_{\P} \weakex.\]
\end{theorem}
\begin{proof}

The reduction proceeds as follows. By the first assumption, one can compute $n(x)$ in polynomial time. Now, since $M'$ is a polynomial-time-bounded Turing machine, and $n(x)$ is a polynomially bounded function, there must exist a polynomial-time computable and polynomially bounded function $T$ such that $T(|x|) \geq \ttime(M'_x(y))$ for all $y \in \{0, 1\}^{n(x)}$.

Now, the reduction needs to take as input $x \in \{0, 1\}^*$ and produce as output the $\weakex$ instance corresponding to $G(C_x, D_x, M'_x)$. By Lemma~\ref{lem:lossy-to-weakex}, any solution to this $\weakex$ instance is a solution to the $\lossyc$ instance $(C_x, D_x)$.

Observe that, because $M'$ is an $\ipFL$ machine and $n(x) \leq \poly(|x|)$, the state size of~$M'$, say~$\ell$, is $O(\log(|x|))$. Thus, every part of the $\weakex$ instance corresponding to $G(C_x, D_x, M'_x)$, except the circuit computing $\ee$ queries, can certainly be computed in polynomial time. In particular, the dimension of the graph $n(x)$ and the bit-length of the labels $\ell+1$ can both be produced in unary. Furthermore, the index table $\ind:\{0, 1\}^{\ell + 1} \rightarrow [n(x)]$ can also be produced. This table has $\exp(\ell) \cdot \log(n(x))$ size. Since $\ell = O(\log (|x|))$ and $n(x) \leq \poly(|x|)$, this is also $\leq \poly(|x|)$. Therefore, all parts of the $\weakex$ instance of $G(C_x, D_x, M'_x)$ except the circuit answering $\ee$ queries can indeed be produced in $\poly(|x|)$ time.

Finally, for the circuit answering $\ee$ queries on $G(C_x, D_x, M'_x)$, we refer to Observation~\ref{obs:ee-from-ipt}. This circuit must check two conditions, and then simply take the logical and of their values. We will show that both conditions can be checked in polynomial time, and thus that a polynomial-size circuit which answers $\ee$ queries can be produced in polynomial time.

The second condition is easier: it simply requires simulating $M'_x$ for one step, and checking whether this step changed the string on the input tape. Clearly the circuit which decides this can be computed in polynomial time.

The first condition is more complicated. Fix an input $v \in \{0, 1\}^{n(x)}$ and $l \in \{0, 1\}^{\ell + 1}$ to $\ee$. The first condition requires checking whether there exists a string $y \in \nonsol(C_x, D_x)$ such that $M'_x$, when executed on input $C_x(y) \circ l_{\ell + 1}$, reaches the configuration $(v, l_{\leq \ell})$. Of course, running a brute-force search over all strings $y \in \nonsol(C_x, D_x)$ would be too expensive; it would be easier to solve $\lossyc$ directly. Instead, we do the following. We simulate the machine $M'_x$ from the configuration $(v, l_{\leq \ell})$ until it terminates. If it does not terminate within $T(|x|)$ steps, we can already reject, so let's assume that it terminated. Let $z \in \{0, 1\}^{n(x)}$ be the string on the read-write input tape upon termination. If there indeed exists a string $y$ which makes the condition true, it must be the case that $y = z$. Thus, we now need to check only two things: whether $z \in \nonsol(C_x, D_x)$, and whether the machine $M'_x$, when executed on the input $C_x(z) \circ l_{\ell + 1}$, reaches the configuration $(v, l_{\leq \ell})$ at any point. Since $M'$ and $M$ are both polynomial-time machines, both of these facts can be checked in polynomial time. Thus, the first condition can also be checked in polynomial time. Since this check is uniform, it means that a circuit which decides the first condition can also be produced in polynomial time.

Combining the circuits corresponding to the two conditions provides a single circuit which answers $\ee$ queries. This completes the input to the $\weakex$ problem, and thus completes the proof.
\end{proof}

\subsection{\texorpdfstring{$\SCL$ Reduces to $\upd$}{SearchCL Reduces to UNPRED}}
The following theorem is proven by Cook, Li, Mertz, and Pyne~\cite{CookLiMertzPyne25}. That paper was written before the one by Cook, Ghentiyala, Mertz, Pyne, and Sheffield~\cite{CookGhentiyalaMertzPyneSheffield25} which introduced $\ipFL$ and related classes. Thus, the theorem statement uses different language. However, our restatement can be inferred directly by unpacking and repackaging definitions. This is a reduction from $\fct$ to $\lossyc$, on the type of $\lossyc$ instances that Theorem~\ref{thm:lossy-to-weakex} efficiently reduces to $\weakex$.
\label{sec:cl-to-upd-conc}
\begin{theorem}[Theorem 3.1, \cite{CookLiMertzPyne25}]
\label{thm:clmp}
Let $M$ be a catalytic logspace machine which, on an input of length $n$, uses at most $p(n)$ catalytic space. There exists a logspace-computable polynomially bounded function $p_M:\mathbb{N} \rightarrow \mathbb{N}$ and a family of $\lossyc$ instances $(C_x, D_x)_{x \in \{0, 1\}^*}$ such that:
\begin{enumerate}
    \item $C_x:\{0, 1\}^{n(x)} \rightarrow \{0, 1\}^{n(x)-1}$ and $D_{x}:\{0, 1\}^{n(x) - 1} \rightarrow \{0, 1\}^{n(x)}$, where the function $n$ is polynomially bounded and logspace computable.
    \item There exists a polynomial-time Turing machine $M_1$ which takes as input $x \in \{0, 1\}^*$ and $y \in \{0, 1\}^{n(x)}$, and produces as output $C_x(y)$. 
    \item There exists a polynomial-time $\ipFL$ Turing machine $M_2$ which, given read-only access to advice $x \in \{0, 1\}^*$ and read-write input $y \in \{0, 1\}^{n(x)}$, computes $D_x^{\ext}(y)$. 
    \item For any $x \in \{0, 1\}^*$, and for any solution $y$ to $\lossyc$ on $(C_x, D_x)$, $y_{\leq p(|x|)}$ is a solution to $\fct(M, p_M)$ on input $x$.
\end{enumerate}
\end{theorem}

This completes all of our intermediate reductions. We can now chain them together into the following concluding theorems.

\begin{theorem}
\label{thm:fct-to-upd}
For every catalytic logspace machine $M$, there exists a logspace-computable polynomially bounded function $p_M$ such that
\[\fct(M, p_M) \leq_{\P} \upd.\]
\end{theorem}
\begin{proof}
The function $p_M$ is the one from the statement of Theorem~\ref{thm:clmp}. Theorem~\ref{thm:clmp} reduces $\fct(M, p_M)$ to $\lossyc$, where the family of $\lossyc$ output instances satisfies the constraints of Theorem~\ref{thm:lossy-to-weakex}. Theorem~\ref{thm:lossy-to-weakex} reduces such $\lossyc$ instances to $\weakex$. Finally, Theorem~\ref{thm:weakex-to-upd} reduces $\weakex$ to $\upd$. 

Therefore, through this chain, we reduce $\fct(M, p_M)$ to $\upd$.
\end{proof}

\begin{theorem}
\label{thm:cl-to-upd}
\[\upd \in \FP \implies \SCL \subseteq \FP \implies \CL \subseteq \P.\]
\end{theorem}
\begin{proof}
This follows directly as a corollary from Theorem~\ref{thm:fct-to-upd}.
\end{proof}
\section{\texorpdfstring{Lower Bounds for $\updo$ and $\nbupdc{\PO}$}{Lower Bounds for UNPRED[O\_P] and NextBitUNPRED[O\_P]}}
\label{sec:unpred-hardness}
In this section we will show two results. The first is that there does not exist a deterministic polynomial-time algorithm for $\updo$. This is in contrast to $\nbupdc{\PO}$, for which deterministic polynomial-time algorithms are known~\cite{GoldreichWigderson00,LiPyneTell24}. The second result shows that the sequential algorithms of~\cite{GoldreichWigderson00,LiPyneTell24} are, in some sense, optimal. We prove this by showing that there does not exist a deterministic parallel algorithm for $\nbupdc{\PO}$ which has truly sublinear depth and polynomial work. Specifically, any truly sublinear depth algorithm must have superpolynomial work. Both of these results are unconditional.

For both problems we do this by explicitly constructing an adaptive adversary. Specifically, we construct predictor oracles $\PO$ which force any (parallel, in the case of $\nbupdc{\PO}$) algorithm either to make a superpolynomial number of queries or to produce an output set of superpolynomial size. Section~\ref{sec:lower-bound-for-updo} contains the lower bound for $\updo$ and Section~\ref{sec:lower-bound-for-nbupdc} contains the lower bound for $\nbupdc{\PO}$.

Both sections proceed in the same way: we begin by modelling the problem that this adversary needs to solve. In both cases, this is an \emph{online} problem. We imagine that the queries made by an $\updo$ (or $\nbupdc{\PO}$) algorithm are the input to an online algorithm, which needs to make irrevocable decisions in response to these queries. Specifically, on each query, it needs to decide the value of a predictor on the queried string. In both cases, this online problem will be a graph balancing problem. We first provide a sufficiently good online algorithm for these problems, in Sections~\ref{sec:hypercube-orientation} and~\ref{sec:prefix-tree-balancing}, and then show how these online algorithms can be used by the adversaries to obtain the final lower bounds in Sections~\ref{sec:unpred-oracle-hardness} and~\ref{sec:next-bit-round-lower-bound}.

For both problems, what we specifically show is a superpolynomial lower bound on the total number of queries made by the algorithm plus its output size. Naturally, the running time (or total work) of any algorithm is lower bounded by the sum of these values. We note that it is not possible to remove this dependence on output size: the naive algorithm which makes no queries and produces as output the entire space $\{0, 1\}^n$ is an algorithm which makes $0$ queries but solves $\updo$. We capture such naive algorithms through the dependence on output size.

\subsection{\texorpdfstring{Lower Bound for $\updo$}{Lower Bound for UNPRED[O\_P]}}
\label{sec:lower-bound-for-updo}
In this section we will prove our lower bound on $\updo$. We restrict ourselves to instances over $\{0, 1\}^n$ where, for each $i \in [n]$, there is exactly one predictor $P_i$ at index $i$. Let $s \in \{0, 1\}^{n-1}$ be a query to the predictor $P_i$. $P_i$ will answer either $0$ or $1$.  If the predictor answers $0$, then it fails on the string $s^{i, 1}$ (which is used to denote the string $s$ extended to an $n$-bit string by placing a $1$ in its $i^{th}$ coordinate). Else, if the predictor answers $1$, then it fails on the string $s^{i, 0}$. Thus, we can view the response to this query as the adversary irrevocably deciding which of the two strings the predictor $P_i$ will fail on.  By Lemma~\ref{lem:yao-for-unpred}, any valid output $S \subseteq \{0, 1\}^n$ for $\updo$ contains a string $s \in S$ such that at least $(1/2 - \epsilon)n$ predictors fail on $s$. We will design the adversary to answer queries in a way which delays this from happening as long as possible, in order to force the algorithm to make many queries. 

On each query $s \in \{0, 1\}^{n-1}$ to $P_i$, the adversary must reveal that $P_i$ fails on either $s^{i, 0}$ or $s^{i, 1}$. This can be viewed as giving an orientation to the hypercube edge between $s^{i, 0}$ and $s^{i, 1}$. If the predictor answers $0$, then it fails on $s^{i, 1}$, and we direct the edge from $s^{i, 0}$ to $s^{i, 1}$. Else, we direct the edge from $s^{i, 1}$ to $s^{i, 0}$. Rephrasing what we said in the previous paragraph, any valid output $S \subseteq \{0, 1\}^n$ to $\updo$ must contain a string $s$ which has in-degree at least $(1/2 - \epsilon)n$ in this graph (i.e., at least $(1/2 - \epsilon)n$ predictors must fail on it). Thus, the online problem for the adversary is as follows: given a stream of hypercube edges, assign an orientation to each edge. The goal is to minimize the maximum in-degree of any vertex in the subgraph formed by the queried edges. Specifically, we would like to obtain an online algorithm with the following guarantee: if any vertex in the subgraph formed by queried edges has in-degree $\geq x$, then it must be the case that at least $f(x)$ many edges have been queried. We want $f$ to be as large as possible. This will imply that any algorithm for $\updo$ needs to make at least $f((1/2 - \epsilon)n)$ many queries in order to produce a valid output $S$. We will show that there exists an online algorithm for this problem with $f(x) = \exp(\Omega(\sqrt{x}))$. This will suffice for our lower bound. 

Note that this problem only guarantees that the subgraph of \emph{queried} edges does not contain a valid output. What if the algorithm outputs a multiset of strings $S \subseteq \{0, 1\}^n$ that it never queried? In order to handle this, we add an additional post-processing step to the algorithm which simply sends the query string $s_{-i}$ to $P_i$ for every $s \in S$ and $i \in [n]$. This adds an additional $|S| \cdot n$ queries to the algorithm. This is exactly how the output size appears in the lower bound.
\subsubsection{Online Hypercube Graph Balancing Problem}
\label{sec:hypercube-orientation}
For $n \in \mathbb{N}$, let $Q_n$ denote the undirected Boolean hypercube on vertex set $\{0, 1\}^n$, where two vertices are adjacent if and only if they differ in exactly one coordinate.
\begin{definition}[Online Hypercube Graph Balancing Problem]
The online hypercube graph balancing problem is the online problem defined as follows:
\begin{enumerate}
    \item \textbf{Input:} $n \in \mathbb{N}$, followed by a sequence of distinct edges $(e_1, e_2, \dots)$ of $Q_n$ \textbf{revealed one at a time}. Each $e_i$ is a query.
    \item \textbf{Output:} For every queried edge $e_i$, the algorithm must orient $e_i$ before seeing $e_{i+1}$. That is, it must make the edge $e_i$ directed by choosing a direction for it. This choice is irrevocable, and in particular cannot be changed after seeing $e_{i+1}$.
\end{enumerate}
The goal of the algorithm is to minimize the maximum in-degree of any vertex. Formally, fix a non-empty input edge sequence $(e_1, \dots, e_t)$. Let $G$ be the directed graph over vertex set $\{0, 1\}^n$ and edge set $\{e_1, \dots, e_t\}$, with directions as chosen by the algorithm. Let $f:\mathbb{R}_{\geq 0} \rightarrow \mathbb{R}_{\geq 0}$ be a non-decreasing function. The algorithm has in-degree bound $f$ if
\[t \geq f(\max_{v \in V(G)} \indeg_{G}(v))\]
holds for every such sequence $(e_1, \dots, e_t)$.
\end{definition}
\begin{restatable}[]{lemma}{lemhypercubebalancingalgo}
\label{lem:hypercube-balancing-algo}
There exists a deterministic algorithm for the online hypercube graph balancing problem with in-degree bound $f$, where $f(h) \geq \exp(\Omega(\sqrt{h}))$.
\end{restatable}

We defer the proof of Lemma~\ref{lem:hypercube-balancing-algo} to Section~\ref{sec:appendix-hypercube-balancing} of the appendix. The work of Hart and Nisan~\cite{HartNisan18} can be adapted to obtain a deterministic algorithm with in-degree bound $\exp(\Omega(h))$, which is stronger than the $\exp(\Omega(\sqrt{h}))$ bound we present here. On the other hand, our algorithm is very simple and suffices for our purpose.

\subsubsection{A Lower Bound for $\updo$ via Hypercube Balancing}
\label{sec:unpred-oracle-hardness}

\begin{theorem}
\label{thm:exp-lb-det-unpred}
Let $\mathcal{A}$ be a deterministic algorithm for $\updo$. Let $f:\mathbb{R}_{\geq 0} \rightarrow \mathbb{R}_{\geq 0}$ be a non-decreasing function such that there exists a deterministic algorithm for the online hypercube graph balancing problem with in-degree bound $f$. Then for every $n \in \mathbb{N}$ and $0 <\epsilon < 1/2$, there exists an instance of $\updo$,  $I = (n, m = n, \epsilon, \ind_1, \dots, \ind_n)$, and a predictor oracle $\mathcal{O}_P$ such that
    \[Q + |S| \ge \frac{f((1/2 - \epsilon)n)}{n},\]
where $Q$ denotes the number of queries $\mathcal{A}$ makes to the oracle $\mathcal{O}_P$ on input instance $I$, and $S$ denotes its output multiset. Moreover, this instance satisfies $\ind_i = i$ for all $i \in [n]$.
\end{theorem}

\begin{proof}
We modify $\mathcal{A}$ so that, after it outputs $S$, it performs a post-processing round where it queries $\PO(i,s_{-i})$ for every $s\in S$ and $i \in [n]$. This leaves the output $S$ unchanged and adds only $n|S|$ new oracle queries.

Since $\mathcal{A}$ is deterministic and fixed, the oracle $\mathcal{O}_P$ can internally simulate $\mathcal{A}$ before answering each query. Thus, we assume while answering each query that $\mathcal{O}_P$ has complete access to all past queries made by $\mathcal{A}$.

For $i \in [n]$ and a string $s \in \{0, 1\}^{n-1}$, we define $s^{i, 0} \in \{0, 1\}^n$ to be the string $s$ with a $0$ added in its $i^{th}$ position, and similarly $s^{i, 1}$ to be the string $s$ with a $1$ added in its $i^{th}$ position. We view a query $(i \in [n], s \in \{0, 1\}^{n-1})$ to $\PO$ as an edge $\{s^{i, 0}, s^{i, 1}\}$ of $Q_n$. In response to a query $(i, s)$, $\PO$ needs to answer either $0$ or $1$. This can be viewed as an orientation of the edge $\{s^{i, 0}, s^{i, 1}\}$: if $\PO$ answers $0$, then the edge is oriented towards $s^{i, 1}$, and if $\PO$ answers $1$, then the edge is oriented towards $s^{i, 0}$. 

Under this viewpoint, we define the oracle $\PO$ to answer queries exactly as the online hypercube graph balancing algorithm which achieves in-degree bound $f$. Since we assumed that $\mathcal{O}_P$ has complete access to all past queries of $\mathcal{A}$, clearly $\PO$ can simulate the answers of the online algorithm. If $\mathcal{A}$ repeats a query, then $\mathcal{O}_P$ returns its previous answer instead of simulating the online algorithm.

For $i \in [n]$ and $s \in \{0, 1\}^{n-1}$ such that $(i, s)$ is never queried by $\mathcal{A}$, arbitrarily define $\PO(i, s) \coloneqq 0$. Let $P_1, \dots, P_n$ be the predictors defined by this oracle $\PO$. By the correctness of $\mathcal{A}$, we have 
    \[
    \adv_{U(S)}(\{P_1,\ldots,P_n\})<\eps.
    \]
By Lemma~\ref{lem:yao-for-unpred}, this implies that there exists $s^* \in S$ with
\begin{equation}
\label{eq:s-star}
    \Pr_{i \in [n]}[P_i \succs s^*] < \frac{1}{2} + \epsilon.
\end{equation}
Let $(e_1, \dots, e_t)$ be the sequence of distinct edges of $Q_n$ queried by $\mathcal{A}$. Consider the directed graph $G$ over vertex set $\{0, 1\}^n$ and edge set $\{e_1, \dots, e_t\}$ (i.e., the subgraph of $Q_n$ formed by the oriented edges) with directions induced by the answers of $\mathcal{O}_P$. Observe that for any query $(i \in [n], s \in \{0, 1\}^{n-1})$ made by $\mathcal{A}$, the edge $\{s^{i, 0}, s^{i, 1}\}$ in $G$ is directed towards $s^{i, 0}$ if and only if $P_i$ fails on $s^{i, 0}$.

The post-processing round of $\mathcal{A}$ ensures that it made every query of the form $(i \in [n], s^*_{-i})$. Therefore, the vertex $s^*$ in $G$ has total degree (in-degree + out-degree) exactly $n$. By Equation~\ref{eq:s-star}, this implies that $s^*$ has in-degree $\geq (\frac{1}{2} - \epsilon)n$ in $G$. Since $\PO$ is an online hypercube graph balancing algorithm with in-degree bound~$f$, this implies that the number of edges $t$ in $G$ must satisfy
\[t \geq f((1/2 - \epsilon)n).\]

In our case, every edge in $\{e_1, \dots, e_t\}$ corresponds to a query made by $\mathcal{A}$ to $\PO$. Due to our post-processing round, we have $t \leq Q + n|S|$. Substituting $t$ with $Q + n|S|$ in the above equation and then dividing both sides by $n$ proves the theorem.
\end{proof}

\begin{corollary}
\label{cor:updob-algo-exp-time}
    Let $\mathcal{A}$ be a deterministic algorithm for $\updo$. Then, on instances with $n$ predictors over $\{0, 1\}^n$ and parameter $\epsilon \coloneqq 1/10$, the worst-case running time of $\mathcal{A}$ satisfies
    \[\ttime(\mathcal{A}) \geq \exp(\Omega(\sqrt{n})).\]
\end{corollary}
\begin{proof}
    By plugging the deterministic algorithm for the online hypercube graph balancing problem from Lemma~\ref{lem:hypercube-balancing-algo} into Theorem~\ref{thm:exp-lb-det-unpred}, we obtain a lower bound on the query complexity plus the output size of $\mathcal{A}$. Of course, $\mathcal{A}$ must use one unit of time for each query it makes and for each string it produces as output. Thus, the stated lower bound on $\ttime(\mathcal{A})$ follows.
\end{proof}

\subsection{\texorpdfstring{Lower Bound for $\nbupdc{\PO}$}{Lower Bound for NextBitUNPRED[O\_P]}}
\label{sec:lower-bound-for-nbupdc}
In this section we will prove our lower bound on parallel algorithms for $\nbupdc{\PO}$. We assume that the reader has read Section~\ref{sec:lower-bound-for-updo}. We further build on the intuition from that section. In particular, we will again focus on instances over $\{0, 1\}^n$ where there is exactly one predictor $P_i$ per index $i \in [n]$.

In the case of $\updo$, a query to a predictor $P_i$ could only force the predictor to reveal one of two strings on which it fails. In the case of $\nbupdc{\PO}$ on the other hand, a single query can force the predictor to reveal $\exp(n)$ many strings on which it fails. For example, consider the empty query $\emptyset$ made to $P_1$. If $P_1$ answers $0$, then it fails on every string whose first bit is $1$. If $P_1$ answers $1$, then it fails on every string whose first bit is $0$. Thus, it immediately reveals $2^{n-1}$ many strings on which it fails. The polynomial-time algorithms of~\cite{GoldreichWigderson00,LiPyneTell24} exploit this property: they force the predictor $P_1$ to reveal which bit it fails on, set the first bit of the string to be equal to that, and then continue with $P_2$ and so on\footnote{These algorithms need to deal with multiple predictors at the same index, and so they are more technical. Nevertheless, they very much follow this simple idea.}. However, these algorithms are inherently sequential: the queries that they make to the predictor $P_i$ depend on the output of the predictor $P_{i-1}$. We exploit this sequential nature of these algorithms in our lower bound.

Let $T_n$ be a complete binary tree of depth $n$. Naturally, each vertex of $T_n$ at depth $i$ corresponds to a string $s \in \{0, 1\}^i$. Specifically, the root corresponds to the string $\emptyset$, and for a non-leaf vertex corresponding to string $s$, the two children of this vertex correspond to the strings $s \circ 0$ and $s \circ 1$.

Let $s^1, \dots, s^Q$ be the strings queried by a parallel algorithm for $\nbupdc{\PO}$ \emph{in a single round}. That is, the algorithm sends all of these strings to the oracle $\PO$ simultaneously. We view each query string as referring to its corresponding internal vertex of $T_n$. The adversary needs to place a \emph{token} on exactly one child of every queried vertex. Placing a token on a vertex corresponding to a string $s \in \{0, 1\}^i$ means that $P_i$ fails on every string whose prefix is $s$. In other words, $P_i$ fails on every string corresponding to a leaf in the subtree of this vertex. Similar to the case of $\updo$, the goal of the adversary is to minimize, after any set of queries, the maximum number of predictors that fail on any given string. This is equivalent to minimizing the maximum number of tokens on any root-to-leaf path. 

Specifically, we would like to design an adversary with the following guarantee: if a set of queries $s^1, \dots, s^Q$ results in the adversary placing $x$ tokens on some root-to-leaf path, then it must be the case that $Q \geq f(x)$. We would like $f$ to be as large as possible. This will imply that any algorithm for $\nbupdc{\PO}$ which runs for $R$ rounds needs to make at least $f((1/2 - \epsilon)n/R)$ many queries in order to produce a valid output $S$. We will show that there exists an algorithm for the adversary with $f(x) = \exp(x)$ (approximately). Thus, any parallel algorithm for $\nbupdc{\PO}$ with $R \leq n^{1 - \delta}$ rounds must make at least $\exp(\Omega(n^{\delta}))$ many queries. This will suffice for our lower bound. In Section~\ref{sec:prefix-tree-balancing}, we phrase this differently as an online problem where, in each step, the online algorithm receives an entire batch of vertices (corresponding to a set of queries made by the $\nbupdc{\PO}$ algorithm). Thus, the requirement on the function $f$ is slightly weaker and phrased differently. Nevertheless, the intuition we have described here is exactly the one that underlies our construction. 

We also note that it is possible to describe the lower bound in Section~\ref{sec:lower-bound-for-updo} in terms of a similar token problem on the hypercube. We chose to instead present it as a hypercube orientation problem only to provide additional ways in which one can view the problem.

\subsubsection{Online Prefix-Tree Balancing Problem}
\label{sec:prefix-tree-balancing}
For $n \in \mathbb{N}$, let $T_n$ denote the binary prefix tree over $\{0, 1\}^n$. That is, the vertex set of $T_n$ is $\bigcup_{i=0}^n \{0, 1\}^i$. The empty string $\emptyset$ is the root of $T_n$. A vertex $s \in V(T_n)$ with $|s| < n$ has two children: the vertex $s \circ 0$ and the vertex $s \circ 1$. 

\begin{definition}[Online Prefix-Tree Balancing Problem]
The online prefix-tree balancing problem is the online problem defined as follows:
\begin{enumerate}
    \item \textbf{Input:} $n \in \mathbb{N}$, followed by a sequence of pairwise disjoint batches $(B_1, B_2, \dots)$, where each batch $B_i$ is a subset of the non-leaf vertices of $T_n$. These batches are \textbf{revealed one at a time}.
    \item \textbf{Output:} For every queried batch $B_i$, the algorithm must irrevocably place a token on exactly one child of each vertex $v \in B_i$ before seeing $B_{i+1}$.
\end{enumerate}
The goal of the algorithm is to minimize the maximum number of tokens on any root-to-leaf path. Formally, fix a batch sequence $(B_1, \dots, B_t)$. For a leaf vertex $v \in \{0, 1\}^n$, let $\load(v)$ denote the number of tokens placed by the algorithm on the path from the root $\emptyset$ to the leaf $v$. Let $B$ denote $\max_{i \in [t]} |B_i|$. For a non-decreasing function $f:\mathbb{R}_{\geq 0} \rightarrow \mathbb{R}_{\geq 0}$, the algorithm has token bound $f$ if
\[t \lfloor \log(B+1) \rfloor \geq f(\max_{v \in \{0, 1\}^n} \load(v))\]
holds for every such sequence $(B_1, \dots, B_t)$.
\end{definition}
\begin{restatable}[]{lemma}{lemtreebalancingalgo}
\label{lem:tree-balancing-algo}
There exists a deterministic algorithm for the online prefix-tree balancing problem with token bound $f(h) = h$.
\end{restatable}
We defer the proof of this statement to Section~\ref{sec:appendix-online-prefix-tree-balancing} of the appendix.

\subsubsection{A Parallel Lower Bound for $\nbupdc{\PO}$ via Prefix-Tree Balancing}
\label{sec:next-bit-round-lower-bound}
In this section, we show a lower bound on the complexity of deterministic parallel algorithms for $\nbupdc{\PO}$. We work with the following general structure of a parallel algorithm: the algorithm has $R$ rounds. In each round, it makes $\leq Q$ queries to the oracle $\PO$. A query made in the $i^{th}$ round can depend upon the answer of $\PO$ to queries made in previous rounds, but not on the answer to other queries made in the $i^{th}$ round.

On a $\nbupdc{\PO}$ instance over $\{0, 1\}^n$ with predictors $P_1, \dots, P_m$, the algorithms of~\cite{GoldreichWigderson00,LiPyneTell24} run in $n$ rounds, and make polynomially many queries per round. In the following theorem, we show that these algorithms are essentially tight. That is, any algorithm which is guaranteed to run in a truly sublinear number of rounds must have superpolynomial work.

\begin{theorem}
\label{thm:next-bit-round-lower-bound}
    Let $\mathcal{A}$ be a deterministic parallel algorithm for $\nbupdc{\PO}$. Let $f:\mathbb{R}_{\geq 0} \rightarrow \mathbb{R}_{\geq 0}$ be a non-decreasing function such that there exists a deterministic algorithm for the online prefix-tree balancing problem with token bound $f$. Then, for every $n \in \mathbb{N}$ and $0 < \epsilon \leq 1/2$, there exists an instance of $\nbupdc{\PO}$, $I = (n, m = n, \epsilon, \ind_1, \dots, \ind_m)$, and a predictor oracle $\PO$ such that
    \[(R+1)\lfloor\log(\max(Q, |S|n) + 1)\rfloor \geq f((\frac{1}{2}-\eps)n),\]
    where $R$ denotes the number of rounds of $\mathcal{A}$ on input $I$ with oracle $\mathcal{O}_P$, $Q$ denotes the maximum number of queries $\mathcal{A}$ makes in any round, and $S$ denotes its output multiset. Moreover, the instance~$I$ satisfies $\ind_i = i$ for all $i \in [n]$.
\end{theorem}

\begin{proof}

Similar to what we did in the proof of Theorem~\ref{thm:exp-lb-det-unpred}, we add a post-processing round to $\mathcal{A}$ where it queries $(i, s_{<i})$ for every $i \in [n]$ and $s \in S$. This adds only one round with $n|S|$ many queries. Again, since $\mathcal{A}$ is a fixed deterministic algorithm, we can assume that $\PO$ has access to all past queries of $\mathcal{A}$.

A query $(i \in [n], s \in \{0, 1\}^{i-1})$ to $\PO$ corresponds to the non-leaf vertex $s$ of $T_n$. Thus, the queries $(i_1, s_1), \dots, (i_k, s_k)$ that $\mathcal{A}$ makes in one round correspond to a subset of non-leaf vertices of $T_n$. In order to answer the queries in each round, the oracle $\PO$ simply treats the non-leaf vertices corresponding to the queries as a batch, and simulates a deterministic online prefix-tree balancing algorithm with token bound $f$. For a specific query $(i_j, s_j)$, if the prefix-tree balancing algorithm chose to place the token of $s_j$ at $s_j \circ 0$, then the oracle $\mathcal{O}_P$ answers this query with $1$ (and vice versa). Of course, if $\mathcal{A}$ repeats a query then $\PO$ responds with the same answer it gave earlier instead of using the online algorithm. For each tuple $(i, s)$ which is never queried by $\mathcal{A}$, we arbitrarily define $\PO(i, s) \coloneqq 0$.

Now, let $P_1, \dots, P_n$ be the predictors corresponding to this oracle $\mathcal{O}_P$. By the correctness of $\mathcal{A}$, we have
\[\adv_{U(S)}(\{P_1, \dots, P_n\}) < \epsilon.\]
By Lemma~\ref{lem:yao-for-unpred}, this implies that there exists $s^* \in S$ with
\[\Pr_{i \in [n]}[P_i \succs s^*] < \frac{1}{2} + \epsilon.\]

Let $(i, s)$ be a query made by $\mathcal{A}$, where $s$ is a prefix of $s^*$. Then, by definition, the online algorithm placed a token on the vertex $s \circ s^*_i$ in $T_n$ if and only if $P_i$ fails on $s^*$. Due to the post-processing round, for every $i \in [n]$, $\mathcal{A}$ must have queried $(i, s^*_{<i})$. Therefore, upon termination of $\mathcal{A}$, the root-to-leaf path of vertex $s^*$ in $T_n$ must contain at least $(1/2 - \epsilon)n$ many tokens. Then, since $\mathcal{O}_P$ is an algorithm with token bound $f$, we have
\[(R+1)\lfloor\log(\max(Q, |S|n) + 1)\rfloor \geq f((\frac{1}{2}-\eps)n).\]
This completes the proof.
\end{proof}

\begin{corollary}
\label{cor:nbupd-exp-lb}
Let $0 < \delta \leq 1$. Let $\mathcal{A}$ be a deterministic parallel algorithm for $\nbupdc{\PO}$ which, on instances of $\nbupdc{\PO}$ with $n$ predictors over $\{0, 1\}^n$ and parameter $\epsilon \coloneqq 1/10$, satisfies $\depth(\mathcal{A}) \leq O(n^{1 - \delta})$. Then the worst-case work of $\mathcal{A}$ on such instances satisfies
\[\work(\mathcal{A}) \geq \exp(\Omega(n^{\delta})).\]
\end{corollary}
\begin{proof}
    Simply note that the work of $\mathcal{A}$ is lower bounded by both the maximum number of queries that it makes in any round, and its output size. Thus, using the notation of Theorem~\ref{thm:next-bit-round-lower-bound}, $\work(\mathcal{A}) \geq \max(Q, |S|n)$. Plugging the algorithm from Lemma~\ref{lem:tree-balancing-algo} into Theorem~\ref{thm:next-bit-round-lower-bound} gives
    \[(R+1)\lfloor \log( \work(\mathcal{A}) + 1) \rfloor \geq (\frac{1}{2} - \epsilon)n.\]
    By assumption we have
    \[R+1 \leq O(n^{1 - \delta}).\]
    Thus,
    \[\work(\mathcal{A}) \geq \exp(\Omega(n^{\delta})).\]
\end{proof}
\section{\texorpdfstring{$\StCL$ versus $\SCL$}{StableCL versus SearchCL}}
\label{sec:stablecl}
In this section, we consider the problem of whether $\SCL$ reduces to $\CL$. For example, does $\CL \subseteq \Logspace$ imply that $\SCL \subseteq \FL$? To see why this is not immediately true, recall Definitions~\ref{def:cat-decision} and~\ref{def:compute-search-problem}. The standard reduction from search to decision is as follows: let's say we have a Turing machine $M$ which solves a search problem $P \subseteq \{0, 1\}^* \times \{0, 1\}^*$. Typically, one can define a Turing machine $M'$ which accepts on input $(x \in \{0, 1\}^*, i \in \mathbb{N})$ if and only if the $i^{th}$ bit of $M(x)$ is equal to $1$. This sort of reduction is generally successful if $M$ and $M'$ are simple time-bounded or space-bounded Turing machines. However, this approach has a clear flaw when $M$ and $M'$ are more complicated machines such as randomized or catalytic. If $M$ is a catalytic machine, then it takes as input an additional catalytic string $\tau$, and the simulating machine $M'$ must also take this string as input. Now, the natural generalization of the approach would be to define $M'(x, i, \tau)$ to accept if and only if the $i^{th}$ bit of $M(x, \tau)$ is equal to $1$. However, note that Definition~\ref{def:compute-search-problem} allows $M$ to have different outputs for different catalytic tapes. That is, there may exist $\tau$ and $\tau'$ such that $M(x, \tau) \neq M(x, \tau')$. At the same time, Definition~\ref{def:cat-decision} requires that a catalytic machine $M'$ solving a decision problem must have the same output for all catalytic tapes. Thus, $M'(x, i, \tau)$ must be equal to $M'(x, i, \tau')$ for all $\tau, \tau'$. It is thus clear that this reduction cannot work naively.

This brings us to the following problem.

\begin{problem}
\label{prob:flcl-equal-scl}
    Is $\FL^{\CL} = \SCL$? 
\end{problem}

For readers familiar with complexity theory, this is similar to the problem of $\SBPP$ vs $\FP^\BPP$. In that case, it is known that $\FP^\BPP$ is equivalent to $\psdBPP$~\cite{GatGoldwasser11}. This is a subclass of $\SBPP$ where the randomized Turing machine is required to output the \textit{same} solution for any given input with high probability. Thus, it eliminates the additional instability in search problems caused by the randomness. 

This approach naturally extends to our case. We define a subclass of $\SCL$ known as $\StCL$, where the catalytic Turing machine is required, for a fixed input $x$, to produce the same output $y$ for all initial catalytic tape configurations. Thus, it eliminates the additional instability in search problems caused by the initial catalytic tape.

\begin{definition}[$\StCL$]
A catalytic Turing machine $M$ which uses $c(n)$ catalytic space and computes a search problem $P \subseteq \{0, 1\}^* \times \{0, 1\}^*$ is stable if for every $x \in \{0, 1\}^n$ and $\tau, \tau' \in \{0, 1\}^{c(n)}$, $M(x, \tau) = M(x, \tau')$.
$\StCL$ is the class of search problems computable by stable catalytic Turing machines with $O(\log n)$ work space and $\poly(n)$ catalytic space.
\end{definition}

In this section, we will show that the question of whether $\StCL = \SCL$ is equivalent to the question of whether $\SCL$ is equal to $\FL^\CL$. Moreover, since $\updc{\TCo} \in \SCL$ (due to Corollary~\ref{cor:tco-unpred-in-searchcl}), the question of whether $\updc{\TCo} \in \StCL$ is a special case of this problem. We show that any algorithm which follows the same idea as the one in Corollary~\ref{cor:tco-unpred-in-searchcl} cannot work. In particular, Corollary~\ref{cor:tco-unpred-in-searchcl} additionally shows that $\updc{\TCo}$ is in $\RSCL$. If we additionally define a stable version of $\RSCL$, then we can show that it is unconditionally impossible for any search problem with output size $\omega(\log n)$ to be computed in stable $\RSCL$.  

We begin by showing that $\StCL = \FL^{\CL}$. Since an $\FL$ machine simply does not possess linear read-write space, we use the classical oracle tape model here.

\begin{lemma}
    $\StCL \subseteq \FL^{\CL}.$
\end{lemma}
\begin{proof}
Let $P$ be a search problem in $\StCL$ and $M$ be the stable catalytic logspace Turing machine which computes it. For any $x \in \{0, 1\}^*$, we define $M(x) \coloneqq M(x, 0\dots0)$. Since $M$ is stable, note that $M(x) = M(x, \tau)$ for any initial catalytic tape configuration $\tau$. We define the languages $P_{\len}$ and $P_{\ind}$ as follows:
\begin{enumerate}
    \item $(x \in \{0, 1\}^*, i \in \mathbb{N}) \in P_{\len}$ if and only if $|M(x)| \geq i$.
    \item $(x \in \{0, 1\}^*, i \in \mathbb{N}) \in P_{\ind}$ if and only if $(x, i) \in P_{\len}$ and the $i^{th}$ bit of $M(x)$ is equal to $1$.
\end{enumerate}

It remains to be proven that $P_{\len}$ and $P_{\ind}$ can be decided by catalytic logspace machines. First, as discussed in Section~\ref{sec:CL-equiv}, a catalytic logspace machine runs in polynomial time on average. Thus, the length of $M(x)$ must be at most polynomial. Therefore, in order to decide $P_{\len}$, a catalytic logspace machine can simply simulate $M$, keep track of its output length on the work tape, and accept if and only if the output length is at least $i$. Similarly, to decide $P_{\ind}$, a catalytic logspace machine can simulate $M$, remember the $i^{th}$ bit of its output, and then simply accept if and only if this bit is $1$. If $M$ terminates before reaching the $i^{th}$ bit then the machine should, of course, reject. Thus, $P_{\len}$ and $P_{\ind}$ are in $\CL$.

Now, simply note that an $\FL$ machine with oracle access to $P_{\ind}$ and $P_{\len}$ can easily compute $P$ as follows: on input $x$, it iterates a counter $i$ from $1$ in ascending order. It first uses $P_{\len}$ to check whether $i$ is at most the length of $M(x)$. If this is false, then it terminates. Else, it uses $P_{\ind}$ to compute the $i^{th}$ bit of $M(x)$, and writes this on the output tape. It then increments $i$. Clearly this machine solves the search problem $P$.

This completes the proof.
\end{proof}

\begin{lemma}
    $\FL^{\CL} \subseteq \StCL$.
\end{lemma}
\begin{proof}
Simply note that any $\FL^{\CL}$ machine can be simulated by a single catalytic logspace machine, which uses its own catalytic tape as the catalytic tape for all $\CL$ machines used by the $\FL^{\CL}$ machine as oracles\footnote{Replacing an oracle with another space-bounded machine is a standard technique in space-bounded algorithms.}. All $\CL$ machines used as oracles are stable, and the $\FL$ machine itself does not use a catalytic tape at all; thus, this simulating machine is also stable. This completes the proof.
\end{proof}

\begin{corollary}
\label{cor:flcl-equals-stcl}
 $\FL^{\CL} = \StCL$.
\end{corollary}

Thus, the following problem is equivalent to Problem~\ref{prob:flcl-equal-scl}.

\begin{problem}
Is $\StCL = \SCL$?
\end{problem}

We note that Agarwala, Alekseev, and Vinciguerra~\cite{AgarwalaAlekseevVinciguerra26} briefly discuss the need for $\StCL$ independently of Problem~\ref{prob:flcl-equal-scl}. Their $\SCL$ algorithm for linear matroid intersection uses the $\SCL$ algorithm for bipartite matching by Agarwala and Mertz~\cite{AgarwalaMertz25} as a subroutine. In fact, they use this subroutine multiple times across their algorithm, even as their algorithm mutates the string on the catalytic tape. Thus, they have to deal with the fact that the bipartite matching algorithm may return different matchings across the course of their algorithm, even when queried on the same graph. This is evidence of the fact that the notion of stability (and pseudodeterminism) is particularly important for subroutines in the low-space setting, where an algorithm cannot afford to store the entire output of a search subroutine on its work tape, and thus relies on stability of the subroutine across multiple calls.

Now, essentially all of the breakthrough $\SCL$ algorithms of recent years (bipartite matching, linear matroid intersection, general matching, graph sparsification) are \textit{not} stable. Furthermore, it is extremely unclear how they could be turned into stable algorithms. There are tricks, such as those in~\cite{AgarwalaVarma26}, that can be exploited to make certain $\SCL$ algorithms stable, such as those for $s$-$t$ minimum cut and global minimum cut (which arise as corollaries of the bipartite matching algorithm). However, these are highly problem-specific and it is unclear whether any general strategy exists. We note that in the related $\SBPP$ vs $\FP^{\BPP}$ case, it is known under standard circuit-hardness assumptions~\cite{ImpagliazzoWigderson97} that efficiently verifiable randomized search problems admit pseudodeterministic polynomial-time algorithms. We can draw no such conclusion here. Thus, it is not only unclear how a proof of $\StCL = \SCL$ would go, but it is also very unclear whether this fact is true at all.

In this section, we show a major barrier towards proving $\StCL = \SCL$. Note that the algorithm for $\updc{\TCo}$ is not only a $\SCL$ algorithm, but also a $\RSCL$ algorithm. This means that the algorithm can explicitly write down the correct output on a prefix of the catalytic tape itself. Now, imagine a variant of $\RSCL$ which additionally requires the algorithm to be stable. That is, we require the correct output to be written on a prefix of the catalytic tape and to be the same across all instantiations of the catalytic tape. We show that such an algorithm is unconditionally impossible for any search problem whose output has length $\omega(\log n)$. Thus, any restorable catalytic logspace algorithm for $\updc{\TCo}$ \textit{cannot} be stable. This implies that any $\StCL$ algorithm for $\updc{\TCo}$ must be substantially different from the one in Corollary~\ref{cor:tco-unpred-in-searchcl}. This also shows why the $\SCL$ algorithms of recent years seem difficult to make stable - because, as we show in this work, they all can be reinterpreted as reductions to $\updc{\TCo}$. 

We begin by defining $\SRSCL$.

\begin{definition}[$\SRSCL$]
    A restorable catalytic Turing machine $M$ is defined to be stable if for all $x \in \{0, 1\}^*$ and all initial catalytic tape configurations $\tau, \tau'$, we have $M(x, \tau) = M(x, \tau')$. A search problem $P$ is in $\SRSCL$ if it is computed by a stable restorable catalytic logspace machine.
\end{definition}

The following lemma shows that, up to additive logarithmic terms, any stable restorable catalytic machine which computes a search problem $P$ must use at least as much work space as the size of the output. 
\begin{lemma}
\label{lem:restorable-stable-needs-large-space}
Let $P \subseteq \{0, 1\}^* \times \{0, 1\}^*$ be a search problem. Fix $x \in \{0, 1\}^*$. Let $M$ be a stable restorable catalytic Turing machine which has a finite-control automaton consisting of $z$ states and, on an input $x \in \{0, 1\}^n$, uses $s$ work space, $c$ catalytic space, and produces as output $y$ such that $(x, y) \in P$. Then
\[s + \lceil\log n\rceil + \lceil\log s\rceil + \lceil\log c\rceil + \lceil\log z\rceil \geq |y|.\]
\end{lemma}
\begin{proof}
Note that the entire configuration of $M$ on an input $x$ with any initial catalytic tape string at any point can be described using $s + c + \lceil\log n\rceil + \lceil\log s\rceil + \lceil\log c\rceil + \lceil\log z\rceil$ many bits. The bits are required to store the locations of the tape heads, the strings on the work tape and catalytic tape, and the current state in the finite control of $M$. 

Consider the map $\Pi_{M}^x:\{0, 1\}^c \rightarrow \{0, 1\}^{s + c + \lceil\log n\rceil + \lceil\log s\rceil + \lceil\log c\rceil + \lceil\log z\rceil}$ which maps an initial catalytic tape $\tau \in \{0, 1\}^c$ to the complete configuration of $M$ when the $\compute_M$ procedure terminates. Observe that, from $\Pi_M^x(\tau)$, the procedure $\restore_M$ recovers $\tau$. Thus, $\Pi^x_M$ must be injective. Now, if $M$ is stable, then the first $|y|$ bits of the catalytic tape when $\compute_M$ terminates are the same for all initial catalytic tapes - they are the string $y$. Thus, the image of $\Pi_M^x$ has size at most $2^{{s + c - |y| + \lceil\log n\rceil + \lceil\log s\rceil + \lceil\log c\rceil + \lceil\log z\rceil}}$. On the other hand, the domain has size $2^c$. Injectivity implies the size of the image is at least the size of the domain. Thus,
\[s + \lceil\log n\rceil + \lceil\log s\rceil + \lceil\log c\rceil + \lceil\log z\rceil \geq |y|.\]
This completes the proof.
\end{proof}

\begin{corollary}
\label{cor:restorable-algos-cant-be-stable}
A restorable catalytic logspace algorithm for $\updc{\TCo}$ cannot be stable.
\end{corollary}
\begin{proof}
This follows directly from Lemma~\ref{lem:restorable-stable-needs-large-space}. It is very easy to see that there are families of $\updc{\TCo}$ instances where any valid output has size $\Omega(n)$. Thus, Lemma~\ref{lem:restorable-stable-needs-large-space} would imply that any stable restorable catalytic algorithm for $\updc{\TCo}$ using $\poly(n)$ catalytic space would require $\Omega(n)$ work space. Such an algorithm would not be a catalytic logspace algorithm.
\end{proof}

\section{Next-Bit Predictors, Pessimistic Estimators, and Graph Sparsification}
\label{sec:nbp-and-ess-est}
In this section we will show that, up to small changes in parameters, the problem of fooling next-bit predictors ($\nbupdc{}$) is equivalent to the problem of computing an approximate solution to a \emph{pessimistic estimator} instance (Definition~\ref{def:pess-est}). In particular, this will imply that the graph sparsification algorithm by Marcussen, Pyne, and Rubinfeld~\cite{MarcussenPyneRubinfeld26} can be restated as a $\TCo$ reduction to $\nbupdc{\TCo}$.

The underlying method of pessimistic estimators was introduced by Raghavan~\cite{Raghavan88}. We begin by defining pessimistic estimator instances, and what an approximate solution to such an instance means.
\begin{definition}[Pessimistic Estimator]
\label{def:pess-est}
    Let $n \in \mathbb{N}$, and define $S_n \coloneqq \bigcup_{i=0}^n \{0, 1\}^i$. A pessimistic estimator instance over $\{0, 1\}^n$ consists of a function $F:S_n \rightarrow \mathbb{R}_{>0}$ with the property that for any string $s \in S_{n-1}$, 
    \begin{equation}
        \label{eqn:pessimistic-estimator-def}
        \frac{F(s \circ 0) + F(s \circ 1)}{2} \leq F(s).
    \end{equation}
For $\alpha \in \mathbb{R}_{>0}$, a string $s \in \{0, 1\}^n$ is an $\alpha$-approximate solution to the pessimistic estimator instance~$F$ if
\[F(s) \leq \alpha \cdot F(\emptyset).\]
\end{definition}

An approach that has been fairly successful in the field of derandomization is this: in order to obtain a string $s \in S \subseteq \{0, 1\}^n$, define a pessimistic estimator $F$ such that any (possibly even approximate) solution to $F$ belongs to $S$.

Observe that, given oracle access to the function $F$, computing an exact solution (i.e., a $1$-approximation) to a pessimistic estimator instance is easy in polynomial time: begin with the string $s = \emptyset$. By Equation~\ref{eqn:pessimistic-estimator-def}, you are guaranteed that there exists $b \in \{0, 1\}$ such that $F(s \circ b) \leq F(s)$. Set $s \leftarrow s \circ b$, and iterate this process until $s$ has length $n$. This process is highly sequential. In fact, it is possible to show (though we will not do it here) that the problem of computing an exact solution to a pessimistic estimator instance $F$ is $\P$-complete under logspace reductions, even when~$F$ is represented explicitly by an $\NCo$ circuit given as input.

Thus, while computing an exact solution to a pessimistic estimator instance is easy in polynomial time, the fact that it is $\P$-complete means we have to settle for approximations in weaker computational models. Observe that the expected value of $F(s)$ over a uniformly random string $s \in \{0, 1\}^n$ is at most $F(\emptyset)$. Thus, for any $\alpha > 1$, a uniformly random string $s \in \{0, 1\}^n$ is an $\alpha$-approximate solution to a pessimistic estimator instance $F$ with probability $\geq 1 - 1/\alpha$. This follows from the Markov bound. This implies that, for $\alpha \geq 1 + \Omega(1/\poly(n))$, a set of $\poly(n)$ uniformly random strings from $\{0, 1\}^n$ contains an $\alpha$-approximate solution to $F$ with high probability. Such a solution can be verified using just two queries to $F$. Thus, the problem of producing an $\alpha$-approximate solution to a pessimistic estimator instance is a total derandomization problem.

Marcussen, Pyne, and Rubinfeld~\cite{MarcussenPyneRubinfeld26} use pessimistic estimators to solve the graph spectral sparsification problem ($\graphspars$, formally defined in Definitions~\ref{def:spectral-approx} and~\ref{def:graphspars}). Specifically, they showed that $\graphspars$ has a $\LUTCo$ reduction to the problem of computing an $n$-approximation to a pessimistic estimator instance $F$, where $F$ is explicitly given as a $\TCo$ circuit. They concluded that $\graphspars$ belongs to $\SCL$ by giving a compress-or-random algorithm which computes an approximate solution to the pessimistic estimator instance.

In Section~\ref{sec:pess-est-to-nbp} we will show that the problem of computing an approximate solution to a pessimistic estimator instance reduces to $\nbupdc{}$. In Section~\ref{sec:graph-spars-to-pess-est} we will state the reduction in~\cite{MarcussenPyneRubinfeld26} from $\graphspars$ to the problem of computing an approximate solution to a pessimistic estimator instance. We will stitch this with our reduction from Section~\ref{sec:pess-est-to-nbp} to show that $\graphspars \leq_{\LUTCo} \nbupdc{\TCo}$. Finally, in Section~\ref{sec:nbp-to-pess-est}, we will show that $\nbupdc{}$ itself reduces to the problem of computing an approximate solution to a pessimistic estimator instance. Thus, Sections~\ref{sec:pess-est-to-nbp} and~\ref{sec:nbp-to-pess-est} together show an equivalence between $\nbupdc{}$ and the problem of computing an approximate solution to a pessimistic estimator instance, up to parameters. 

Before we proceed further, we address a technicality. We will deal with problems where the function $F$ is explicitly computed by a circuit. Note that the input to $F$ is not of fixed length. Thus, when we refer to a circuit $C$ computing $F$, we mean that $C$ has $n + \lceil\log (n+1)\rceil$ input gates, where the first $\lceil\log(n + 1)\rceil$ bits of any input are interpreted as an integer $i \in \{0, 1, \dots, n\}$. For an input $(i, s \in \{0, 1\}^n)$ to $C$, we have $C(i, s) = F(s_{\le i})$. That is, $C$ evaluates the function $F$ on the first $i$ bits of the string $s$. Moreover, in all of our computational theorems, each output of $C$, and thus of $F$, will be a positive rational encoded as a pair of numerator-denominator integers, instead of arbitrary positive reals. We will follow this convention for the entire section.

Now, the following two definitions together define the problem of $\graphspars$. Note that, in the end, we will largely just cite the reduction of~\cite{MarcussenPyneRubinfeld26}. Thus, it is not strictly necessary to understand these definitions. Nevertheless, we include them for the interested reader. 

\begin{definition}[Spectral Approximation]
\label{def:spectral-approx}
    Let $G$ be an undirected graph with Laplacian $L_G$, and let $\epsilon>0$. Let $H$ be a weighted undirected graph such that $H$ has the same vertex set as $G$ and the edge weights of $H$ are nonnegative. $H$ is defined to be an $\epsilon$-spectral approximation of $G$ if
    \[(1-\epsilon)L_G\preceq L_H\preceq(1+\epsilon)L_G,\]
    where $\preceq$ denotes the Loewner order on real symmetric matrices: $A\preceq B$ if and only if $B-A$ is positive semidefinite.
\end{definition}

\begin{definition}[$\graphspars$]
\label{def:graphspars}
The graph spectral sparsification problem is a search problem defined as follows:
\begin{enumerate}
    \item $\mathbf{Input:}$ An unweighted undirected simple graph $G$ on $n$ vertices and $m$ edges, and a parameter $\delta \in \mathbb{R}_{> 0}$.
    \item $\mathbf{Output:}$ A weighted graph $H$ with $O(n \delta^{-2}\log n)$ edges such that $H$ is a $\delta$-spectral approximation of $G$.
\end{enumerate}
\end{definition}

\subsection{\texorpdfstring{Computing Approximate Solutions to Pessimistic Estimator Instances Reduces to $\nbupdc{}$}{Computing Approximate Solutions to Pessimistic Estimator Instances Reduces to NextBitUNPRED}}
\label{sec:pess-est-to-nbp}

We begin by defining, for any pessimistic estimator instance $F$, the family of next-bit predictors that will be the result of our reduction. This is done in Definition~\ref{def:nbp-from-pess-est}. 

The intuition behind this definition is as follows. We are given a pessimistic estimator instance~$F$, and we want to define a set of next-bit predictors $\mathcal{P}(F)$ such that, for any multiset $S$ which does \emph{not} contain an approximate solution to~$F$, the predictors $\mathcal{P}(F)$ perform well on $S$. In order to do this, we define one predictor $P_i$ for each index $i \in [n]$. Given as input $s_{<i}$ for a string $s \in \{0, 1\}^n$,~$P_i$ must predict $s_i$. $P_i$ works as follows: for each $b \in \{0, 1\}$, it produces as output the bit $b$ with probability $F(s_{<i} \circ b)/(F(s_{<i} \circ 0) + F(s_{<i} \circ 1))$. 

To see why this works, let's consider a string $s \in \{0, 1\}^n$ such that $s$ is far from a good solution to the pessimistic estimator instance $F$. Consider all prefixes of $s$: $s_{\leq 0} = \emptyset, s_{\leq 1}, \dots, s_{\leq n-1}, s_{\leq n} = s$. Since $F(s)$ is much larger than $F(s_{\leq 0})$, there must exist $i \in [n]$ such that $F(s_{\leq i})$ is much larger than $F(s_{\leq {i-1}})$. Now consider the predictor $P_i$ as defined above, acting on the string $s$. The definition of $P_i$ naturally makes its prediction lean towards the bit $b$ which maximizes $F(s_{\leq i-1} \circ b)$. Since $F(s_{\leq i}) > F(s_{\leq {i-1}})$, this bit $b$ is exactly $s_i$. Thus, $P_i$ succeeds on $s$ with good probability. 

This is exactly the intuition that we will formalize in Lemma~\ref{lem:pess-est-reduces-to-nbp}. We note that this is quite similar to Yao's hybrid argument.

Note that the predictors described above are randomized, while we require deterministic predictors. Thus, we will actually define a family $\mathcal{P}(F, T)$ where $T$ is an integer parameter. $T$ should be thought of as a discretization parameter, which is used to define deterministic variants of the randomized predictors $\mathcal{P}(F)$ described above. Larger values of $T$ will result in better predictors at the cost of a larger family.

\begin{definition}[Next-Bit Predictors Corresponding to a Pessimistic Estimator]
\label{def:nbp-from-pess-est}
Let $n \in \mathbb{N}$, let $F:S_n \rightarrow \mathbb{R}_{>0}$ be a pessimistic estimator instance over $\{0, 1\}^n$ as in Definition~\ref{def:pess-est}, and let $T \in \mathbb{N}$ be a parameter.

We define the family $\mathcal{P}(F, T) \coloneqq \{P_{ij}\mid i \in [n], j \in [T]\}$ of next-bit predictors, where $P_{ij}$ is a next-bit predictor at index $i$ defined as follows: For $s \in \{0, 1\}^{i-1}$,
\[P_{ij}(s) \coloneqq \mathbbm{1}\left[j \leq \lfloor \frac{1}{2} +  \frac{T \cdot F(s \circ 1)}{F(s \circ 0) + F(s \circ 1)} \rfloor\right].\]
\end{definition}

Now, we show that the predictors $\mathcal{P}(F, T)$ indeed constitute a valid reduction.
\begin{lemma}[]
\label{lem:pess-est-reduces-to-nbp}
Let $n \in \mathbb{N}$ and let $F:S_n \rightarrow \mathbb{R}_{>0}$ be a pessimistic estimator instance over $\{0, 1\}^n$ as in Definition~\ref{def:pess-est}. Let $\alpha, \epsilon \in \mathbb{R}_{>0}$, and $T \in \mathbb{N}$ be parameters such that
\begin{equation}
\label{eqn:pess-estimator-to-nbp-parameter-constraints}
\epsilon \leq \frac{\ln \alpha}{2n} - \frac{1}{2T}.
\end{equation}
Let $S \subseteq \{0, 1\}^n$ be a multiset such that $\adv_{U(S)}(\mathcal{P}(F, T)) < \epsilon$. There exists $s \in S$ such that $s$ is an $\alpha$-approximate solution to the pessimistic estimator instance $F$.    
\end{lemma}
\begin{proof}
Fix $i\in[n]$ and a prefix $s\in\{0,1\}^{i-1}$, and write
\[
p\coloneqq\frac{F(s\circ1)}{F(s\circ0)+F(s\circ1)},
\qquad
r\coloneqq\left\lfloor Tp+\frac{1}{2}\right\rfloor.
\]
By definition, $P_{ij}(s)=1$ if and only if $j\le r$. Since $j$ is an integer, we get
\[P_{ij}(s) = 1 \iff F(s \circ 1) \geq \frac{(2j  - 1)}{2T} \cdot (F(s \circ 0) + F(s \circ 1)).\]
Note that exactly $r$ of the predictors $\{P_{ij}\}_{j=1}^{T}$ predict $P_{ij}(s)=1$. Moreover, rounding gives
\[
Tp-\frac{1}{2}<r\le Tp+\frac{1}{2}
\quad\implies\quad
\left|\frac{r}{T}-p\right|\le\frac{1}{2T}.
\]
Next, let $s\in\{0,1\}^i$, with $p$ and $r$ defined as above for the prefix $s_{<i}$. If $s_i=1$, the success probability of a uniformly random predictor from $\{P_{ij}\}_{j=1}^{T}$ on $s$ is $\frac{r}{T}\ge p-\frac{1}{2T}$. If $s_i=0$, it is $1-\frac{r}{T}\ge1-p-\frac{1}{2T}$. Therefore,
\begin{align}
    \Pr_{j \in [T]}[P_{ij}(s_{<i}) = s_i] &\geq \frac{F(s)}{F(s_{<i} \circ 0) + F(s_{<i} \circ 1)} - \frac{1}{2T} \notag\\
    &\geq \frac{F(s)}{2F(s_{<i})} - \frac{1}{2T}. &&(\text{by Equation}~\ref{eqn:pessimistic-estimator-def}) \label{eqn:nbp-equality-lb}
\end{align}
Now, let $s \in \{0, 1\}^n$ be a string which is \textbf{not} an $\alpha$-approximate solution to $F$. That is,
\[F(s) > \alpha \cdot F(\emptyset).\]
We have 
\begin{align*}
    \Pr_{i \in [n], j \in [T]}[P_{ij}(s_{<i}) = s_i] &= \frac{1}{n} \sum_{i=1}^n \Pr_{j \in [T]}[P_{ij}(s_{<i}) = s_i]\\
    &\geq \left(\frac{1}{n} \sum_{i=1}^n \frac{F(s_{\le i})}{2F(s_{<i})}\right) - \frac{1}{2T} &&(\text{By Equation}~\ref{eqn:nbp-equality-lb})\\
    &\geq \left(\frac{F(s)}{F(\emptyset)2^n}\right)^{1/n} - \frac{1}{2T} &&(\text{By AM-GM Inequality})\\
    &\geq \left(\frac{\alpha}{2^n}\right)^{1/n} - \frac{1}{2T}\\
    &= \frac{\alpha^{1/n}}{2} - \frac{1}{2T}\\
    &\geq \frac{1}{2} + \frac{\ln(\alpha)}{2n} - \frac{1}{2T}&&(\text{By } e^x \geq 1+x)\\
    &\geq \frac{1}{2} + \epsilon.&&(\text{By Equation}~\ref{eqn:pess-estimator-to-nbp-parameter-constraints})
\end{align*}
This implies that for any multiset $S \subseteq \{0, 1\}^n$ such that $\forall s \in S$, $s$ is \textbf{not} an $\alpha$-approximate solution to $F$, we have
\[\Pr_{s \in S, i \in [n], j \in [T]}[P_{ij} \succs s] \geq \frac{1}{2} + \epsilon.\]
This implies that there exists $P_{ij} \in \mathcal{P}(F, T)$ such that
\[\Pr_{s \in S}[P_{ij} \succs s] \geq \frac{1}{2} + \epsilon\]
\[\implies \adv_{U(S)}(\mathcal{P}(F, T)) \geq \epsilon.\]
Thus, for any multiset $S$ with $\adv_{U(S)}(\mathcal{P}(F, T)) < \epsilon$, there must exist $s \in S$ such that $s$ is an $\alpha$-approximate solution to $F$. This completes the proof.
\end{proof}

Finally, we show that this reduction can be performed on explicit circuits without blowing up their size or depth.

\begin{theorem}
\label{thm:pess-est-to-nbp}
There exists a logspace algorithm which takes as input $n \in \mathbb{N}$, a circuit $F$ which computes a pessimistic estimator instance over $\{0, 1\}^n$, and a parameter $T \in \mathbb{N}$ encoded in unary. It produces as output the set of predictors $\mathcal{P}(F, T)$ from Definition~\ref{def:nbp-from-pess-est} in the form of circuits, such that:
\begin{enumerate}
    \item Each predictor $P_{ij} \in \mathcal{P}(F, T)$ has
    \begin{equation*}
        \size(P_{ij}) \leq \poly(\size(F) + \log T) \qquad \text{and} \qquad \depth(P_{ij}) \leq O(\depth(F)).
    \end{equation*}
    \item For any parameters $0 < \epsilon < 1$ and $\alpha \geq 1$ such that
    \[\epsilon \leq \frac{\ln\alpha}{2n} - \frac{1}{2T},\]
    and any multiset $S \subseteq \{0, 1\}^n$ such that $\adv_{U(S)}(\mathcal{P}(F, T)) < \epsilon$, there exists $s \in S$ such that $s$ is an $\alpha$-approximate solution to the pessimistic estimator instance $F$.
\end{enumerate}
\end{theorem}
\begin{proof}
$\mathcal{P}(F, T)$ is of course exactly the set of predictors defined in Definition~\ref{def:nbp-from-pess-est}. The size and depth bounds of the first part are easy to see: the predictor $P_{ij}$ on an input $s$ simply computes the values $F(s \circ 0)$ and $F(s \circ 1)$ in parallel, and then produces as output the result of a simple comparison involving the aforementioned values, $T$, and $j$. The second part follows directly from Lemma~\ref{lem:pess-est-reduces-to-nbp}.
\end{proof}

\subsection{\texorpdfstring{Graph Sparsification Reduces to $\nbupdc{\TCo}$}{Graph Sparsification Reduces to NextBitUNPRED[TC1]}}
\label{sec:graph-spars-to-pess-est}

In this section we use the reduction by Marcussen, Pyne, and Rubinfeld~\cite{MarcussenPyneRubinfeld26}, which shows that $\graphspars$ reduces to the problem of computing an approximate solution to a $\TCo$ pessimistic estimator instance, and our reduction from Section~\ref{sec:pess-est-to-nbp}, to show that $\graphspars$ reduces to $\nbupdc{\TCo}$.

The following theorem was proven by Marcussen, Pyne, and Rubinfeld~\cite{MarcussenPyneRubinfeld26}. It is mainly Theorem 4.1 of their paper, but we combine it with additional details from other parts to obtain a clean statement that will suffice for our reduction.
\begin{theorem}[\cite{MarcussenPyneRubinfeld26}]
\label{thm:sparsification-reduces-to-pess-est}
There exists a logspace algorithm which takes as input a simple unweighted undirected graph $G$ on $n$ vertices and $m$ edges and a parameter $0 < \epsilon < 1/2$, and produces as output $\TCo$ circuits $\mathcal{A}$, $\mathcal{B}$, and $\mathcal{C}$ such that:
\begin{enumerate}
        \item $\mathcal{A}$ produces as output a nonnegative integer $A(G, \epsilon) = \poly(n, m)$.
        \item $\mathcal{B}$ takes as input a binary string $s$ of length at most $A(G,\epsilon)$. It produces as output a value denoted by $F_G^{\epsilon}(s) \in \mathbb{R}_{>0}$ such that the function
        \[F_G^{\epsilon}:S_{A(G, \epsilon)} \rightarrow \mathbb{R}_{>0}\]
        is a pessimistic estimator instance over $\{0, 1\}^{A(G, \epsilon)}$.
        \item $\mathcal{C}$ takes as input a string $s \in \{0, 1\}^{A(G, \epsilon)}$. It produces as output a simple edge-weighted undirected graph $H(G, s, \epsilon)$ such that:
        \begin{enumerate}
            \item (1) $V(H(G, s, \epsilon)) = V(G)$, (2) $E(H(G, s, \epsilon))  \subseteq E(G)$, and (3) all edge weights of $H(G, s, \epsilon)$ lie in $[n]$.
            \item For every integer $k \in \mathbb{Z}_{>0}$, if $s$ is an $n^k$-approximate solution to the pessimistic estimator instance $F_G^{\epsilon}$, the subgraph $H(G, s, \epsilon)$ has $O((1+2^k\epsilon)n \epsilon^{-2} \log n)$ edges and is a $2^k\epsilon$-spectral approximation of $G$.
        \end{enumerate}
    
    \end{enumerate}
\end{theorem}

\noindent
Therefore, $\graphspars$ on an input $(G, \delta)$ with $0<\delta<1$ reduces in $\LUTCo$ to the problem of computing an $n$-approximate solution to a pessimistic estimator instance. The pessimistic estimator instance is over $\{0, 1\}^{\poly(n, m)}$ and the pessimistic estimator function is computable by a $\TCo$ circuit. This follows by invoking Theorem~\ref{thm:sparsification-reduces-to-pess-est} with parameters $k \coloneqq 1$ and $\epsilon \coloneqq \delta / 2$.

\begin{theorem}
\label{thm:sparsification-reduces-to-nbp}
    \[\graphspars \leq_{\LUTCo} \nbupdc{\TCo}.\]
\end{theorem}
\begin{proof}
Let $(G, \delta)$ be a $\graphspars$ instance, where $G = (V, E)$, $|V| = n$, $|E| = m$, and $\delta \in \mathbb{R}_{>0}$. Theorem~\ref{thm:sparsification-reduces-to-pess-est} reduces $\graphspars$ to the problem of computing an $n$-approximate solution to a pessimistic estimator instance $F$ over $\{0, 1\}^{A(G, \delta/2)}$, where $A(G, \delta/2) = \poly(n, m)$ and $F$ is a circuit with $\depth(F) = O(\log(n+m))$ and $\size(F) = \poly(n)$.

We first handle some edge cases: if $\delta \geq 1$, we can return the empty graph on vertex set $V(G)$. Else, if $n \leq 2$, we can return $G$. Finally, if $A(G, \delta/2) = 0$, we can simply return $\mathcal{C}(\emptyset).$ From here on, we assume that $\delta < 1$, $n \geq 3$, and $A(G, \delta/2) \geq 1$.

Set $T \coloneqq 2A(G, \delta/2) = \poly(n, m)$ and $\epsilon \coloneqq 0.25/A(G, \delta/2)$. By Theorem~\ref{thm:pess-est-to-nbp}, there exists a logspace algorithm which takes as input $F$ and $T$, and produces as output circuits computing a set of next-bit predictors $\mathcal{P}(F, T)$, such that the circuits have $\size = \poly(n, m)$ and $\depth = O(\log(n+m))$. Theorem~\ref{thm:pess-est-to-nbp} guarantees that for any $S \subseteq \{0, 1\}^{A(G, \delta/2)}$ which is a solution to the $\nbupdc{\TCo}$ instance given by $\mathcal{P}(F, T)$ and $\epsilon$, there exists $s \in S$ which is an $n$-approximate solution to the pessimistic estimator instance $F$.

Thus, given a polynomial-size solution $S$ to this $\nbupdc{\TCo}$ instance, an $\LUTCo$ algorithm could simply check in parallel for each $s \in S$ whether $F(s) \leq n \cdot F(\emptyset)$. By Theorem~\ref{thm:sparsification-reduces-to-pess-est}, such an $s$ suffices to solve $\graphspars$ in $\LUTCo$.

This completes the proof.
\end{proof}

\subsection{\texorpdfstring{$\nbupdc{}$ Reduces to Computing Approximate Solutions to Pessimistic Estimator Instances}{NextBitUNPRED Reduces to Computing Approximate Solutions to Pessimistic Estimator Instances}}
\label{sec:nbp-to-pess-est}

In this section, we show that $\nbupdc{}$ itself also reduces to the problem of computing approximate solutions to pessimistic estimator instances. The construction here is very similar to those in Wigderson and Xiao~\cite{WigdersonXiao08}, and was directly inspired by Li~\cite{Li21}.

We begin by formally defining the pessimistic estimator instance that will be the result of our reduction. We are given a $\nbupdc{}$ instance over $\{0, 1\}^n$. Let's say the predictors are $P_1, \dots, P_m$, where $P_i$ predicts index $\ind_i \in [n]$. For such an instance, the pessimistic estimator $F$ will need to be over much larger strings of length $kn$, for a parameter $k$. We will then interpret a string $S \in \{0, 1\}^{kn}$ as the concatenation of $k$ strings $s^1, \dots, s^k \in \{0, 1\}^n$. Similar to Section~\ref{sec:upd-in-cl}, we will interpret $S$ to contain $s^1, \dots, s^k$ in column-major order. That is, the first $k$ bits of $S$ are the first bit of each string, and so on. 

Now, to define the pessimistic estimator itself. The estimator $F$ is the sum of $m$ estimators, one for each predictor. Let us describe the estimator $F_i$ corresponding to just one predictor $P_i$. Let $S'$ be a prefix of $S \in \{0, 1\}^{kn}$. We will define $F_i(S')$. In order to do this, the estimator needs two values in particular: $x$ and $j$. $x$ is the number of strings $s \in S$ such that the bits of $s_{\leq \ind_i}$ have already been completed in $S'$, and they result in $P_i$ succeeding. The second value, $j$, is simply the number of strings $s \in S$ whose first $\ind_i$ bits have already been filled into $S'$. This is similar to $x$, but also includes strings on which $P_i$ is now guaranteed to fail. For a fixed parameter $\epsilon$, we define
\[F_i(S') \coloneqq (1 + \epsilon)^x\left(\frac{2+\epsilon}{2}\right)^{k-j}.\]
We emphasize that $x$ and $j$ depend on $S'$. It is a simple exercise to verify that this is indeed a pessimistic estimator. Claim~\ref{claim:upper-bound-on-initial-estimator} shows that the initial value $F_i(\emptyset)$ of this estimator is $\leq \exp(k \epsilon/2)$. Theorem~\ref{thm:nbp-reduces-to-pess-est} will show that if the multiset corresponding to a string $S \in \{0, 1\}^{kn}$ is not a solution to $\nbupdc{}$ with parameter $\epsilon/2$, then $F_i(S) \geq \exp(k\epsilon(1+\epsilon)/(2+\epsilon))$ for some $i \in [m]$. Thus, non-solutions $S$ result in pessimistic estimator values $F_i(S)$ that are larger than $F_i(\emptyset)$. We will use this fact to then complete the reduction in Corollary~\ref{cor:nbp-reduces-to-pess-est}.

Note that this reduction depends on parameters $k$ and $\epsilon$. Thus, the result of our reduction will be called $F(\mathcal{P}, k, \epsilon)$, and $F(P, k, \epsilon)$ for a fixed predictor $P \in \mathcal{P}$.

\begin{definition}[Pessimistic Estimators Corresponding to Next-Bit Predictors]
\label{def:pess-est-from-nbp}
Let $\mathcal{P} = \bigcup_{i=1}^n \mathcal{P}_i$ be a non-empty set of next-bit predictors over $\{0, 1\}^n$ where $\mathcal{P}_i$ is the set of predictors at index $i$.

For a parameter $k \in \mathbb{N}$, a string $s \in S_{kn}$, and $i \in [n]$, let 
\[j_{\max}(k, s, i) \coloneqq \min\{k, \max\{0, |s| - k(i-1)\}\}.\]
Furthermore, for $1 \leq j \leq j_{\max}(k, s, i)$, we define
\[\str(k, s, i, j) \coloneqq s_j \circ s_{k + j} \circ s_{2k + j} \circ \dots \circ s_{(i-1)k + j}.\]
For a predictor $P \in \mathcal{P}_i$ and a parameter $\epsilon \in \mathbb{R}_{>0}$, we define a pessimistic estimator $F(P, k, \epsilon)$ over $\{0, 1\}^{kn}$ as follows: For $s \in S_{kn}$, let 
\[X(P, k, s) \coloneqq \sum_{j=1}^{j_{\max}(k, s, i)} \mathbbm{1}[P(\str(k, s, i, j)_{<i}) = \str(k, s, i, j)_{i}],\]
and
\[F(P, k, \epsilon)(s) \coloneqq (1 + \epsilon)^{X(P, k, s)}\left(\frac{2 + \epsilon}{2}\right)^{k - j_{\max}(k, s, i)}.\]
Finally, we define a single pessimistic estimator instance $F(\mathcal{P}, k, \epsilon)$ corresponding to the entire set of predictors $\mathcal{P}$, as follows:
\[F(\mathcal{P}, k, \epsilon)(s) \coloneqq \sum_{P \in \mathcal{P}} F(P, k, \epsilon)(s).\]
\end{definition}

It is a simple exercise to verify that each $F(P, k, \epsilon)$ satisfies Equation~\ref{eqn:pessimistic-estimator-def}, and is therefore indeed a pessimistic estimator. It is equally simple to verify that the sum of pessimistic estimators (i.e., $F(\mathcal{P}, k, \epsilon)$) is itself a pessimistic estimator. In the following claim we show an upper bound on $F(P, k, \epsilon)(\emptyset)$.

\begin{claim}
\label{claim:upper-bound-on-initial-estimator}
Let $F$ be as in Definition~\ref{def:pess-est-from-nbp}.
Let $\mathcal{P}$ be a set of next-bit predictors over $\{0, 1\}^n$. Let $k \in \mathbb{N}$ and $\epsilon \in \mathbb{R}_{>0}$. Then, for all $P \in \mathcal{P}$,
\[F(P, k, \epsilon)(\emptyset) \leq \exp(\frac{k \epsilon}{2}).\]
\end{claim}
\begin{proof}
Note that $j_{\max}(k, \emptyset, i) = 0$ for all $k \in \mathbb{N}$ and $i \in [n]$. Therefore, $X(P, k, \emptyset) = 0$. This implies that
\begin{align*}
    F(P, k, \epsilon)(\emptyset) &= \left(\frac{2 + \epsilon}{2}\right)^{k}\\
    &= \exp(k\ln(1 + \frac{\epsilon}{2}))\\
    &\leq \exp(\frac{k\epsilon}{2}). &&(\text{Because }1+x \leq e^x)
\end{align*}
This completes the proof.
\end{proof}
In the following theorem we show that any string $s \in \{0, 1\}^{kn}$ such that $F(\mathcal{P}, k, \epsilon)(s)$ is sufficiently small must represent a solution to the $\nbupdc{}$ instance $\mathcal{P}$ with parameter $\epsilon/2$.
\begin{theorem}
\label{thm:nbp-reduces-to-pess-est}
Let $F$ and $\str$ be as in Definition~\ref{def:pess-est-from-nbp}.
Let $\mathcal{P}$ be a set of next-bit predictors over $\{0, 1\}^n$. Let $k \in \mathbb{N}$, $\epsilon \in \mathbb{R}_{>0}$, and $s \in \{0, 1\}^{kn}$ be such that $F(\mathcal{P}, k, \epsilon)(s) < \exp(k\epsilon(1+\epsilon)/(2+\epsilon))$. Then, the multiset $S \coloneqq \{ \str(k, s, n, j) \mid j \in [k]\}$ satisfies
\[\adv_{U(S)}(\mathcal{P}) < \frac{\epsilon}{2}.\]
\end{theorem}
\begin{proof}
Assume to the contrary that 
\[\adv_{U(S)}(\mathcal{P}) \geq \frac{\epsilon}{2}.\]
This implies that there exists $P \in \mathcal{P}$ such that
\[\adv_{U(S)}(P) \geq \frac{\epsilon}{2}.\]
Let's say that $P \in \mathcal{P}_i$. Simply observe that $j_{\max}(k, s, i) = k$. Therefore,
\begin{align*}
    X(P, k, s) &= \sum_{j=1}^k \mathbbm{1}[P \succs \str(k, s, n, j)]\\
    &\geq \frac{k}{2}(1 + \epsilon).
\end{align*}
Now,
\begin{align*}
    F(P, k, \epsilon)(s) &= (1 + \epsilon)^{X(P, k, s)}\left(\frac{2 + \epsilon}{2}\right)^{k - j_{\max}(k, s, i)} \\
    &= (1 + \epsilon)^{X(P, k, s)}\\
    &\geq (1+\epsilon)^{k(1+\epsilon)/2} && (\text{Because }X(P, k, s) \geq k(1+\epsilon)/2)\\
    &= \exp(\frac{k(1+\epsilon)}{2} \ln(1+\epsilon))\\
    &\geq \exp(\frac{k\epsilon(1+\epsilon)}{2+\epsilon}). && (\text{Because }\ln(1+\epsilon) \geq \frac{2\epsilon}{2+\epsilon})
\end{align*}
Therefore, this implies that $F(\mathcal{P}, k, \epsilon)(s) \geq F(P, k, \epsilon)(s) \geq \exp(k\epsilon(1+\epsilon)/(2+\epsilon))$, which is a contradiction. This completes the proof.
\end{proof}

In the following corollary, for an appropriate choice of the parameter $k$, we show that the lower bound for non-solutions from Theorem~\ref{thm:nbp-reduces-to-pess-est} is more than twice the upper bound on $F(\mathcal{P}, k, 2\epsilon)(\emptyset)$ obtained from Claim~\ref{claim:upper-bound-on-initial-estimator}. Thus, any $2$-approximate solution to the pessimistic estimator instance $F(\mathcal{P}, k, 2\epsilon)$ corresponds to a solution to the $\nbupdc{}$ instance with parameter $\epsilon$.

\begin{corollary}
\label{cor:nbp-reduces-to-pess-est}
Let $F$ and $\str$ be as in Definition~\ref{def:pess-est-from-nbp}.
Let $\mathcal{P}$ be a set of $m$ next-bit predictors over $\{0, 1\}^n$. Let $0 < \epsilon < 1/2$, and $k \in \mathbb{N}$ such that $k \geq \ln(2.01m)(1 + \epsilon)/\epsilon^2$. Let $s \in \{0, 1\}^{kn}$ be a $2$-approximate solution to the pessimistic estimator instance $F(\mathcal{P}, k, 2\epsilon)$. Then, the multiset $S \coloneqq \{ \str(k, s, n, j) \mid j \in [k]\}$ satisfies
\[\adv_{U(S)}(\mathcal{P}) < \epsilon.\]
\end{corollary}
\begin{proof}
    For all $P \in \mathcal{P}$, due to Claim~\ref{claim:upper-bound-on-initial-estimator}, we have
    \[F(P, k, 2\epsilon)(\emptyset) \leq \exp(k\epsilon).\]
    Thus, 
    \[F(\mathcal{P}, k, 2\epsilon)(s) \leq 2m\exp(k\epsilon).\]
    Note that
    \begin{align*}
    \frac{2m\exp(k\epsilon)}{\exp(2k\epsilon(1+2\epsilon)/(2+2\epsilon))} &= 2m\exp(k\epsilon - \frac{k\epsilon(1+2\epsilon)}{1+\epsilon})\\
    &= 2m\exp(\frac{-k\epsilon^2}{1+\epsilon})\\
    &\leq 2/2.01 &&(\text{By plugging in the lower bound on }k.)\\
    &< 1.
    \end{align*}
    Thus, we have \[F(\mathcal{P}, k, 2\epsilon)(s) < \exp(2k\epsilon(1+2\epsilon)/(2+2\epsilon)).\]
    Then, by Theorem~\ref{thm:nbp-reduces-to-pess-est}, the corollary follows.
\end{proof}

In the following observation we show that the circuit corresponding to the pessimistic estimator~$F$ can be constructed efficiently, the circuit itself incurs only a polynomial blow-up in size compared to the predictors, and only a constant factor blow-up in depth.

\begin{observation}
\label{obs:nbp-to-pess-est-reduction-is-efficient}
There exists a logspace algorithm which takes as input $n \in \mathbb{N}$, a set of $m$ next-bit predictors $\mathcal{P} = \bigcup_{i=1}^n \mathcal{P}_i$ over $\{0, 1\}^n$ in the form of their circuits, a unary-encoded integer $k \in \mathbb{N}$, and a unary-encoded parameter $0 < \epsilon \leq 1$. The algorithm produces as output a circuit $C_F$ computing the function $F(\mathcal{P},k,\epsilon)$ from Definition~\ref{def:pess-est-from-nbp} such that
\[ \size(C_F) = \poly(n, m, k, \epsilon^{-1}, \sum_{P \in \mathcal{P}} \size(P))\]
and
\[\depth(C_F) \leq O(\max_{P\in\mathcal{P}}\depth(P)).\]
\end{observation}
\begin{proof}
Let $i \in [n]$ and $P \in \mathcal{P}_i$. Note that the circuit which computes $F(P, k, \epsilon)$ on a string $s \in S_{kn}$ is straightforward: it first computes $X(P, k, s)$ by querying $P$ on $\str(k, s, i, j)_{<i}$ for each $j \in [j_{\max}(k, s, i)]$ in parallel. Once it has computed $X(P, k, s)$, then computing $F(P, k, \epsilon)(s)$ only requires simple arithmetic which can be performed by a polynomial-size constant-depth circuit~\cite{HesseAllenderMixBarrington02}. The final step to computing $F(\mathcal{P}, k, \epsilon)$ is just summing these values over all $P \in \mathcal{P}$, which can again be done by a polynomial-size constant-depth circuit.

This completes the proof.
\end{proof}

Therefore, by Observation~\ref{obs:nbp-to-pess-est-reduction-is-efficient} and Corollary~\ref{cor:nbp-reduces-to-pess-est}, we have a reduction from $\nbupdc{}$ to the problem of computing an approximate solution to a pessimistic estimator instance, where the pessimistic estimator is computed by a circuit of size at most polynomial in $n$, the sum of sizes of the predictor circuits and $\epsilon^{-1}$, and depth at most a constant factor larger than the maximum depth of the predictor circuits. Combined with the reduction from Section~\ref{sec:pess-est-to-nbp}, this shows that the problem of computing an approximate solution to a pessimistic estimator instance is, up to a small change in parameters, equivalent to $\nbupdc{}$.

\section{\texorpdfstring{$\blossyc$ Reduces to $\upd$}{BlockLossyCoding Reduces to UNPRED}}
\label{sec:block-lossy}
In this section we will show that the catalytic logspace algorithms for  bipartite matching~\cite{AgarwalaMertz25}, linear matroid intersection~\cite{AgarwalaAlekseevVinciguerra26}, and general matching~\cite{ChakrabortyDattaKusreMukhopadhyaySinhababu26} reduce in $\TCo$ to $\upd^{\TCo}$. We will prove this fact by identifying a common structural property in all three of these algorithms: these algorithms can be interpreted as reductions to a special form of the $\lossyc$ problem (as defined in Definition~\ref{def:lossy-coding}), that we call $\blossyc$.

An arbitrary $\lossyc$ instance has no promised structure in its compression and decompression. This lack of structure is exactly what makes it a difficult problem.  The compress-or-random algorithms that we mentioned above, on the other hand, use compression in a very specific way. In all of these cases, the input to the compressor is interpreted as the concatenation of $n$ strings, each of length $\ell$. On any input string, the compressor selects a specific block, and then \emph{only compresses the selected block}. The other blocks are left unchanged. Note that the selected block must be compressed enough to also store its index, which means that we must have $\ell > \log n$. We call such an instance of $\lossyc$ an $(n, \ell)$ $\blossyc$ instance.

We will show that, when $\ell$ is logarithmic in the input size, $(n, \ell)$ $\blossyc$ admits a simple reduction to $\upd$. When the compressor and decompressor are both $\TCo$ circuits, this reduction will result in $\TCo$ predictors. The catalytic search-to-decision reduction for the isolation lemma by Arvind, Chakraborty, and Datta~\cite{ArvindChakrabortyDatta26} follows the same form but, for the sake of succinctness, we will only prove the $\upd^{\TCo}$ reduction for the three algorithms stated above.

We begin by formally defining $\blossyc$.

\begin{definition}[$\blossyc$]
\label{def:block-lossy}
An $(n, \ell)$ $\blossyc$ instance is a $\lossyc$ instance
\begin{enumerate}
    \item $C:(\{0, 1\}^\ell)^n \rightarrow \{0, 1\}^{\lceil \log(n+1) \rceil} \times \{0, 1\}^{k} \times (\{0, 1\}^\ell)^{n-1}$ and
    \item $D:\{0, 1\}^{\lceil \log(n+1) \rceil} \times \{0, 1\}^{k} \times (\{0, 1\}^\ell)^{n-1} \rightarrow (\{0, 1\}^\ell)^n$,
\end{enumerate}
where $k \coloneqq \ell-1-\lceil \log(n+1) \rceil \geq 0$. The instance must satisfy the following two conditions:
\begin{enumerate}
    \item For all $x=(x_1, \dots, x_n) \in (\{0, 1\}^{\ell})^n$,
    \[(C(x))_1 \in [n].\]
    We define $\ind_C(x) \coloneqq (C(x))_1 \in [n]$ and $\advice_C(x) \coloneqq (C(x))_2 \in \{0, 1\}^k$.
    \item For all $x=(x_1, \dots, x_n) \in (\{0, 1\}^{\ell})^n$,
    let $C(x) = (\ind_C(x) , \advice_C(x), (y_1, \dots, y_{n-1}))$. Then
    \[(y_1, \dots, y_{n-1}) = (x_1, \dots, x_{\ind_C(x)-1}, x_{\ind_C(x)+1}, \dots, x_n).\]
\end{enumerate}
\end{definition}

We briefly explain the connection between this definition and our informal description. The idea is that the $\ind_C(x)^{th}$ block of the input string $x$ is compressed down to the strings $\ind_C(x)$ and $\advice_C(x)$, whose combined length is strictly less than $\ell$. A possible misconception is that one could try to solve $\blossyc$ by interpreting it as $\lossyc$ on a fixed block. A block consists of only $\ell$ bits. A brute-force search over all length $\ell$ strings could then solve the problem efficiently. The reason this does not work is that $\ind_C(x)$ itself depends on $x$. Thus, it is not possible to restrict the problem to only one block, because the choice of the block itself depends on the input string.

Now, we prove our reduction from $\blossyc$ to $\upd$. We begin by defining the predictors that will appear as a result of this reduction. Intuitively, one should imagine the predictors to be the result of a two-step reduction.

First, imagine that the predictors, instead of acting over a string of length $n\ell$ with the alphabet $\{0, 1\}$, are acting over a string of length $n$ with the alphabet $\{0, 1\}^{\ell}$. That is, instead of guessing the value of a single bit, each predictor guesses the value of a whole block. Then, we could define a set of predictors $\{P_{it} \mid i \in [n], t \in \{0, 1\}^k\}$ where $P_{it}$ guesses the $\ell$-bit string in the $i^{th}$ block. That is, given a string $s_1, s_2, \dots, s_{i-1}, s_{i+1}, \dots, s_n$ where each $s_i \in \{0, 1\}^{\ell}$, $P_{it}$ tries to predict $s_i$. To do this, it uses the decompressor $D$, plugging in the string $t \in \{0, 1\}^k$ as the advice, and $i \in [n]$ as the index. Thus, 
\[P_{it}(s_1, \dots, s_{i-1}, s_{i+1}, \dots, s_n) \coloneqq D(i, t, s_1, \dots, s_{i-1}, s_{i+1}, \dots, s_n)_i.\]
By definition of $\blossyc$, for any string $s \in \nonsol(C, D)$, there must be $i \in [n]$ and $t \in \{0, 1\}^k$ such that $P_{it}$ succeeds on $s$ (where success means guessing the whole block correctly). In particular, this occurs when $i = \ind_C(s)$ and $t = \advice_C(s)$. Therefore, the strong pigeonhole principle implies that for a multiset $S \subseteq \nonsol(C, D) \subseteq (\{0, 1\}^{\ell})^n$, there must exist $i \in [n]$ and $t \in \{0, 1\}^k$ such that $P_{it}$ succeeds on $s \sim U(S)$ with probability at least $1/(n2^k)$. Plugging in the definition of $k$, this success probability is $\geq 2/2^{\ell}$. For bit predictors, this is not a very high success probability. However, note that we are not predicting just a single bit correctly, but rather an entire block. The probability of random guessing being correct on an entire block of $\ell$ bits is $1/2^{\ell}$. Thus, these predictors do indeed beat random guessing by a margin of $1/2^{\ell}$. 

The second step is simply to transfer this $1/2^{\ell}$ margin of success over random guessing into bit predictors. To do this, the idea is as follows: instead of predicting the entire block at once, we define predictors $P_{itj}$, where $i$ and $t$ are as above, and $j \in [\ell]$. The predictor $P_{itj}$ will try to predict only the $j^{th}$ bit of the $i^{th}$ block. Of course, if $P_{itj}$ just simulated $P_{it}$ and output its $j^{th}$ bit, then it would not necessarily achieve a better success probability. Instead, it works as follows: $P_{itj}$ simulates $P_{it}$, but then also checks whether the output of $P_{it}$ has succeeded on the first $j-1$ bits of the block. If any of those bits were predicted wrongly by $P_{it}$, then $P_{itj}$ just produces as output a uniformly random bit. Else, it produces as output the $j^{th}$ bit of $P_{it}$. In order to simulate this random guessing, we will actually define two predictors for each $i, t,$ and $j$: $P_{itj0}$ and $P_{itj1}$.

The following definition formally defines these predictors, and then Lemma~\ref{lem:pred-from-block-lossy} will prove that they indeed correspond to a valid reduction.

\begin{definition}[Predictors Corresponding to a $\blossyc$ Instance]
\label{def:pred-from-block-lossy}
Let $(C, D)$ be an $(n, \ell)$ $\blossyc$ instance, and let $k \coloneqq \ell-1-\lceil \log(n+1) \rceil$. We define
\[\mathcal{P}(C, D) \coloneqq \{P_{itjb} \mid i \in [n], t \in \{0, 1\}^k,j \in [\ell], b \in \{0, 1\}\}.\]
$P_{itjb}$ is a middle-bit predictor over $\{0, 1\}^{n\ell}$ at index $(i-1)\ell+j$. For $s=s_1\circ\dots\circ s_n$ with each $s_i\in\{0, 1\}^{\ell}$, let
\[d_{it}(s) \coloneqq D(i, t, (s_1, \dots, s_{i-1}, s_{i+1}, \dots, s_n))_i.\]
Then
\[P_{itjb}(s_{-((i-1)\ell+j)}) \coloneqq
\begin{cases}
d_{it}(s)_j, & \text{if } (s_i)_{<j} = d_{it}(s)_{<j},\\
b, & \text{otherwise.}
\end{cases}\]
\end{definition}

\begin{lemma}
\label{lem:pred-from-block-lossy}
Let $(C, D)$ be an $(n, \ell)$ $\blossyc$ instance, and let $\mathcal{P}(C, D)$ be the predictors from Definition~\ref{def:pred-from-block-lossy}. For any non-empty multiset $S \subseteq (\{0, 1\}^{\ell})^{n}$ such that
\[\adv_{U(S)}(\mathcal{P}(C, D)) < \frac{1}{2^{\ell+1}},\]
$S$ contains a solution to the $\blossyc$ instance $(C, D)$.
\end{lemma}
\begin{proof}
Let $k \coloneqq \ell-1-\lceil \log(n+1) \rceil$ and $\epsilon \coloneqq 1/2^{\ell+1}$.  Let $S \subseteq (\{0, 1\}^{\ell})^n$ such that
\[\adv_{U(S)}(\mathcal{P}(C, D)) < \epsilon.\]
Assume for the sake of contradiction that $S \subseteq \nonsol(C, D)$. Then, for every $s \in S$, the pair $i \coloneqq \ind_C(s)$ and $t \coloneqq \advice_C(s)$ satisfies
\[d_{it}(s) = D(C(s))_i = s_i.\]
By the strong pigeonhole principle, there must exist a fixed pair $i \in [n]$ and $t \in \{0, 1\}^k$ such that
\[\Pr_{s \sim U(S)}[d_{it}(s) = s_i] \geq \frac{1}{n2^k} > \frac{2}{2^\ell}.\]
For this pair, we define
\[\chi_j \coloneqq \Pr_{s \sim U(S)}[(s_i)_{\le j} = d_{it}(s)_{\le j}]\] for $j \in \{1, \dots, \ell\}$, and $\chi_0 \coloneqq 1$. The strong pigeonhole principle argument showed that \[\chi_{\ell} > \frac{2}{2^{\ell}}.\]
Now, fix any $j \in [\ell]$. Observe that the predictors $P_{itj0}$ and $P_{itj1}$ on a string $s \in (\{0, 1\}^{\ell})^n$ have two cases: if $d_{it}(s)_{<j} = (s_i)_{< j}$, then both predictors enter the first case of their definition and guess $d_{it}(s)_j$. They \emph{both} succeed exactly when $(s_i)_{\leq j} = d_{it}(s)_{\leq j}$. This happens with probability $\chi_j$. On the other hand, if $(s_i)_{< j} \neq d_{it}(s)_{< j}$, then the predictors enter the second case of their definition and guess $0$ and $1$ respectively, and exactly one of them succeeds. This happens with probability $(1 - \chi_{j-1})$. Therefore, for any $j \in [\ell]$,
\[\frac{\succc_{U(S)}(P_{itj0})+\succc_{U(S)}(P_{itj1})}{2}
=\chi_j+\frac{1-\chi_{j-1}}{2}<\frac{1}{2}+\epsilon.\]
The last inequality follows due to the assumption that $\adv_{U(S)}(\mathcal{P}(C, D)) < \epsilon$. Therefore, 
\[\chi_j \leq \frac{\chi_{j-1}}{2}+\epsilon.\]
Since $\chi_0 = 1$, this implies that
\[\chi_{\ell} \leq 2^{-\ell}+2\epsilon
=\frac{2}{2^\ell}.\]
This contradicts the lower bound on $\chi_\ell$. Thus, $S$ contains a solution to $(C, D)$.
\end{proof}

\begin{lemma}
\label{lem:block-lossy-to-unpred}
There exists an algorithm which takes as input $n, \ell \in \mathbb{N}$, along with an $(n, \ell)$ $\blossyc$ instance $(C, D)$ in circuit form. The algorithm uses $O(\ell + \log n + \log(\size (C)) + \log(\size (D)))$ work space. The algorithm produces as output, in circuit form, a set of $\leq \ell 2^{\ell}$ middle-bit predictors, $\mathcal{P}(C, D)$, over $\{0, 1\}^{n\ell}$, and $\epsilon \coloneqq 1/2^{\ell+1}$, such that:
\begin{enumerate}
    \item For each predictor $P \in \mathcal{P}(C, D)$, $\size(P) \leq \size(D) + O(\ell)$, and $\depth(P) \leq \depth(D) + O(\log\ell)$.
    \item For any non-empty multiset $S \subseteq \{0, 1\}^{n\ell}$ such that
    \[\adv_{U(S)}(\mathcal{P}(C, D)) < \epsilon,\]
    $S$ contains a solution to the $\blossyc$ instance $(C, D)$.
\end{enumerate}
\end{lemma}
\begin{proof}
Let $k \coloneqq \ell-1-\lceil \log(n+1) \rceil$. The set $\mathcal{P}(C, D)$ from Definition~\ref{def:pred-from-block-lossy} contains at most $\ell n2^{k+1} \leq \ell2^{\ell}$ predictors.

Each predictor $P_{itjb}$ contains the circuit $D$ with its first $\ell - 1$ input bits fixed (to $i \circ t$), and the output restricted to only $\ell$ of its $\ell n$ output bits, together with $O(\ell)$ gates of depth $O(\log\ell)$ to check the prefix and select the output. These circuits can be produced in the stated work space by enumerating $i, t, j, b$ and copying $D$. The unary encoding of $1/\epsilon = 2^{\ell+1}$ can also be produced using $O(\ell)$ work space. Thus, the first part of the lemma follows. The second part follows from Lemma~\ref{lem:pred-from-block-lossy}.

This completes the proof.
\end{proof}

Finally, Agarwala and Mertz~\cite{AgarwalaMertz25} implicitly prove the following reduction from $\bipmatch$ to $\blossyc$.

\begin{lemma}[Adapted from Theorem 6.3 of~\cite{AgarwalaMertz25}]
\label{lem:AM25}
There exists a logspace algorithm which takes as input an undirected bipartite graph $G = (V, E)$ with $n$ vertices and $m$ edges, and produces three circuits, each of which has $\size = \poly(n+m)$ and $\depth = O(\log(n+m))$:
\begin{enumerate}
    \item $C_G$ and $D_G$. This is an $(m, \ell = O(\log (n+m)))$ block lossy coding instance.
    \item $Sol_G$. This circuit takes as input any $s \in (\{0, 1\}^{\ell})^m$ which is a solution to the $\lossyc$ instance $(C_G, D_G)$, and produces as output a maximum matching of $G$.
\end{enumerate}
\end{lemma}
\begin{corollary}
\label{cor:bip-match-to-upd}
    \[\bipmatch \leq_{\LUTCo} \updc{\TCo}.\]
\end{corollary}
\begin{proof}
The reduction works as follows: it takes as input a bipartite graph $G$ with $n$ vertices and $m$ edges. First, it simulates the logspace algorithm from Lemma~\ref{lem:AM25} to produce the $(m, O(\log(n+m)))$ block lossy instance $C_G$, $D_G$, where both $C_G$ and $D_G$ have $\size = O(\poly(n+m))$ and $\depth = O(\log(n+m))$. Then, it uses the reduction from Lemma~\ref{lem:block-lossy-to-unpred} to obtain, using $O(\log(n+m))$ work space, a set of $\leq \poly(n, m)$ predictors $\mathcal{P}(C_G, D_G)$ and $\epsilon = 1/2^{\ell+1} \geq 1/\poly(n+m)$. Each predictor has $\size = O(\poly(n+m))$ and $\depth = O(\log(n+m))$. Together, these predictors define an $\updc{\TCo}$ instance.

Let $S$ be a polynomial-size solution to the $\updc{\TCo}$ instance. Then the reduction could iterate over all $s \in S$ in parallel and test whether $D_G(C_G(s)) = s$, i.e., whether $s \in \nonsol(C_G, D_G)$. For the first $s$ it finds such that $s \notin \nonsol(C_G, D_G)$ (by Lemma~\ref{lem:block-lossy-to-unpred}, such an $s$ must exist), it could feed this $s$ to the circuit $Sol_G$, obtained from Lemma~\ref{lem:AM25}, to obtain a maximum matching of $G$. 

This completes the reduction.
\end{proof}

The algorithm of Agarwala, Alekseev, and Vinciguerra~\cite{AgarwalaAlekseevVinciguerra26} can be reinterpreted as the exact same statement as Lemma~\ref{lem:AM25}, but for linear matroid intersection. Similarly, the algorithm of Chakraborty, Datta, Kusre, Mukhopadhyay, and Sinhababu~\cite{ChakrabortyDattaKusreMukhopadhyaySinhababu26} can be reinterpreted as the same statement as Lemma~\ref{lem:AM25} but for general matching. Unfortunately, they do not state this as explicitly as Agarwala and Mertz~\cite{AgarwalaMertz25}. It would obfuscate our contribution to recreate their algorithms here. Thus, we simply state the results we get by reinterpreting their algorithms as reductions to block lossy coding. The \textit{exact} same proof as Corollary~\ref{cor:bip-match-to-upd} then implies the following statements. 

\begin{corollary}[\cite{AgarwalaAlekseevVinciguerra26}]
\label{cor:lmi-to-upd}
    \[\lmi \leq_{\LUTCo} \updc{\TCo}.\]
\end{corollary}

\begin{corollary}[\cite{ChakrabortyDattaKusreMukhopadhyaySinhababu26}]
\label{cor:match-to-upd}
    \[\match \leq_{\LUTCo} \updc{\TCo}.\]
\end{corollary}

\section{Artificial Intelligence Disclosure}
\label{sec:ai-discussion}
This work was the result of approximately 16 months of research, much of which was spent before LLM models possessed the strength that they now do.

Except for Section~\ref{sec:CL-equiv}, LLMs played no role in this paper beyond proofreading, cleaning text, generating tables, and generating TikZ code after being given hand-drawn figures. Section~\ref{sec:CL-equiv} of this paper, in particular the reduction from the restricted version of $\lossyc$ to $\upd$, was obtained in collaboration with ChatGPT 5.6 Sol Extra High. However, this was extremely far from a one-shot prompt. 

To elaborate: at this point, the authors had already obtained every other result in this paper. In particular, we were able to prove that every problem \emph{currently known} to be in $\CL$ reduces to $\upd^{\TCo}$. However, there was exactly one exception: a restricted version of $\lossyc$, which was shown to be in $\SCL$ by Cook et al.~\cite{CookGhentiyalaMertzPyneSheffield25}. This was a far more restricted version of $\lossyc$ than the one we study in this work: one where both the compressor and decompressor must be computable by $O(\log n)$ depth, $n + O(\log n)$ width, and fan-in $2$ circuits. 

We first gave ChatGPT access to all of our proofs (in a Notion workspace), and asked it to prove that $\fct$ reduces to $\upd$. ChatGPT claimed to have proved this. Upon analyzing ChatGPT's proof, we realized that its proof was too strong: we believed that it could be extended to show that all of $\ZPP$ reduces to $\upd$. The authors, along with Edward Pyne, spent significant time understanding whether this was possible. This led us to uncover a minor mathematical error in a previous paper, which ChatGPT had used as a hammer in its incorrect proof. We have already contacted the authors responsible for that paper, and thus we will not name it publicly. 

We concluded that this problem was too hard for ChatGPT. Thus, we tempered our expectations and asked it to prove that the extremely restricted version of $\lossyc$ described above reduces to $\upd^{\TCo}$. Once again, we gave it access to all of the other proofs in this work. On this prompt, ChatGPT succeeded.

ChatGPT produced a proof that was extremely clever, yet opaque. We initially expected it to be erroneous, as its last attempt was. Thus, we spent a significant amount of time understanding and verifying this proof. In this process, we realized that if one abandoned the $\TCo$ requirement, then the proof that ChatGPT came up with relies primarily on the in-place computation of the decompressor. We realized that the decompressor of Cook et al.~\cite{CookLiMertzPyne25} also has this structure. From here, the authors were able to generalize ChatGPT's idea to obtain the full result of Section~\ref{sec:CL-equiv}. 

We would like to remark that we put significant effort into the exposition of these results. In particular, ChatGPT did not structure its proof as a reduction to $\weakex$, or even formally define a graph\footnote{It did occasionally use graph terminology, such as walks and flow, which was comprehensible around 10\% of the time.}. The reduction to $\weakex$ that we present in this work came as a result of several iterations of this paper that aimed to make the exposition as readable as possible.

The chat session described above consisted of a total of 36 prompts from us, and ChatGPT worked for a total of 35 hours, 33 minutes, and 15 seconds. We gave a link to the entire chat session, along with this AI disclosure section, to another ChatGPT 6.1 Sol Ultra session. It confirmed that this disclosure is a fair and accurate representation of events.

\ifblind
\section*{Acknowledgements}
\addcontentsline{toc}{section}{Acknowledgements}
Both authors thank Ian Mertz, James Cook, and in particular Edward (Ted) Pyne for extensive discussions. The first author thanks Danupon Nanongkai for extensive discussions and supervision. Both authors thank Julian Casablancas for his underrated album \textit{First Impressions of Earth}, without which this work would not have been possible.
\fi

\formatbibliography
\bibliographystyle{alphaurl}
\newpage

\phantomsection
\addcontentsline{toc}{section}{\refname}
\bibliography{bibliography}
\newpage 
\appendix
\section{Online Hypercube Graph Balancing Problem}
\label{sec:appendix-hypercube-balancing}

In this section, we prove the following lemma:
\lemhypercubebalancingalgo

Consider the following simple greedy algorithm: At each step, upon receiving a query edge $\{u, v\}$, the algorithm orients $\{u,v\}$
toward the endpoint which has lower in-degree in the directed graph induced by the edges queried so far, with ties broken arbitrarily.

For the rest of this section, fix an edge query sequence $(e_1, \dots, e_t)$. Let $G$ be the directed graph over vertex set $\{0, 1\}^n$ and edge set $\{e_1, \dots, e_t\}$, with directions as chosen by the greedy algorithm.

The following observation shows that any vertex in $G$ with high in-degree has several in-neighbours, each of which has reasonably high in-degree.
\begin{observation}
\label{obs:neighbour-counter-prop}
Let $G$ be as fixed above.
For every vertex $v \in V(G)$ and integer $j \in \mathbb{Z}_{\geq 0}$ such that $0 \leq j < \indeg_G(v)$, $v$ has at least $\indeg_G(v) - j$ distinct in-neighbours in $G$, each of which has in-degree $\geq j$ in $G$. 
\end{observation}
\begin{proof}
Consider the last $\indeg_G(v)-j$ edges that were oriented towards $v$ by the greedy algorithm. In the step when each of these edges was queried, the in-degree of $v$ must have already been at least $j$. The greedy algorithm always orients an edge towards the vertex with lower in-degree. Thus, the fact that the algorithm chose to orient these edges towards $v$ implies that the in-degree of the other endpoint of each of these edges must have been at least $j$. These are the in-neighbours of $v$ with in-degree at least $j$. This completes the proof.
\end{proof}

\begin{lemma}
\label{lem:hypercube-large-counter}
Let $G$ and $t$ be as fixed above.
For every vertex $v \in V(G)$, 
\[t \geq \frac{1}{2e^2} \exp\left(\sqrt{\frac{2 \cdot \indeg_G(v)}{3}}\right).\]
\end{lemma}

\begin{proof}
    Let $h \coloneqq \indeg_G(v)$. When $h \leq 1$, the lemma statement follows trivially from the fact that $t\ge1$. For the rest of this proof assume that $h \ge 2$. We define
    \[
    b \coloneqq\left\lfloor\sqrt{\frac{2h}{3}}\right\rfloor \qquad \text{and}
    \qquad
    R_d \coloneqq h-db-\binom{d}{2}
    \quad (\text{for }0\le d\le b).
    \]
    Since $h \geq 2$, we have $b\ge1$ and
    \[
    R_b=h-\frac{3b^2-b}{2}\ge\frac{b}{2}>0.
    \]

    For a vertex $u \in V(G)$, we define its \textit{level} $\in \{0, \dots, n\}$ to be its Hamming distance from $v$. Furthermore, we call a vertex $u$ at level $d \in \{0, \dots, b\}$ \textit{good} if $\indeg_G(u) \geq R_d$. Observe that the vertex $v$ is good at level zero.

    Let $0 \leq d < b$, and $u \in V(G)$ be a good vertex at level $d$. Due to Observation~\ref{obs:neighbour-counter-prop} with $j \coloneqq R_{d+1}$, $u$ must have at least $R_d - R_{d+1} = b+d$ in-neighbours $u'$ with $\indeg_G(u') \geq R_{d+1}.$ Since the Hamming distance of $u$ from $v$ is exactly $d$, at most $d$ of these vertices can be at level $d-1$. The remaining $\geq b$ of these in-neighbours must be at level $d+1$.

    Consequently, there are at least $b^b$ paths in $G$, all of which end at the vertex $v$, have length exactly $b$, decrease their Hamming distance from $v$ at each step, and visit only good vertices. Any vertex of $G$ can be the source of at most $b!$ such paths. We conclude that there are at least $b^b/b!$ distinct good vertices at level $b$. Therefore,
    \[
    t = \sum_{v \in V(G)} \indeg_G(v) \ge R_b\frac{b^b}{b!}.
    \]
    The following upper bound on $\ln(b!)$
    \[
    \ln(b!)
    \le \int_1^b\ln x\,dx+\ln b
    =b\ln b-b+1+\ln b
    \]
    implies that
    \[t \geq R_b \frac{e^{b \ln b}}{e^{b \ln b - b + 1 + \ln b}} = R_b \frac{e^{b-1}}{b}.\]
    Together with $R_b\ge b/2$, this implies that
    \[
    t
    \ge \frac b2\cdot\frac{e^{b-1}}b
    =\frac{1}{2}e^{b-1}
    \ge
    \frac{1}{2e^2}
    \exp\!\left(\sqrt{\frac{2h}{3}}\right).
    \]
    This completes the proof.
\end{proof}

\noindent
Lemma~\ref{lem:hypercube-large-counter} shows that the greedy algorithm has in-degree bound 
\[f(h) \coloneqq \frac{1}{2e^2} \exp\left(\sqrt{\frac{2h}{3}} \right).\] Clearly, $f(h) = \exp(\Omega(\sqrt{h}))$. This proves Lemma~\ref{lem:hypercube-balancing-algo}.
\section{Online Prefix-Tree Balancing Problem}
\label{sec:appendix-online-prefix-tree-balancing}
In this section, we prove the following lemma:
\lemtreebalancingalgo

Consider the following simple greedy algorithm: At step $i$, let $B_i = \{v_1, \dots, v_k\}$ be the queried batch with $|v_1| \geq |v_2| \geq \dots \geq |v_k|$. The algorithm iterates $j$ from $1$ to $k$. The algorithm places the token of $v_j$ on the child whose subtree contains fewer tokens placed \textbf{during the current batch}, with ties broken arbitrarily.

For the rest of this section, fix a query sequence $(B_1, \dots, B_t)$.

\begin{observation}
\label{obs:batch-lower-bound}
Let $T_n$, $(B_1, \dots, B_t)$, and the greedy algorithm be as above.
    Let $v \in \{0, 1\}^n$ be a leaf vertex of $T_n$, and $B_i$ be a batch in which the greedy algorithm places $h$ tokens on the path from the root $\emptyset$ to the leaf $v$. Then $|B_i| \geq 2^h - 1$.
\end{observation}
\begin{proof}
    We prove this via induction on $h$. First, for $h \leq 1$, the statement follows from the fact that $|B_i| \geq h$. We will now prove this for $h > 1$. Assume the inductive hypothesis that this statement is true for all $h' < h$. 

    Let $B_i = \{v_1, \dots, v_k\}$ with $|v_1| \geq |v_2| \geq \dots \geq |v_k|$. Let $v_j \in B_i$ be the $h^{th}$ vertex whose token the greedy algorithm places on the path from the root to $v$. Let's say that this token is placed on vertex $v_j \circ b$. Let $B'_i$ denote the vertices in $\{v_1, \dots, v_{j-1}\}$ which are contained in the subtree of $v_j \circ b$. Observe that vertices outside $B'_i$ do not affect the placement of the tokens of vertices in $B'_i$. Exactly $h-1$ tokens from $B'_i$ are placed by the algorithm on the path from the root to $v$. Thus, by induction, we have $|B'_i| \geq 2^{h-1} - 1$. 

    Now, the greedy algorithm always places the token on the subtree with \textit{fewer} tokens already placed on it. The fact that the algorithm chose to place the token on $v_j \circ b$ instead of $v_j \circ (\lnot b)$ implies that the subtree of $v_j \circ (\lnot b)$ already had at least $2^{h-1} - 1$ tokens placed on it. All of these tokens must come from queries in $\{v_1, \dots, v_{j-1}\} \setminus B'_i$. Thus, we have
    \[|\{v_1, \dots, v_{j-1}\}| \geq |\{v_1, \dots, v_{j-1}\} \setminus B'_i| + |B'_i| \geq 2^h - 2.\]
    Therefore, 
    \[|B_i| \geq |\{v_1, \dots, v_j\}| = |\{v_1, \dots, v_{j-1}\}| + 1 \geq 2^h - 1.\]
    This completes the proof.
\end{proof}

\begin{proof}[Proof of Lemma~\ref{lem:tree-balancing-algo}]
Simply observe that for any $v \in \{0, 1\}^n$, there must exist a batch $B_i$ in which the algorithm places $h \geq \load(v)/t$ many tokens on the path from the root to the leaf $v$. Observation~\ref{obs:batch-lower-bound} implies that 
\[|B_i| \geq 2^{h} - 1.\] 
This implies that
\[t \lfloor \log(|B_i| + 1) \rfloor \geq ht \geq \load(v).\]
Since for every vertex $v$, there exists a batch $B_i$ for which this equation holds, taking the $\max$ over all $B_i$ and over all vertices $v$ proves the lemma.
\end{proof}
\section{\texorpdfstring{A Uniform Version of $\upd$}{A Uniform Version of UNPRED}}
\label{sec:uniform-unpred}
Let $L$ be a fixed language. We define a version of $\upd$ known as $\updc{L}$ as follows.

\begin{definition}[$\updc{L}$]
$\updc{L}$ is the search problem defined as follows:
\begin{enumerate}
    \item \textbf{Input:} $n, m \in \mathbb{N}$, $0 < \epsilon \leq 1$, strings $x_1, \dots, x_m$, and $\ind_1, \dots, \ind_m \in [n]$.
    
    \noindent
    For all $i \in [m]$, $P_i:\{0, 1\}^{n-1} \rightarrow \{0, 1\}$ is a middle-bit predictor over $\{0, 1\}^n$ at index $\ind_i \in [n]$. The value of a predictor $P_i$ on a string $s \in \{0, 1\}^{n-1}$ is defined as $P_i(s) \coloneqq L(x_i, s)$.
    \item \textbf{Output:} A non-empty multiset $S \subseteq \{0, 1\}^n$ such that
    \[\adv_{U(S)}(\{P_1, \dots, P_m\}) < \epsilon.\]
\end{enumerate}
\end{definition}

In this definition, we think of the strings $x_1, \dots, x_m$ as advice. If $M$ is a Turing machine which decides the language $L$, then in order to obtain the value of a predictor $P_i$ for $i \in [m]$ on a string $s$, one would evaluate $M$ on $s$ with advice $x_i$. If we fix $L$ to be the Boolean circuit evaluation problem, and interpret the advice strings $x_1, \dots, x_m$ as circuit descriptions, then we recover the original definition of $\upd$.

Now, we define a model of computation which is similar to $\ipFL$, but allows for read-only access to an advice string.

\begin{definition}[Restorable Partial In-Place Computation]
A restorable partial in-place Turing machine $M$ consists of two procedures: $\compute_M$ and $\restore_M$, which share a read-only input tape, a read-write input tape, and a read-write work tape. $M$ computes $R \subseteq \{0, 1\}^* \times \{0, 1\}^* \times \{0, 1\}^*$ if it satisfies the following two properties when executed on a read-only input $x$ and read-write input $y$: 1) $\compute_M(x, y)$ halts with $z \in \{0, 1\}^*$ such that $(x, y, z) \in R$ written on the first $|z|$ bits of the read-write input tape; and 2) $\restore_M$ starts from the read-write input tape and work tape configuration that $\compute_M(x, y)$ halted with, and halts with $y$ written on the read-write input tape.
\end{definition}

We do not prove the following facts here, but we have done so in a previous draft, and we would be happy to share that draft with anyone interested. 

\begin{theorem}
For any language $L$ such that $L$ can be computed by a restorable partial in-place Turing machine with logarithmic work space, we have
\[\updc{L} \in \SCL.\]
\end{theorem}
Let $M$ be such a machine. This follows directly from Theorem~\ref{thm:oracle-binunpred-in-searchcl}, by replacing each oracle query with an explicit evaluation of $M$. We use the part of the input tape of the catalytic machine which has $x_i$ written on it as the read-only input tape of $M$, and use the catalytic tape which has the query string $s$ written on it as the read-write input tape of $M$. We use fresh bits on the work tape of the catalytic machine as the work tape for $M$. For each query, we run $\compute_M(x_i,s)$, copy the output bit of $M$ to the work tape, and then run $\restore_M$ to restore the catalytic tape.

Finally, the same argument as in Section~\ref{sec:CL-equiv}, with more tedious proofs, shows the following theorem.

\begin{theorem}
There exists a language $L$ that can be computed by a polynomial-time restorable partial in-place Turing machine with logarithmic work space such that
\[\SCL \leq_{\P} \upd^{L}.\]
\end{theorem}

These theorems together close the gap between $\upd^{\TCo}$ and $\upd$ which exists in our simplified presentation.
\end{document}